# Nonlinear Modeling of MEMS Fixed-Fixed Beams

by

**Xi Luo**

Presented to the Graduate and Research Committee
of Lehigh University
in Candidacy for the degree
of Doctor of Philosophy

in

Electrical Engineering

Lehigh University
May, 2016

Approved and recommended for acceptance as a dissertation in partial fulfillment of the requirements for the degree of Doctor of Philosophy.

**Date**

**Dr. James C. M. Hwang**, Dissertation Advisor, Chair

**Accepted Date**

**Committee Members:**

**Dr. Douglas Frey**

**Dr. Svetlana Tatic-Lucic**

**Dr. Herman F. Nied**

**Dr. Richard P. Vinci**

// Acknowledgments

I greatly appreciate my advisor Professor James C. M. Hwang's guidance, support, and encouragement through my Ph.D. study. He provided me a wide range of advanced research topics and also trained me to be an independent researcher by working on these topics. I also would like to thank Professor Herman F. Nied for his valuable help and suggestions on my dissertation. Moreover, I would like to thank Professor Douglas Frey, Professor Svetlana Tatic-Lucic, and Professor Richard P. Vinci for their help and support.

I am grateful to my former and current colleagues at Compound Semiconductor Technology Laboratory (CSTL), particularly, Dr. Subrata Halder, Dr. David Molinero, and Dr. Cristiano Palego who provided valuable contribution to my growth in research. I am also thankful to the help from Dr. Weike Wang, Dr. Laura Jin, Dr. Yaqing Ning, Vahid Gholizadeh, Mohammad Asadi, Xiao Ma, Zhibo Cao and Kevin Xiong, who have made my graduate research at CSTL an enjoyable experience. I would also like to express my gratitude to Dr. Charles Goldsmith at MEMtronics Corp. for providing precious device samples and helpful discussions.

I owe my deepest gratitude to my family members, my wife Jin Wang, my parents and parents-in-law. Without your support and sacrifice, I cannot go this far.

# Table of Contents

# List of Figures

# Abstract

This dissertation studies critical topics associated with MEMS fixed-fixed beams. One of the typical devices of fixed-fixed beams is radio frequency microelectromechanical system (MEMS) capacitive switches. The interesting topic for this device includes the instability at the pull-in voltage; the switches' deformation characteristics when subject to an electrostatic force; nonlinear stretching effects, and the capacitance calculation in small scale. Specifically, the accuracy of parallel-plate theory in calculating the pull-in voltage and capacitance is investigated. The study shows that applying average displacement rather than maximum displacement into parallel-plate theory demonstrates better accuracy. The improvement increases with the bottom stationary electrode to moveable electrode ratio and it reaches 50% when the ratio is equal to 1. Besides average displacement, the nonlinear stretching effect and empirical linear correction coefficients are also added to the parallel-plate model to extend model's validity range. In order to improve the lifetime of RF MEMS capacitive switch, a relationship between switches' geometry and membrane strain is derived, which helps avoid switches operating beyond the elastic region.

Furthermore, this dissertation presents a new coupled hyperbolic electro-mechanical model that is an improvement on the classical parallel-plate approximation. The model employs a hyperbolic function to account for the beam's deformed shape and electrostatic field. Based on this, the model accurately calculates the deflection of a fixed-fixed beam subjected to an applied voltage and the switch's capacitance-voltage characteristics without using parallel-plate assumption. For model validation, the model solutions are compared with ANSYS finite element results and experimental data. It is found that the model works especially well in residual stress dominant and stretching dominant cases. The model shows that the nonlinear stretching significantly increases the pull-in voltage and extend the beam's maximum travel range. Based on the model, a graphene nanoelectromechanical systems (NEMS) resonator is designed and the performance agrees very well with the experimental data. The proposed coupled hyperbolic model demonstrates its capacity to guide the design and optimization of both RF MEMS capacitive switches and NEMS devices.

# Chapter 1 Introduction

One of the typical device of fixed-fixed beams is radio frequency microelectromechanical system (MEMS) capacitive switches. Regarding the switches, solid state switches (PIN diodes, field-effect transistor (FET)), coaxial electro-mechanical (EM) switches, and radio frequency (RF) micro-electro-mechanical systems (MEMS) switches are used extensively in microwave systems for signal routing between instruments and devices under test (DUT). Compared with conventional solid state switches, coaxial EM switches and RF MEMS switches demonstrate superior performance on insertion loss, isolation, linearity, return loss, and Electrostatic discharge (ESD) immunity [1]. In addition, RF MEMS switches are often much smaller in size than coaxial EM switches, which satisfies the demands of integration with other RF components.

The RF MEMS switches show significant improvement in ultra-low insertion loss, low DC consumption and high linearity. The ultra-low insertion loss of RF MEMS switches makes routing of RF signals possible with much lower loss, giving RF systems better noise figure and sensitivity. As most MEMS devices are electrostatically operated, they consume essentially no DC power, which makes them an excellent candidate for battery or hand-held devices, as well as satellite and space systems. The high linearity is beneficial to broadband communications systems and systems where the high dynamic range is required [2]. This dissertation focuses mainly on one RF MEMS capacitive switch, but the conclusion can be applied to other devices with fixed-fixed beams.

## *1.1 RF MEMS Capacitive Switches Background*

Critical topics associated with RF MEMS capacitive switches include the instability at the pull-in voltage; the switches' deform characteristics when subject to an electrostatic force; and the capacitance calculation in small scale. Important physical details include the air damping effects, device reliability, and failure mechanism [3]. In order to effectively investigate the complex electromechanical interactions associated with MEMS devices, it is necessary to use advanced analysis methods. This includes finite element method (FEM), and finite difference method (FDM). The multidisciplinary coupling effects and the nonlinearity of the structure and electrostatic forces make accurate modeling of electrostatically actuated microstructures challenging.

## 1.2 Pull-in Voltage Calculation

The Fig. 1-1 (a) and (b) show a top-view and cross-sectional view of a micro-encapsulated RF MEMS capacitive switches from MEMtronix Corp. An electrostatically actuated MEMS switch generally consists of a movable conductor electrode suspended above a stationary conductor electrode. The applied voltage between a movable and a stationary electrode has an upper limit beyond which the electrostatic force is not balanced by the restoring force. When the movable electrode is imbalanced and snaps down to stationary electrode, this phenomenon is called pull-in instability [4].The instability at pull-in is important for many MEMS applications. For micro-mirrors and micro-resonators, the instability is not desirable. While for switching applications, the effects are exploited to obtain optimum performance [4]. In addition, the accurate prediction of pull-in voltage for a fixed-fixed beam is critical in designing the sensitivity, frequency response and dynamic range of the devices [5]. Due to its multidisciplinary nature and nonlinear electrostatic forces, there is no trivial solution for calculating the pull-in voltage. Various closed-form expressions are proposed to calculate the pull-in voltage for a fixed-fixed beam based on specific assumptions and simplifications. For example, an expression was developed based on parallel-plate assumption [6], which assumed the beam had a linear spring constant, the beam deflection was the same across the entire beam length and the electrostatic force was uniformly distributed on the beam. The nonlinear stretching effects were also neglected. This model predicted that the beam collapses to the stationary electrode when the maximum deflection reached one third of the air-gap height, which agreed well with [7]. In another simplified lumped mass-spring model of the fixed-fixed beam [8], the pull-in voltage was determined when the fundamental frequency of the system drops to zero. In [9], the energy minimum principle of the parallel-plate capacitor was used to determine the pull-in voltage of the structure. The pull-in voltage was determined when the second derivative of total potential energy equaled to zero. Some of the electromechanical effects commonly ignored, such as fringe effects, plane-strain effects and anchor compliance, were considered in [10]. It added the effective width as a first order compensation for the electrostatic field fringe effects and plane-strain effects. Numerical compliance factors were included for non-ideal fixed-fixed boundary conditions. In [11], a closed–form expression for the pull-in voltage of fixed-fixed beams and fixed-free beams is derived. Also, the effects of partial electrode configuration, of axial stress, stretching effects, and fringing fields were considered in a simple lumped spring-mass system.

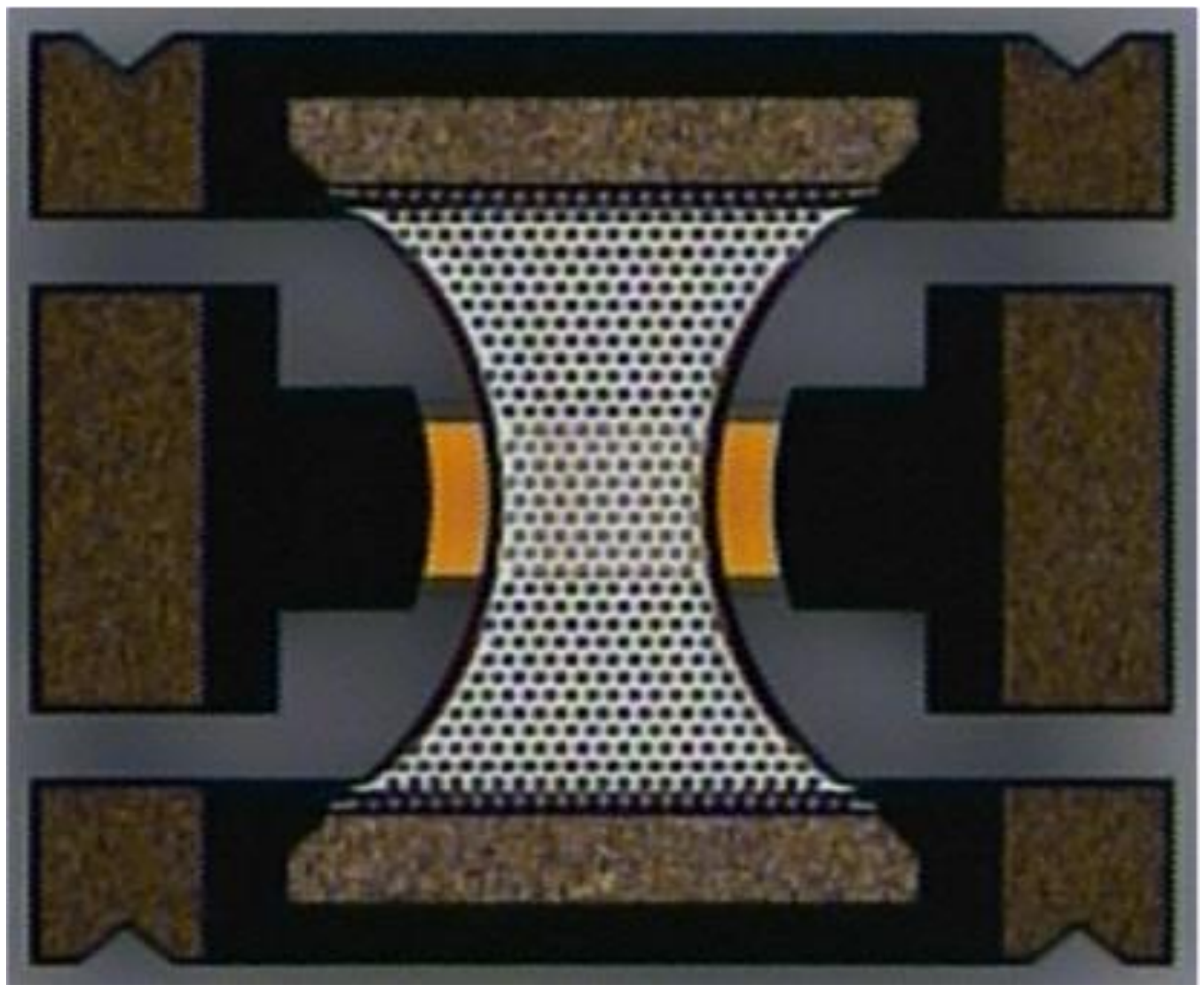

(a)

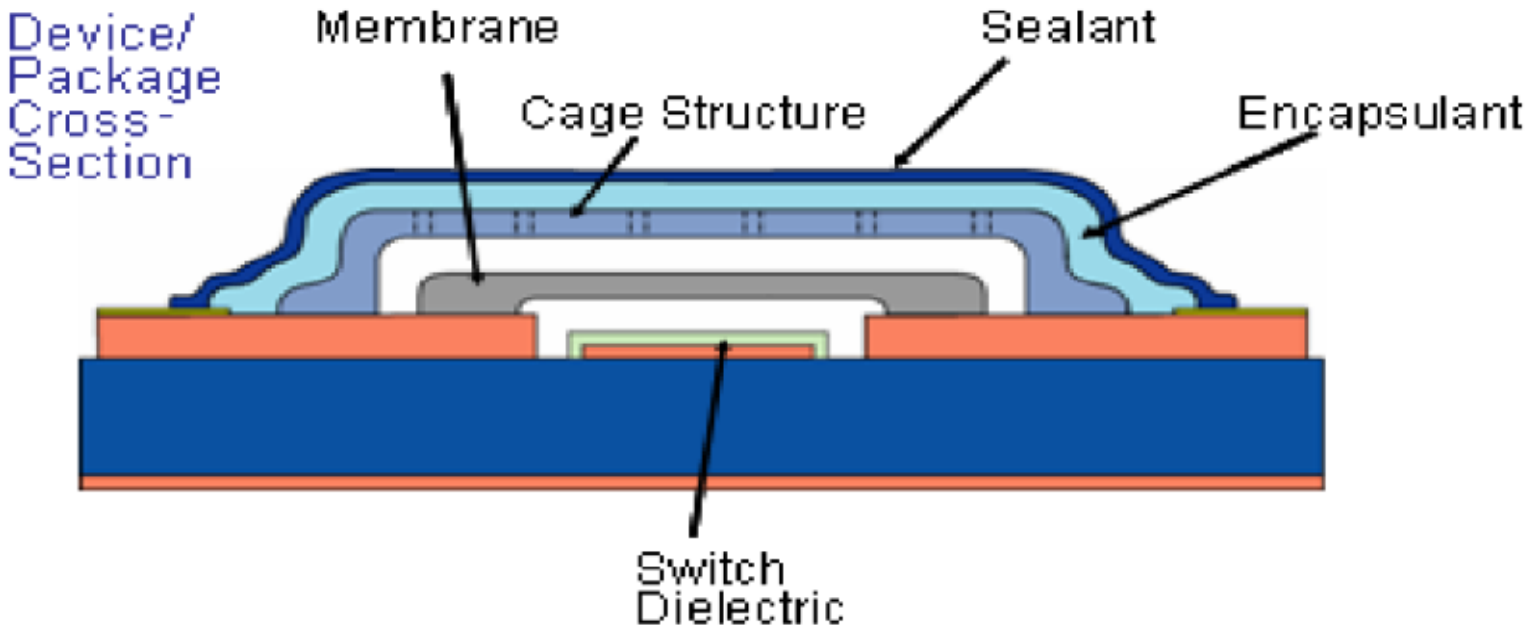

(b)

Fig. 1-1 (a) Top-view and (b) cross-sectional view of a microencapsulated RF MEMS capacitive switch [2].

The approaches used in [6]–[11] are compared with 3-D electromechanical finite element analysis (FEA) and a parametric behavioral model in [12]. It found out that the accuracy of the presented model varied widely depending on the device specifications and modeling parameters. For wide beams, which means beam width $w > 5t$ ($t$ is beam thickness), in the small deflection regime ($t$ > air gap $g_0$), where the fringe field effects and the stress induced stretching is neglected, the performance of all four methods agreed well with FEA results. The maximum 2.6% deviation from CoSolve FEA results was observed [12]. For narrow beams ($w < 5t$), in the small deflection regime ($t > g_0$), the maximum deviation from FEA simulation results was about 20%, which was primarily due to fringe capacitance. For wide beams ($w > 5t$), in the large deflection regime ($t < g_0$), only the approach in [11] shows a small 10% deviation. On the other hand, the pull-in voltage predicted by other four approaches in [6]–[10] gave only one-fourth the values when compared with FEA results. It appeared that the proper modeling of the fringe field and the stretching effects are the key factors to improve the accuracy of a closed-form solution.

Although the fringe field and the stretching effects were considered in the pull-in voltage expression [11], the relationship between the maximum deflection of the beam before pull-in and the nonlinear stretch factor was not explored, which was critical for determining the pull-in voltage accurately. That relationship will be studied in detail in section 2.2.3 of this dissertation.

### *1.3 Nonlinear Stretching Effect*

The pull-in instability limits the travel distance of elastically suspended parallel-plate electrostatic capacitor to about 1/3 of the gap height. In order to extend the travel range before pull-in instability occurs, different approaches were proposed.

One approach was to employ leveraged bending and strain-stiffening (stretching) methods by optimizing the switches' stationary electrode and structural design [13]. It was reported that the leveraged bending effect could be used to achieve full gap travel at the cost of increased actuation voltage. The strain-stiffening effect could be used to achieve a stable travel distance up to about 3/5 of the gap [14].

Another approach employed a series capacitor to provide stabilizing negative feedback and charge control techniques to extend the travel range of a movable electrode. The parasitic and tilting instabilities limit the actuation range [15]. This approach was further improved by using a switched-capacitor configuration [16]–[18]. Through charge control techniques by using current pulses injecting the required amount of charge, the displacement beyond the pull-in point is achieved [18].

The relationship between the stretching effects and maximum travel range will be discussed in section 2.2.3 and 3.1.4, and the optimum design of switches geometry is also suggested in section 2.2.4.

## *1.4 Small Length Scale Effect*

For relatively small gap height to the movable electrode length ratio, the two electrodes could be considered to be locally parallel to each other [4]. This is justified by the small gap height to beam length ratio (air-gap height $g_0$ to movable electrode length $\ell$ around $10^{-2}$–$10^{-3}$) [19]. However, the advancement in fabrication technologies and materials leads to the reduction in the size of electrostatically actuated MEMS, so the $g_0/\ell$ ratio cannot be considered small (on the order of $10^{-1}$–$10^{-2}$) [20] or even larger [21]. For larger $g_0/\ell$ ratio, more accurate estimates could be developed by considering the slope and the curvature of the movable electrode [19],[22]. An improved second order approximation was suggested based on the representation of the electrode surface locally as a cylindrical surface [19], which improved significantly the quality of the approximation and was applicable in cases when the use of the parallel capacitor formula leads to an error.

This dissertation will discuss the approach of representing the movable electrode deformation using a hyperbolic function in section 3.1.2 and calculating the capacitance accordingly. In addition, the fringe effect will be included using inverse cosine conformal mapping techniques in section 3.1.2.

## *1.5 Organization of the Dissertation*

After introducing RF MEMS capacitive switches and associated significant topics in Chapter 1, Chapter 2 discusses the validity of applying parallel-plate model to address these issues, such as calculating pull-in voltage, capacitance, and predict switches' deformation characteristics as a function of an electrostatic force. Chapter 3 proposed a new hyperbolic model to address similar issues without applying parallel-plate theory. The hyperbolic model is validated with experimental data in Chapter 4. Chapter 5 concludes the dissertation and gives suggestions for future study.

## *References*

[1] Agilent Solid State Switches Application Note [Online]. Avaliable: https://www.agilent.com

[2] RF MEMS Switches [Online]. Avaliable: http://www.memtronics.com

[3] Wan-Chun Chuang, Hsin-Li Lee, Pei-Zen Chang and Yuh-Chung Hu, "Review on the modeling of electrostatic MEMS," *Sensors*, vol. 10, no. 6, pp. 6149–6171, Jun. 2010.

[4] R. C. Batra, M. Porfiri, and D. Spinello, "Review of modeling electrostatically actuated microelectromechanical systems," *Smart Mater*. *Struct*., vol. 16, no. 6, pp. 23–31, Oct. 2007.

[5] R. Puers, and D. Lapadatu, "Electrostatic forces and their effects on capacitive mechanical sensors," Sensors and Actuators A: Physical, vol. 56, no. 3, pp. 203–210, Mar. 1996.

[6] P. Osterberg, H. Yie, X. Cai, J. White, and S. Senturia, "Self-consistent simulation and modelling of electrostatically deformed diaphragms," in *IEEE Workshop on Micro Electro Mechanical Systems*, Jan. 1994, pp. 28–32.

[7] H. C. Nathanson, W. E. Newell, R. A. Wickstrom, and J. R. Davis, "The resonant gate transistor," *IEEE Trans. Electron Devices*, vol. 14,no. 3, pp. 117–133, Mar. 1967.

[8] D. J. Ijntema, and H. A.C. Tilmans, "Static and dynamic aspects of an air-gap capacitor," *Sensors and Actuators A: Physical*, vol. 35, no. 2, pp. 121–128, Dec. 1992.

[9] H. A.C. Tilmans, and R. Legtenberg, "Electrostatically driven vacuum-encapsulated polysilicon resonators: Part II. Theory and performance," *Sensors and Actuators A: Physical*, vol. 45, no. 1, pp. 67–84, Jun. 1994.

[10] C. O'Mahony, M. Hill, R. Duane, and A Mathewson, "Analysis of electromechanical boundary effects on the pull-in of micromachined fixed–fixed beams," *J. Micromech. Microeng*., vol. 13, no. 4, pp. 75–80, Jun. 2003.

[11] S. Pamidighantam, R. Puers, K. Baert, and H. A. C. Tilmans, "Pull-in voltage analysis of electrostatically actuated beam structures with fixed–fixed and fixed–free end conditions," *J. Micromech. Microeng*., vol. 12, no. 4, pp. 458–464, Jul. 2002.

[12] S. Chowdhury, M. Ahmadi, and W. C. Miller, "A comparison of pull-in voltage calculation methods for MEMS-based electrostatic actuator design," in *Proc. 1st Int. Conf. Sensing Technology*, Nov. 2005, pp. 112–117.

[13] E. S. Hung and S. D. Senturia, "Extending the travel range of analog-tuned electrostatic actuators," *J. Microelectromech. Syst.*, vol. 8, no. 4, pp. 497–505, Dec. 1999.

[14] Y. Nemirovsky and O. Bochobza-Degani, "A methodology and model for the pull-in parameters of electrostatic actuators," *J. Microelectromech. Syst.*, vol. 10, no. 4, pp. 601–615, Dec. 2001.

[15] E. K. Chan and R. W. Dutton, "Electrostatic micromechanical actuator with extended range of travel," *J. Microelectromech. Syst.*, vol. 9, no. 3, pp. 321–328, Sep. 2000.

[16] J. I. Seeger and B. E. Boser, “Dynamics and control of parallel-plate actuators beyond the electrostatic instability,” in *Proc. 10th Int. Conf. Solid-State Sensors and Actuators*, Jun. 1999, pp. 474–477.

[17] J. I. Seeger and B. E. Boser, "Charge control of parallel-plate, electrostatic actuators and the tip-in instability," *J. Microelectromech. Syst.*, vol. 12, no. 5, pp. 656–671, 2003.

[18] R. Nadal-Guardia, A. Dehe, R. Aigner, and L. M. Castaner, "Current drive methods to extend the range of travel of electrostatic microactuators beyond the voltage pull-in point," *J. Microelectromech. Syst.*, vol. 11, no. 3, pp. 255–263, Jun. 2002.

[19] Slava Krylov and Shimon Seretensky, "Higher order correction of electrostatic pressure and its influence on the pull-in behavior of microstructures," *J. Micromech. Microeng.*, vol. 16, no. 7, pp. 1382–1396, Jun. 2006.

[20] J. Teva, G. Abadal, Z.J. Davis, J. Verd, X. Borrisé, A. Boisen, F. Pérez-Murano, N. Barniol, "On the electromechanical modelling of a resonating nano-cantilever-based transducer," *Ultramicroscopy*, vol. 100, no.3, pp. 225–232, Aug. 2004.

[21] V. Sazonova, Y. Yaish, H. Üstünel, D. Roundy, T. A. Arias and P. L. McEuen, "A tunable carbon nanotube electromechanical oscillator," *Nature*, vol. 431, pp. 284–287, Jul. 2004.

[22] J. A. Pelesko, T. A. Driscoll. "The effect of the small-aspect-ratio approximation on canonical electrostatic MEMS models," *J. Eng. Mathematics*, vol. 53, no. 3–4, pp. 239–252, Dec. 2005.

[23] E. Barke, “Line-to-ground capacitance calculation for VLSI: A comparison,” *IEEE Trans. Comput.-Aided Des. Integr. Circuits Syst.*, vol. 7, no. 2, pp. 295–298, Feb. 1988.

# Chapter 2 Theory and Parallel-plate Models

The important effects that influence the capacitance-voltage (*C*-*V*) correlation are studied in this chapter by employing both analytical and computational approaches. The first effect examined the influence of the stationary electrode to movable electrode length ratio on the conventional parallel-plate theory. The effect of bending, residual stress, and membrane stretch are also studied. All the factors above are critical for understanding the nonlinear capacitance in the suspended and actuated state. In addition to analytical derivations based on various simplifying assumptions, the same geometric configurations are also simulated using ANSYS finite element software. Comparisons between the analytical solutions and computational results quantify the range of validity for the assumed analytical approximations.

## *2.1 Comparison of Analytical and Computational Approaches*

The analytical solution is always the preferred approach because it provides correlation of variables explicitly. However, for the specific problem of a fixed-fixed beam that deforms with applied voltage, a closed-formed solution cannot be achieved due to the coupling between the electrostatic domain and mechanical domain. This is not unique to this particular problem, i.e., closed-form solutions for the electrostatic problems are only available for a limited number of simple prescribed geometries. Therefore, a numerical approach is required [1]–[2]. In this study, ANSYS finite element software will be used to generate the bulk of the numerical solutions.

Two computational methods are available within ANSYS to solve this coupled problem: 1) the Direct Coupling method and 2) the Multi-Field Solver method [3]. Of course, the converged solutions from the two methods should be identical and those results in turn should be compared with available analytical solutions in limiting cases. Once all convergences checks and solutions have been verified, the most efficient solution methodology will be employed to simulate specific geometric configurations of interest.

Consider a fixed-fixed aluminum beam plate capacitor with the top movable electrode length $\ell$ = 300 μm, width $W \rightarrow \infty$, thickness $t$ = 0.6 μm, suspends $g_0$ = 3 μm above the bottom stationary electrode of length $\ell'$ = 100 μm (refer to Fig. 2-2). Assume the stationary electrode thickness is zero and the tensile residual stress ($\sigma$) in the movable beam materials, which is generated during the deposition process, is 50 MPa. This configuration is simulated using the Direct Coupling method and Multi-Field Solver method. The results of both methods on maximum vertical displacement at beam center normalized by gap height ratio $\left(\frac{\Delta z_{MAX}}{g_0}\right)$ and the capacitance normalized by parallel-plate capacitance of overlap region ( $C_0 = \frac{\varepsilon_0 W \ell'}{g_0}$, $\varepsilon_0$ is vacuum permittivity) are shown in Fig. 2-1(a), (b). The bias voltage is normalized by the pull-in voltage. The two methods show maximum 5% difference in displacement and maximum 1% difference in capacitance simulation. Considering it is better solved in a single solution using a coupled formulation when the coupled-field involves strongly coupled-physics [1], the Direct Coupling method will be employed as the computational approach in this thesis.

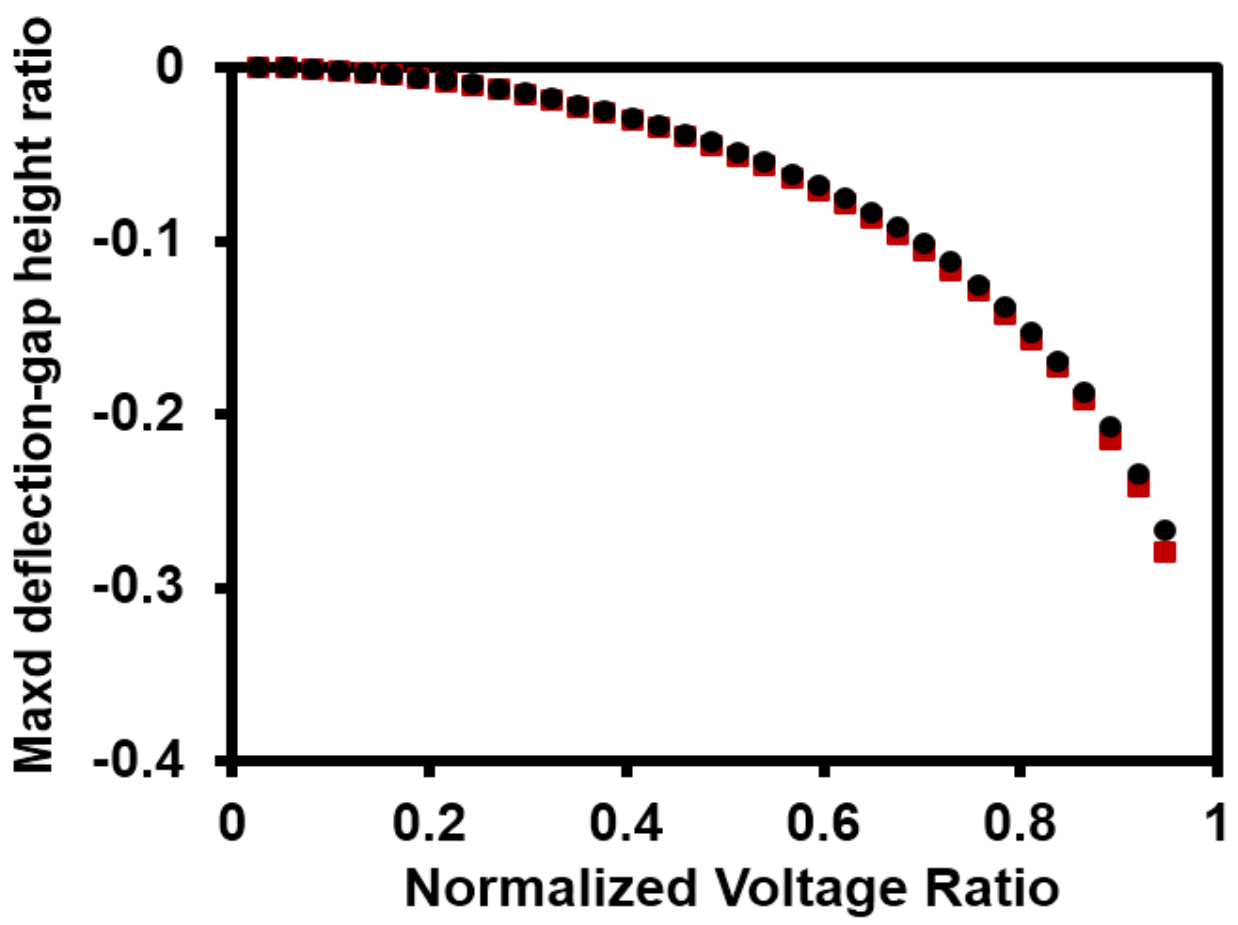

(a)

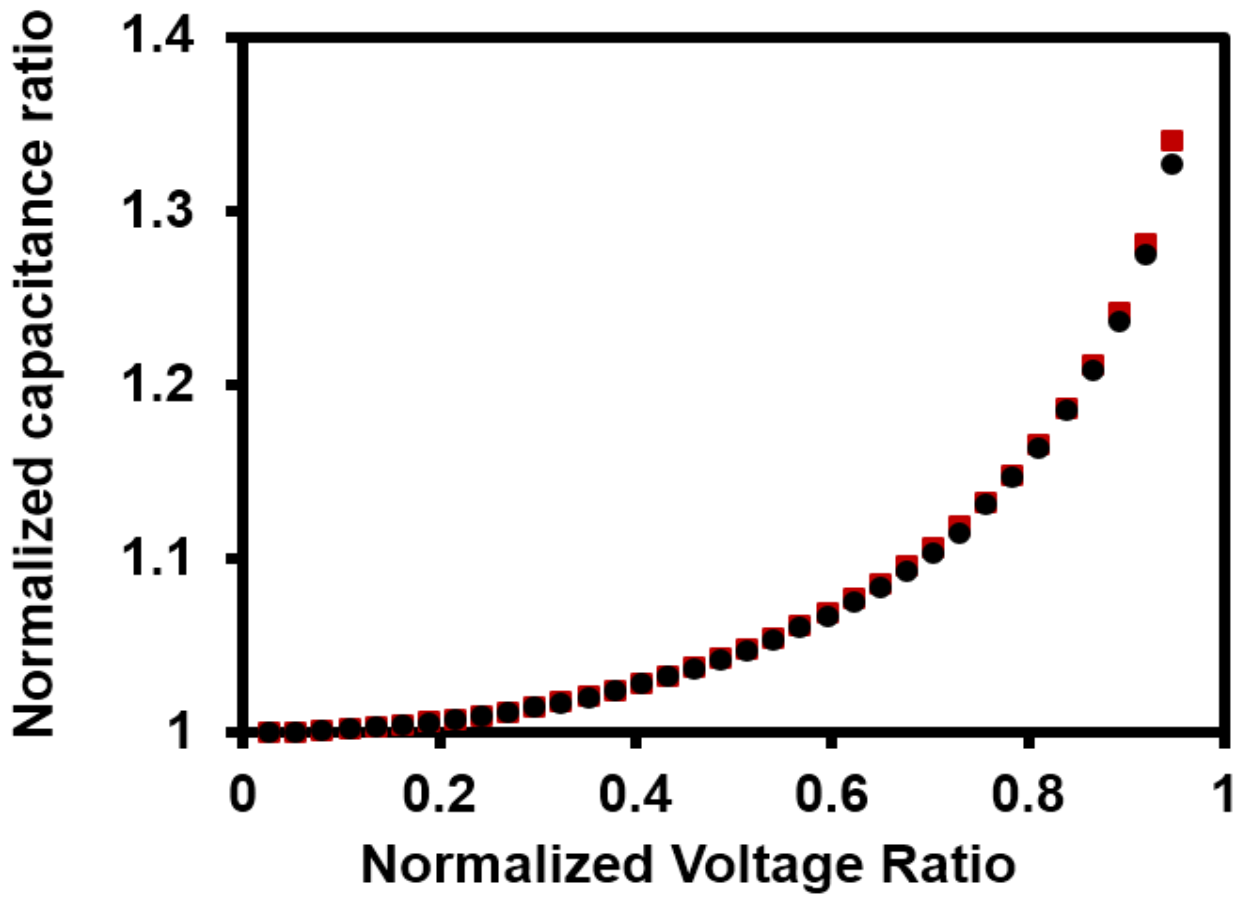

(b)

Fig. 2-1. ANSYS simulation results using Direct Coupling method (■) and Multi-Field Solver method (●). (a) Normalized max deflection as a function of normalized voltage. (b) Normalized capacitance as a function of normalized voltage.

The parallel-plate capacitor assumption neglects the fringe capacitance and beam shape subject to the electrostatic force. This is justified by the small gap height to beam length ratio ($g_0/\ell$ around $10^{-2}$–$10^{-3}$) [4]. However, the advancement in fabrication technologies and materials leads to the reduction in the size of electrostatically actuated MEMS, so the $g_0/\ell$ ratio cannot be considered small (on the order of $10^{-1}$–$10^{-2}$) [5]. The parallel-plate capacitor assumption on large for large $g_0/\ell$ ratio devices is no longer valid [4].

## *2.2 Parallel-plate Assumption in Electromechanical Structure*

### *2.2.1 Parallel-plate Theory and Effect of Length Ratio*

Parallel-plate theory is widely used to model RF MEMS capacitive switches and to determine the pull-in voltage ($V_{PI}$). This is probably due to its simplicity. The theory assumes that the electrostatic force is evenly distributed across the top and bottom electrode's region of overlap. Meanwhile, the vertical displacement within the overlap region ($\ell'$ in Fig. 2-2) is equal to $\Delta z$ at all locations. The cross section of such a parallel-plate capacitor model is shown in Fig. 2-2. The parallel-plate capacitor includes top movable electrode with length $\ell$, and width $W$. It is suspended a distance $g_0$ above the bottom stationary electrode of a length $\ell'$. The load is evenly distributed within the top and bottom electrode overlap region. This simplified model represents the RF MEMS capacitive switch in the suspended state and assumes bottom electrode has zero thickness and only air between the top and bottom electrode. The electro-mechanical behavior of this plate capacitor provides a clear understanding of the dominant switch behavior characteristics.

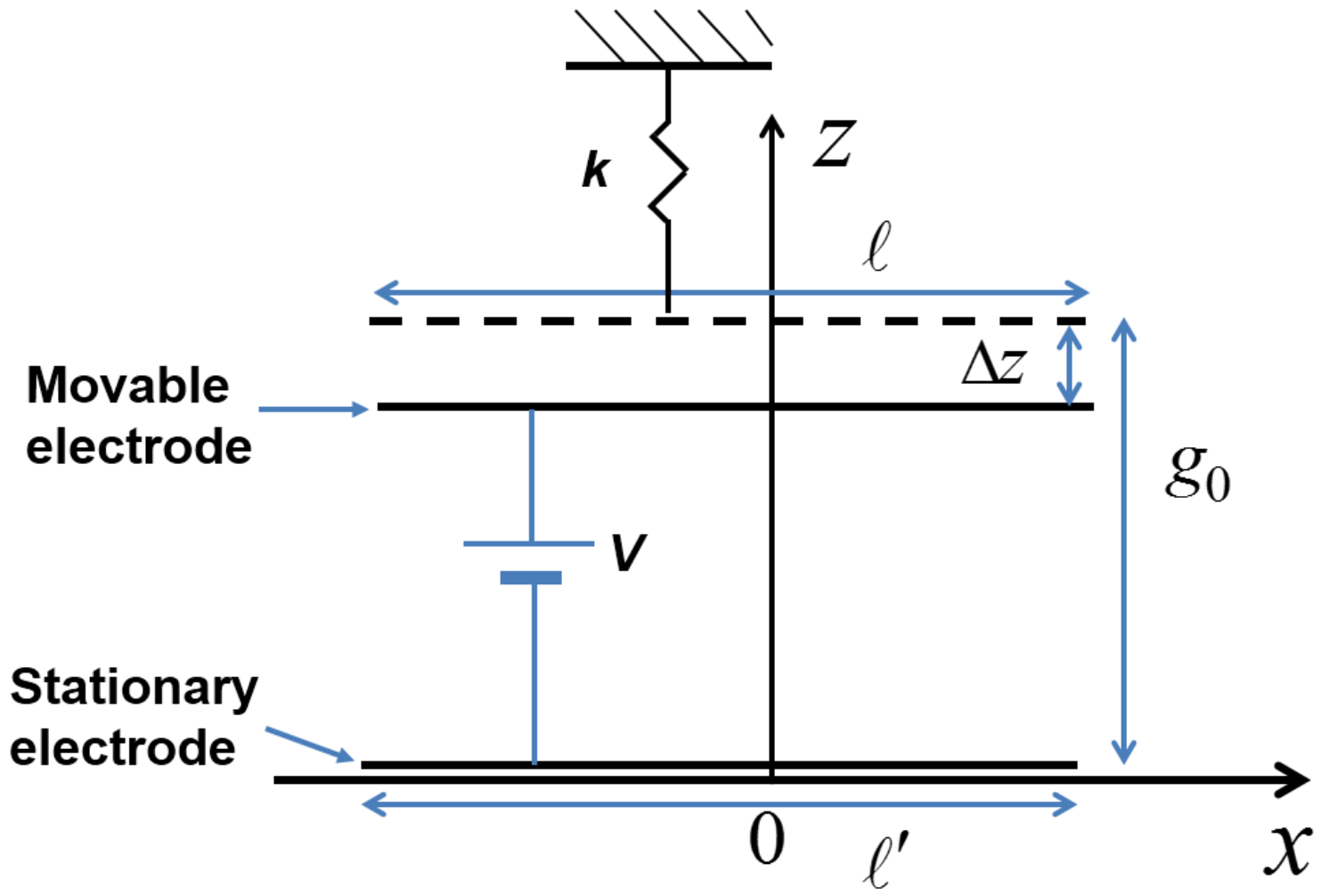

Fig. 2-2. 2-D cross section of parallel plate capacitor. The electrode's width $W$ is not shown in the figure.

Neglects the fringing capacitance, the capacitance of RF MEMS switches based on parallel-plate model for two perfectly flat plates before pull-in is:

$$C = \frac{\varepsilon_0 W \ell'}{g_0 - \Delta z}. \tag{2-1}$$

When a voltage is applied between the top and bottom electrode, an electrostatic force is induced on the beam. In addition, the corresponding mechanical force in the movable electrode will resist the electrostatic force. The mechanical behavior is described in terms of a spring constant $k$ (either linear or nonlinear), and the mechanical restoring force is given by $F = k\cdot\Delta z$, where $\Delta z$ is the vertical deflection of the beam at the center. When the electrostatic force and mechanical restoring force are balanced,

$$\frac{1}{2}\frac{\varepsilon_0 W\ell' V^2}{\left(g_0-\Delta z\right)^2} = k\Delta z. \tag{2-2}$$

It is implicitly assumed in (2-2) that the shape of the deforming electrode does not deviate significantly from a flat surface, and thus the parallel-plate electrostatic solution remains valid. In addition, the fringe capacitance and stretch effect are neglected [6]. It will be shown to what extend this approximation is valid in follows.

Reorganizing (2-2) gives

$$V = \sqrt{\frac{2k}{\varepsilon_0 W\ell'}\Delta z\left(g_0-\Delta z\right)^2}, \tag{2-3}$$

$$\frac{dV}{d\Delta z} = \sqrt{\frac{k}{2\varepsilon_0 W\ell'}}\frac{\left(g_0-3\Delta z\right)\left(g_0-\Delta z\right)}{\sqrt{\Delta z}\left(g_0-\Delta z\right)}. \tag{2-4}$$

For $\frac{dV}{d\Delta z}=0$, the root is given by $\Delta z = \frac{1}{3}g_0$ [7]. In (2-3), it can be seen that the voltage achieves the maximum value at $\Delta z = \frac{1}{3}g_0$. For $0 < \Delta z \le \frac{1}{3}g_0$, $\frac{dV}{d\Delta z} \ge 0$, the voltage increases with the vertical displacement and the electrostatic force and mechanical restoring force is in static equilibrium. In the region $\frac{1}{3}g_0 < \Delta z \le g_0$, $\frac{dV}{d\Delta z} \le 0$, and thus maintain a force balance relationship between two forces, the bias voltage would need to decrease with increasing vertical displacement. Otherwise, the electrostatic force will be greater than the mechanical force and the beam deflection becomes unstable, i.e., a small increase in voltage causes a very large increase in deflection, as shown in Fig. 2-3.

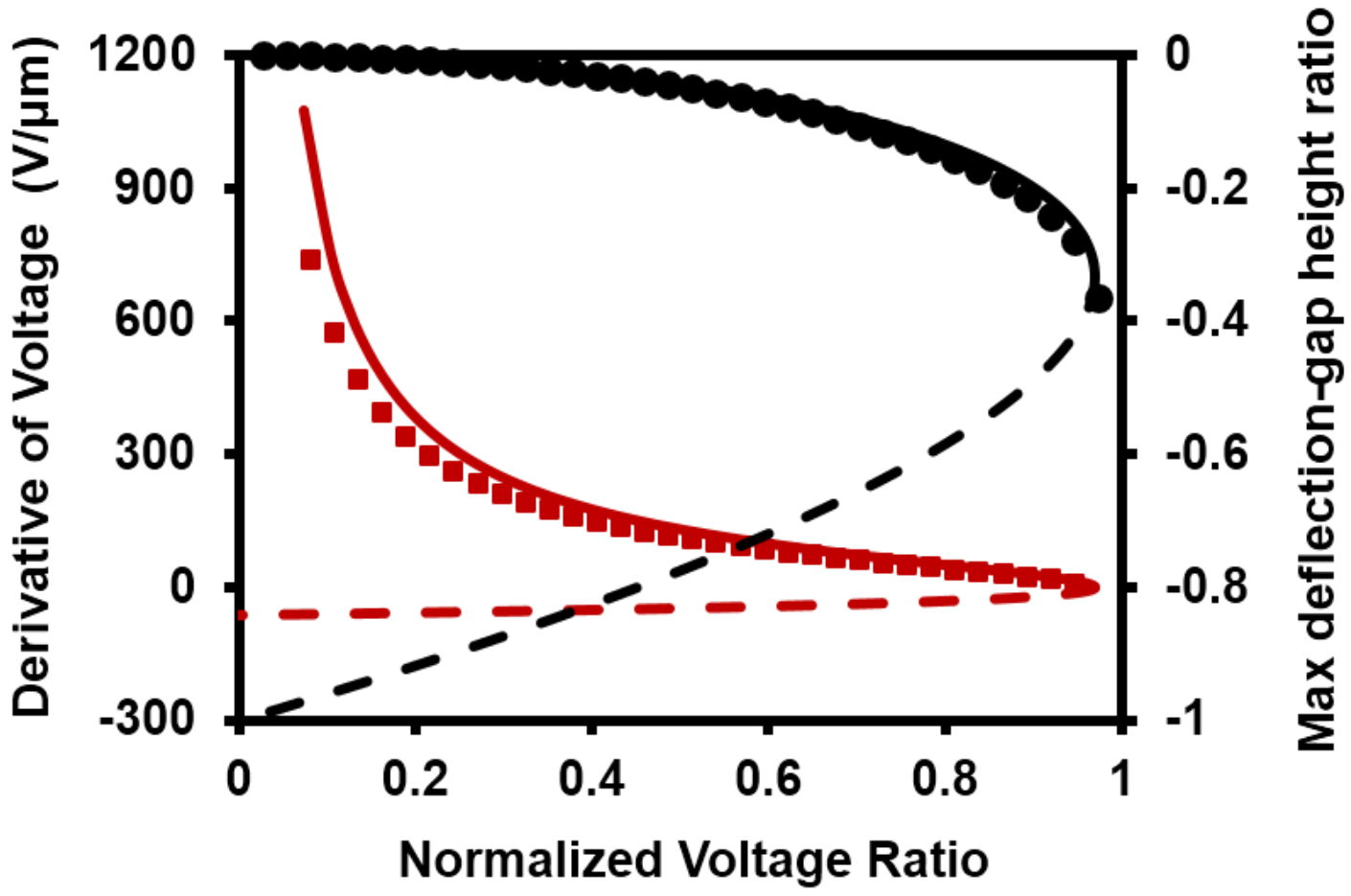

Fig. 2-3. ANSYS simulation results of derivative of voltage (■) and normalized max deflection versus normalized voltage (●) for a typical plate capacitor. The solid line is the analytical solution and the dashed line represents the unstable behavior. The bias voltage is normalized by pull-in voltage.

Therefore, $\Delta z = \frac{1}{3} g_0$ is defined as the critical point between stable and unstable behavior and the maximum voltage at $\frac{dV}{d\Delta z} = 0$ is usually defined as the pull-in voltage $V_{P0}$ (subscript "0" denotes pull-in voltage based on parallel-plate assumption),

$$V_{P0} = \sqrt{\frac{8k}{27\varepsilon_0 W \ell'} g_0^3}. \quad (2\text{-}5)$$

However, in a more realistic numerical simulation, the simulation diverges before $\frac{dV}{d\Delta z}$ actually reaches 0. Therefore, the linear extrapolation is used for $\frac{dV}{d\Delta z}$ curve within short intervals before pull-in. For the numerical results a convenient determination for the pull-in voltage is given by the intersection point of extrapolated curve with the *x* axis.

In fact, as the fixed-fixed top movable electrode starts to deform under the electrostatic force, the vertical displacement is a function of horizontal location (not a constant), which invalidates the parallel-plate assumption. To study how the beam shape and stationary to movable electrode length ratio affects the parallel-plate approximation, the analytical solution is derived assuming the load is uniformly distributed across overlap region. Further, the numerical simulations were carried out for the stationary electrode length to movable electrode length ($\ell'/\ell$), varying in ratio from 1/10 to 1.

### 2.2.2 Effect of Bending and Residual Stress

To predict the beam deflection in vertical direction, the Euler-Bernoulli equation is employed [8]. This static equation does not capture the dynamic behavior of the beam or consider the geometric nonlinearities in the large-deflection regime. However, for the purpose of validating the parallel-plate assumption, the classical static solution is satisfactory. In this study, the static solution is compared with ANSYS finite element analysis, which includes all nonlinear effects, including large-deflection behavior.

The bending effect is the first effect to study. It leads to the linear relationship between the mechanical restoring force and transverse displacement. Consider a fixed-fixed aluminum beam plate capacitor with $\ell$ = 100 μm, $W \rightarrow \infty$, $t$ = 0.6 μm, with gap height $g_0$ = 1 μm. The bottom stationary electrode has a length $\ell'$. Assume bottom electrode with zero thickness and only air between the top and bottom electrode. The residual stress is equal to zero. In this configuration, bending is the dominant deformation behavior.

The Euler-Bernoulli differential equations that govern the transverse deflection are [9]–[10]:

$$\begin{cases} EI\dfrac{d^4\Delta z(x)}{dx^4} = \xi, & -\dfrac{\ell'}{2} < x < \dfrac{\ell'}{2} \\ EI\dfrac{d^4\Delta z(x)}{dx^4} = 0, & -\dfrac{\ell}{2} < x < -\dfrac{\ell'}{2} \text{ and } \dfrac{\ell'}{2} < x < \dfrac{\ell}{2} \end{cases} \tag{2-6}$$

where $E$ is the Young's modulus, $I$ is the moment of inertia. For a rectangular cross section, $I = \dfrac{Wt^3}{12}$, $\xi$ is the uniform load across the overlap region.

The left-side boundary conditions for (2-6) are $\Delta z\left(-\dfrac{\ell}{2}\right) = 0, \left.\dfrac{d\Delta z(x)}{dx}\right|_{x=-\frac{\ell}{2}} = 0$. In addition, the differential equation requires $\dfrac{d^3\Delta z(x)}{dx^3}, \dfrac{d^2\Delta z(x)}{dx^2}, \dfrac{d\Delta z(x)}{dx}$, and $\Delta z(x)$ to be continuous at $x = -\dfrac{\ell'}{2}$. Apply the boundary conditions and continuity requirements for (2-6), taking advantage of symmetry, the solution for is given by

$$\Delta z = \begin{cases} \dfrac{\xi\left[16\ell x^4 + 2\left(\ell^4 - 12\ell^2 x^2\right)\ell' + 24\ell x^2 \ell'^2 - 2\left(\ell^2 + 4x^2\right)\ell'^3 + \ell\ell'^4\right]}{384EI\ell}, & -\dfrac{\ell'}{2} < x < \dfrac{\ell'}{2} \\ \dfrac{\xi\left(\ell + 2x\right)^2 \ell'\left[\ell\left(\ell - 4x\right) - \ell'^2\right]}{192EI\ell}, & -\dfrac{\ell}{2} < x < -\dfrac{\ell'}{2} \\ \dfrac{\xi\left(\ell - 2x\right)^2 \ell'\left[\ell\left(\ell + 4x\right) - \ell'^2\right]}{192EI\ell}, & \dfrac{\ell'}{2} < x < \dfrac{\ell}{2} \end{cases} \tag{2-7}$$

For $x = 0, \Delta z(0) = \dfrac{\xi\left(2\ell^3\ell' - 2\ell\ell'^3 + \ell'^4\right)}{32EWt^3}$, and the deflection at the center of the beam ($x$ = 0) is used to determine the spring constant $k_0'$. For a beam that is subject to a uniformly distributed load, the spring constant is given by

$$k_0' = \frac{\xi\ell'}{\Delta z\left(0, \ell'\right)} = \frac{32EW\left(\dfrac{t}{\ell}\right)^3}{2 - 2\left(\dfrac{\ell'}{\ell}\right)^2 + \left(\dfrac{\ell'}{\ell}\right)^3} \tag{2-8}$$

which agrees with [11]–[12].

The average displacement within the overlap region ($\Delta z_{AVE}$) to maximum displacement at beam center ($\Delta z_{MAX}$) ratio is

$$\frac{\Delta z_{AVE}}{\Delta z_{MAX}} = \frac{\frac{1}{\ell'}\int_{-\frac{\ell'}{2}}^{\frac{\ell'}{2}} \Delta z\left(x,\ell'\right)dx}{\Delta z\left(0,\ell'\right)}$$

$$= \frac{2\left[15-30\left(\frac{\ell'}{\ell}\right)^2+24\left(\frac{\ell'}{\ell}\right)^3-5\left(\frac{\ell'}{\ell}\right)^4\right]}{15\left[2-2\left(\frac{\ell'}{\ell}\right)^2+\left(\frac{\ell'}{\ell}\right)^3\right]} \quad (2\text{-}9)$$

The Fig. 2-4(a) illustrates the trend of analytical solutions (2-9) and ANSYS simulations for $\frac{\Delta z_{AVE}}{\Delta z_{MAX}}$. They both predict that $\frac{\Delta z_{AVE}}{\Delta z_{MAX}}$ decreases along with $\ell'/\ell$ increases and it reaches 0.53 at $\ell'/\ell = 1$. It can be seen that flat plate assumption gradually becomes invalid as $\frac{\Delta z_{AVE}}{\Delta z_{MAX}}$ deviate from 1.

Considering the influence of beam shape, when the maximum/average displacement ratio is applied to parallel-plate assumption based pull-in voltage:

$$V_{PI} = V_{P0}\sqrt{\frac{\Delta z_{MAX}}{\Delta z_{AVE}}}$$

$$= V_{P0}\sqrt{\frac{15\left[2-2\left(\frac{\ell'}{\ell}\right)^2+\left(\frac{\ell'}{\ell}\right)^3\right]}{2\left[15-30\left(\frac{\ell'}{\ell}\right)^2+24\left(\frac{\ell'}{\ell}\right)^3-5\left(\frac{\ell'}{\ell}\right)^4\right]}}. \quad (2\text{-}10)$$

It must be understood that the final position of beam's maximum displacement in a pull-in situation cannot be accurately predicted by ANSYS simulation. When the switch approaches the pull-in situation, an infinitesimal voltage increase leads to a large maximum displacement increase, so it is difficult to obtain the exact position of the beam's maximum deflection in this unstable configuration. Although the exact beam maximum deflection at pull-in cannot be precisely predicted, the pull-in voltage is determined accurately from ANSYS simulations. It can be seen from Fig. 2-4(b) that pull-in voltage results that consider beam shape are in better agreement with ANSYS simulations than those results that only use the parallel-plate assumption. Further, before $\ell'/\ell$ =1/3, the pull-in voltage maximum difference between the two cases that consider and does not consider beam curve, is smaller than 7%. This indicates that the parallel-plate model is still a good approximation when $\ell'/\ell$ is smaller than 1/3.

Other approaches to predict pull-in voltage include the natural frequency approach [13] and the energy methods [14]. However, the parallel-plate approach and other two approaches do not consider fringe capacitance and nonlinear stretch effect, which limits their accuracy and applicable situations.

Using $C_0$, the parallel-plate capacitance with gap height $g_0$, as the normalization constant, the normalized capacitance based on $\Delta z_{AVE}$ is given by

$$\frac{C}{C_0} = \frac{\dfrac{\varepsilon_0 \ell'}{g_0 - \Delta z_{AVE}}}{\dfrac{\varepsilon_0 \ell'}{g_0}} = \frac{1}{1 - \dfrac{\Delta z_{AVE}}{g_0}}. \tag{2-11}$$

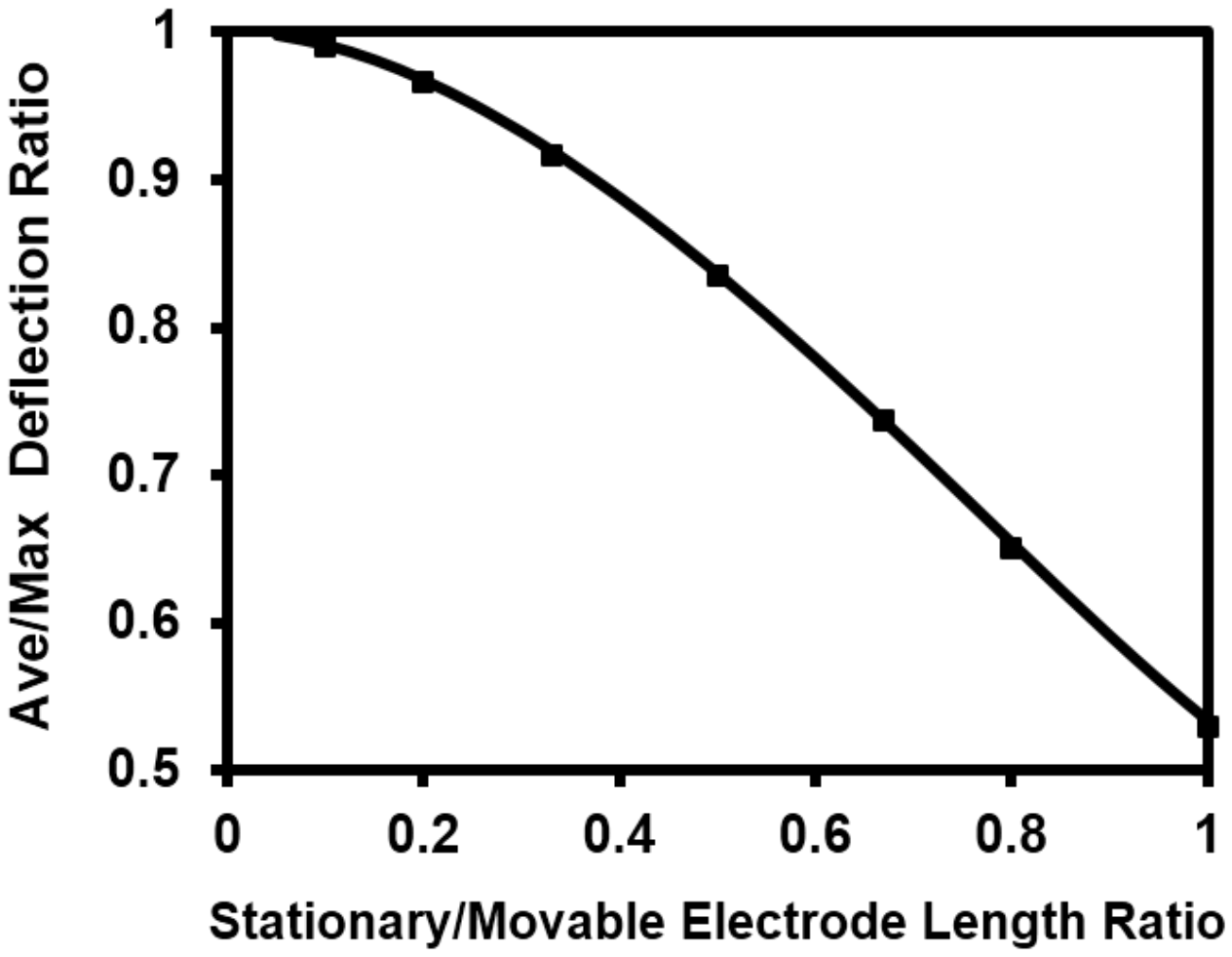

(a)

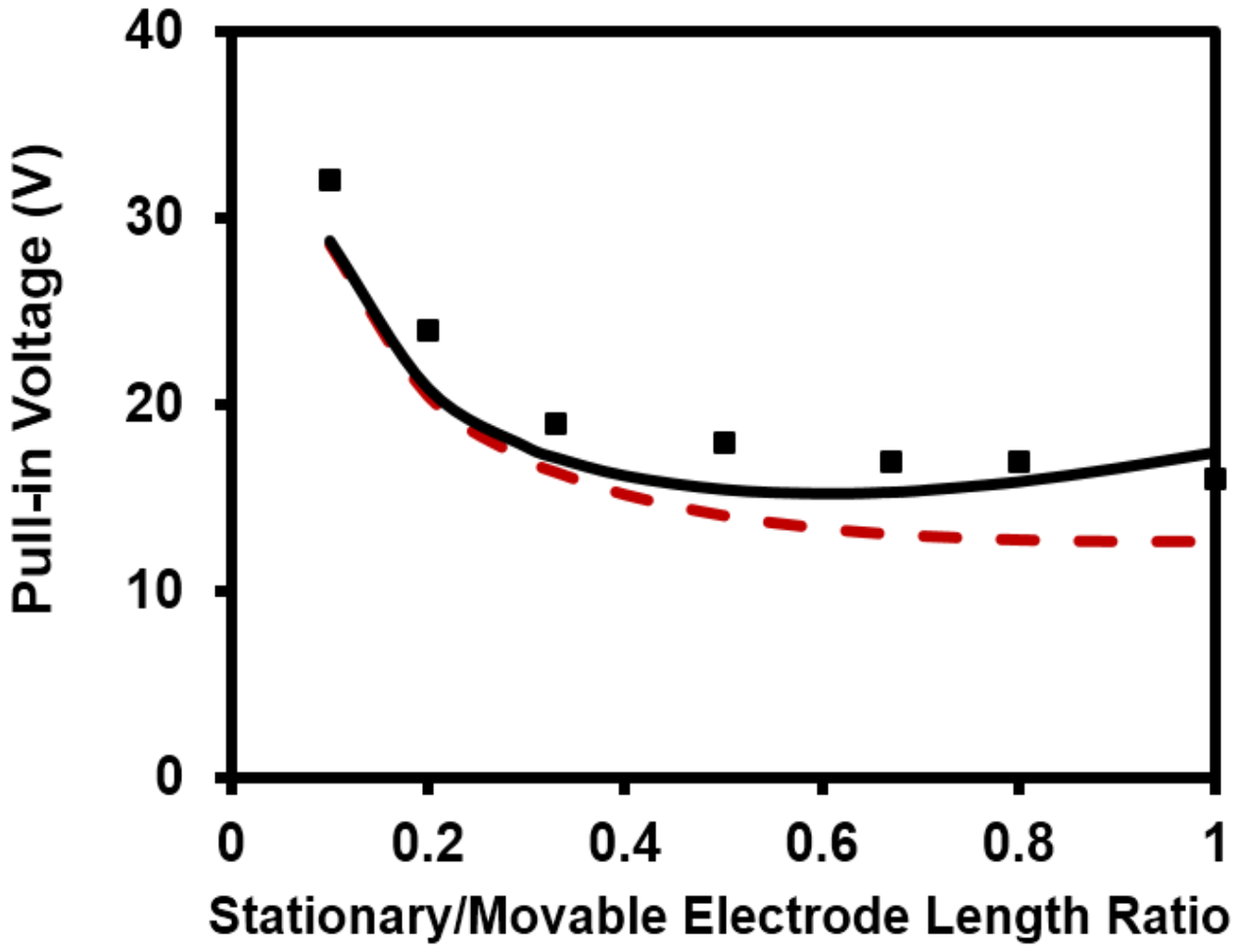

(b)

Fig. 2-4. (a) Analytical solution (——) and ANSYS simulation (■) of average/maximum deflection ratio. (b) Analytical solutions for pull-in voltage based on average deflection (——), maximum deflection (- - -), compared with ANSYS simulation (■).

It can be seen in Fig. 2-5 that both normalized parallel-plate capacitances calculated based on $\Delta z_{MAX}$ and $\Delta z_{AVE}$ agree with ANSYS simulation at ℓ′/ℓ =1/10. This is because $\frac{\Delta z_{AVE}}{\Delta z_{MAX}}$ (0.99) is close to 1 in this case. However, it can be seen that only capacitance obtained from $\Delta z_{AVE}$ is in good agreement with ANSYS simulation at ℓ′/ℓ =1. This is because $\frac{\Delta z_{AVE}}{\Delta z_{MAX}}$ (0.53) deviates significantly from 1.

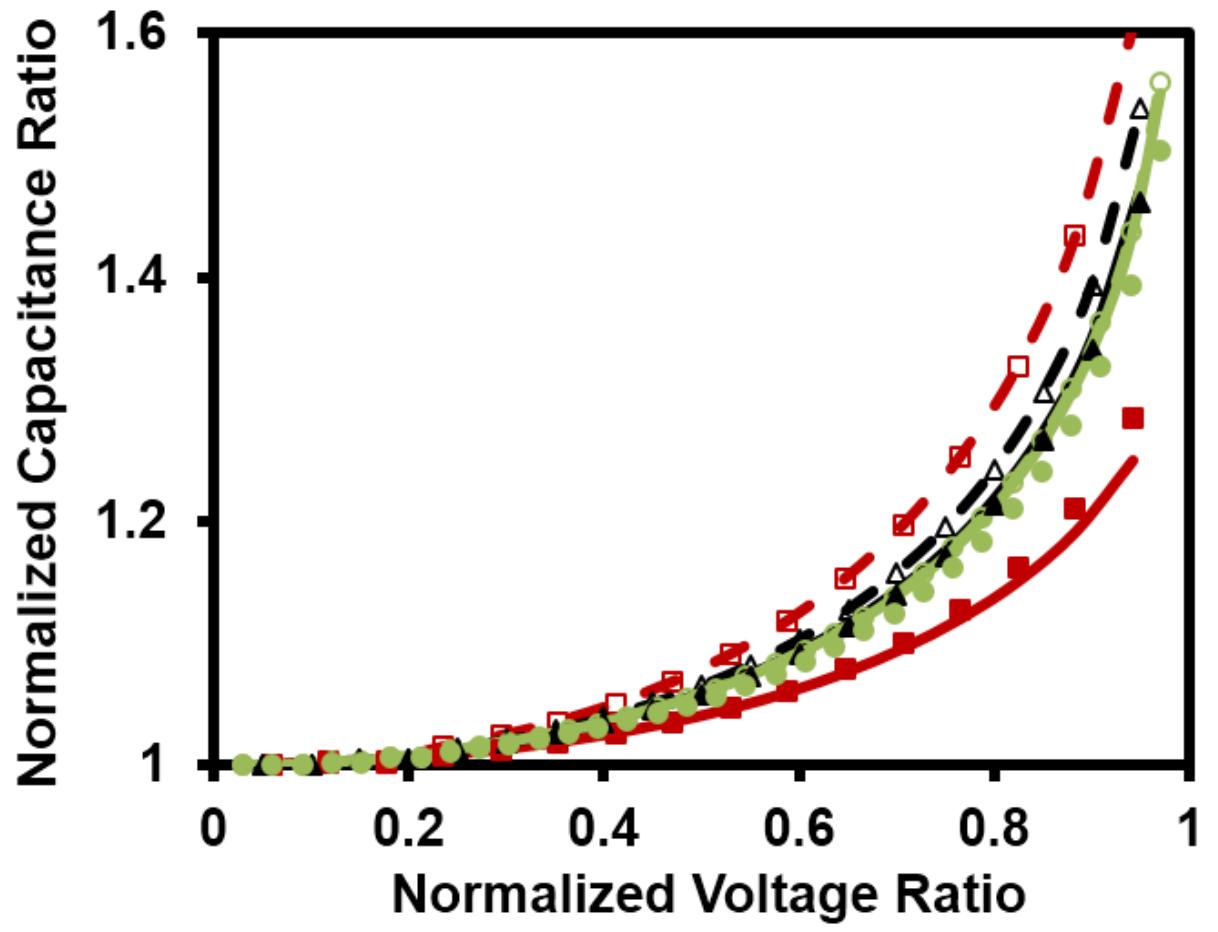

Fig. 2-5. Normalized capacitance ratio based on average deflection when stationary/movable electrode length ratio ℓ′/ℓ = 1/10(——), 1/3(——), 1(——), maximum deflection when ℓ′/ℓ = 1/10(-○-), 1/3(-Δ-), 1 (-□-), compared with ANSYS simulation results when ℓ′/ℓ = 1/10(●), 1/3(▲), 1 (■).

In conclusion, for bending effect dominant case, the parallel-plate approximation invalidates gradually as ℓ′/ℓ increases. The derivation of this approximation is 8% when ℓ′/ℓ = 1/3 and it rises to 50% when ℓ′/ℓ = 1/1. Large derivation leads to great error in pull-in voltage and capacitance estimation. By applying the correction factor, based on average displacement of the beam, the error can be reduced significantly. Therefore, the modified parallel-plate approximation has wider validity range.

However, the correction factor does not solve the problem completely, since the calculation requires analytical approximation that the electrostatic force across the overlap region is uniform. The finite element analysis and numerical calculation is needed for exact solutions.

Similar to the bending effect, the enhanced mechanical restoring force associated with residual stress, depends linearly on the transverse displacement. Although a low residual stress in switches is generally desirable, switches with almost zero residual stress are more subject to problems that include stuck switches, curling, and buckling [15].

Consider a fixed-fixed aluminum beam plate capacitor with $\ell$ = 300 μm, $W \to \infty$, $t$ = 0.6 μm, with gap height $g_0$ = 3 μm. The bottom stationary electrode has a length $\ell'$. The tensile residual stress $\sigma$ = 50 MPa. In this configuration, the residual stress is the dominant deformation behavior.

The differential equations that governs the transverse deflection are [9]–[10]:

$$\begin{cases} -N\dfrac{d^2\Delta z(x)}{dx^2} = \xi, & -\dfrac{\ell'}{2} < x < \dfrac{\ell'}{2} \\ -N\dfrac{d^2\Delta z(x)}{dx^2} = 0, & -\dfrac{\ell}{2} < x < -\dfrac{\ell'}{2} \text{ and } \dfrac{\ell'}{2} < x < \dfrac{\ell}{2} \end{cases} \tag{2-12}$$

where $N = \sigma Wt$, and $\xi$ is the uniform load across the overlap region.

The left-side boundary condition for (2-12) is $\Delta z\left(-\frac{\ell}{2}\right) = 0$. In addition, the differential equation requires $\frac{d\Delta z(x)}{dx}$, and $\Delta z(x)$ to be continuous at $x = \frac{\ell'}{2}$. Apply the boundary conditions and continuity requirements for (2-12), the solution is

$$\Delta z(x,\ell') = \begin{cases} -\dfrac{\xi\left[4x^2 - 2\ell\ell' + (\ell')^2\right]}{8N}, & -\dfrac{\ell'}{2} < x < \dfrac{\ell'}{2} \\ \dfrac{\xi\ell'\left(x + \dfrac{\ell}{2}\right)}{2N}, & -\dfrac{\ell}{2} < x < -\dfrac{\ell'}{2} \\ \dfrac{\xi\ell'\left(-x + \dfrac{\ell}{2}\right)}{2N}, & \dfrac{\ell'}{2} < x < \dfrac{\ell}{2}. \end{cases} \tag{2-13}$$

At $x = 0$, $\Delta z(0,\ell') = \frac{\xi\left(2\ell\ell' - \ell'^2\right)}{8\sigma wt}$, and the corresponding spring constant is

$$k_0^{"} = \frac{\xi\ell'}{\Delta z(0,\ell')} = \frac{8\sigma w\left(\dfrac{t}{\ell}\right)}{2 - \left(\dfrac{\ell'}{\ell}\right)}. \tag{2-14}$$

which agrees with [11]–[12].

The $\frac{\Delta z_{AVE}}{\Delta z_{MAX}}$ in residual stress dominant case is

$$\frac{\Delta z_{AVE}}{\Delta z_{MAX}} = \frac{\frac{1}{\ell'}\int_{-\frac{\ell'}{2}}^{\frac{\ell'}{2}} z\left(x,\ell'\right)dx}{z(0,\ell')} = \frac{\frac{\xi\left(3\ell\ell'-2\ell'^2\right)}{12}}{\frac{\xi\left(2\ell\ell'-\ell'^2\right)}{8}} = \frac{6-4\frac{\ell'}{\ell}}{6-3\frac{\ell'}{\ell}}. \tag{2-15}$$

Both the analytical solution and ANSYS simulation predict that $\frac{\Delta z_{AVE}}{\Delta z_{MAX}}$ decreases along with $\ell'/\ell$ increases and it reaches two third when $\ell'/\ell = 1$. This trend is shown in Fig. 2-6(a).Considering the influence of beam shape, when the maximum/average displacement ratio is applied to parallel-plate assumption based pull-in voltage:

$$V_{PI} = V_{P0}\sqrt{\frac{\Delta z_{MAX}}{\Delta z_{AVE}}} = V_{P0}\sqrt{\frac{6-3\frac{\ell'}{\ell}}{6-4\frac{\ell'}{\ell}}}. \tag{2-16}$$

Fig. 2-6(b) shows that the ANSYS simulation confirms the Pull-in voltage and relationship that obtained from analytical calculation.

Fig. 2-7 shows that, like the bending dominant case, at $\ell'/\ell$ =1/10, the parallel capacitance calculated based on $\Delta z_{AVE}$ and $\Delta z_{MAX}$ agrees well with the ANSYS simulation because $\frac{\Delta z_{AVE}}{\Delta z_{MAX}}$ (0.99) is close to 1 in this case. However, only the capacitance based on $\Delta z_{AVE}$ is in good agreement with the ANSYS simulation as $\ell'/\ell$ increases from 1/3 to 1.

In conclusion, for either the residual stress dominant or the bending dominant cases, the parallel-plate assumption shows similar validity range. The correction factor help achieve better prediction of the pull-in voltage and capacitance.

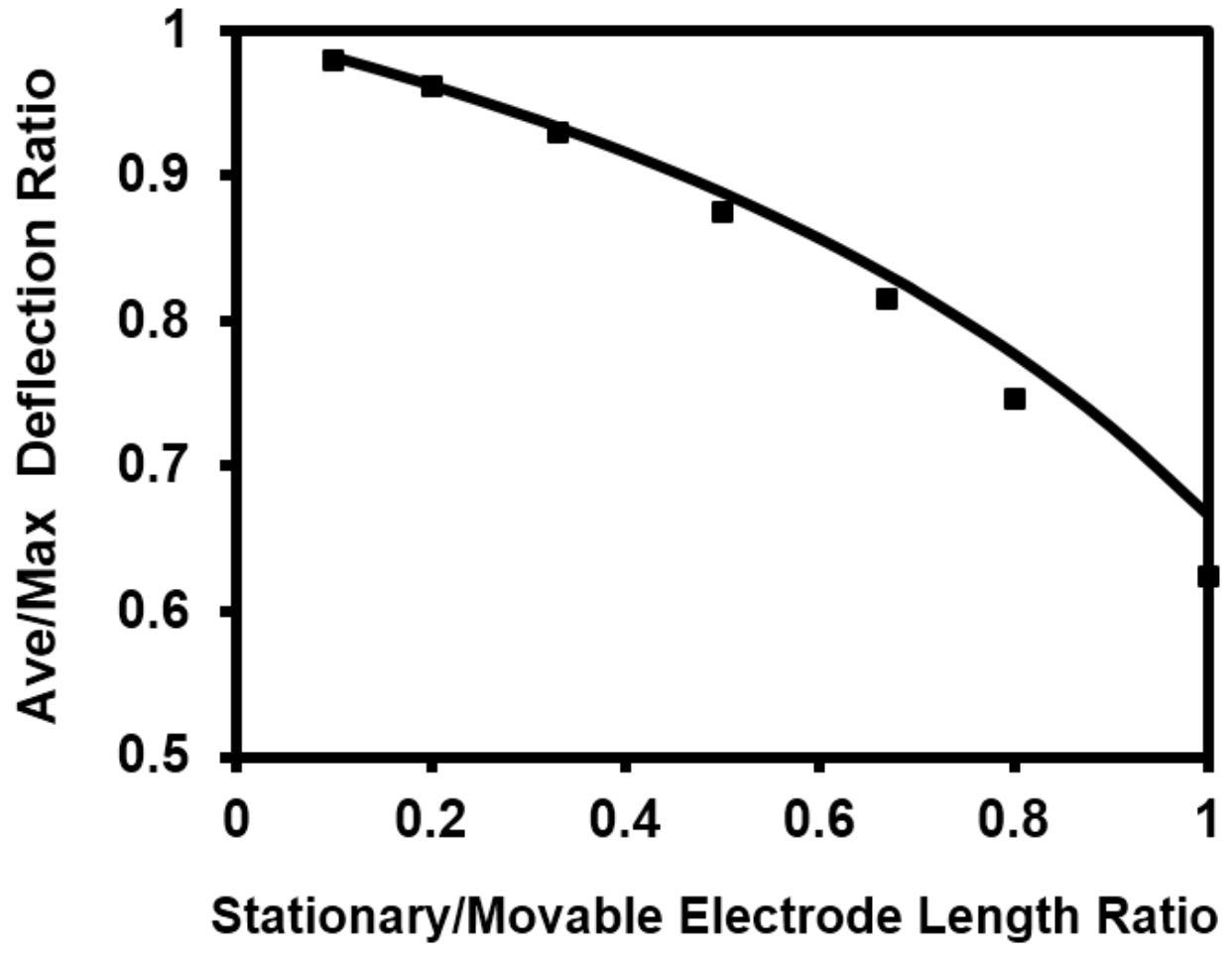

(a)

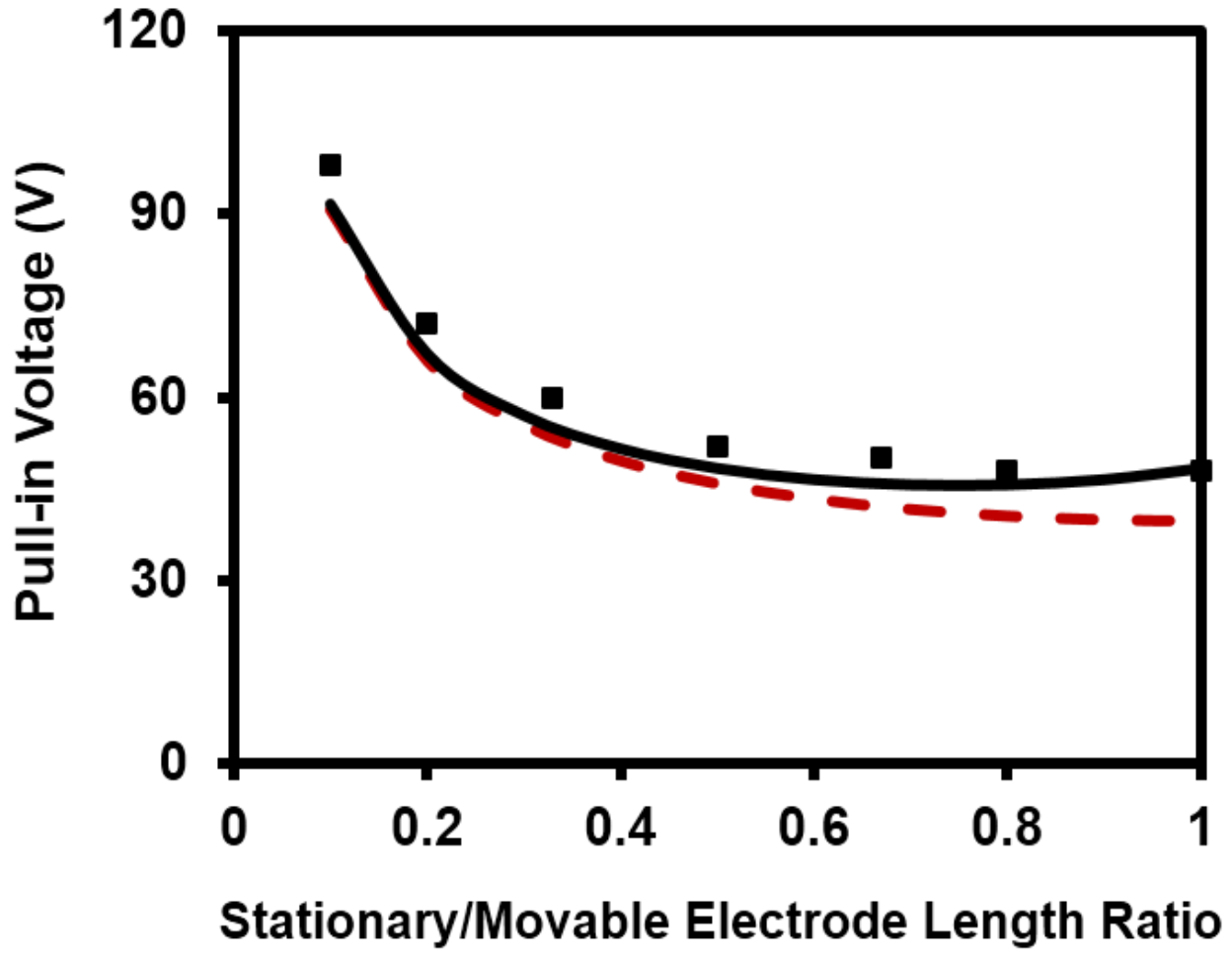

(b)

Fig. 2-6. (a) Analytical solution (——) and ANSYS simulation (symbol) of average/maximum deflection ratio residual stress dominant case. (b) Analytical solutions for pull-in voltage based on average deflection (——), maximum deflection (- - -), compared with ANSYS simulation (symbol).

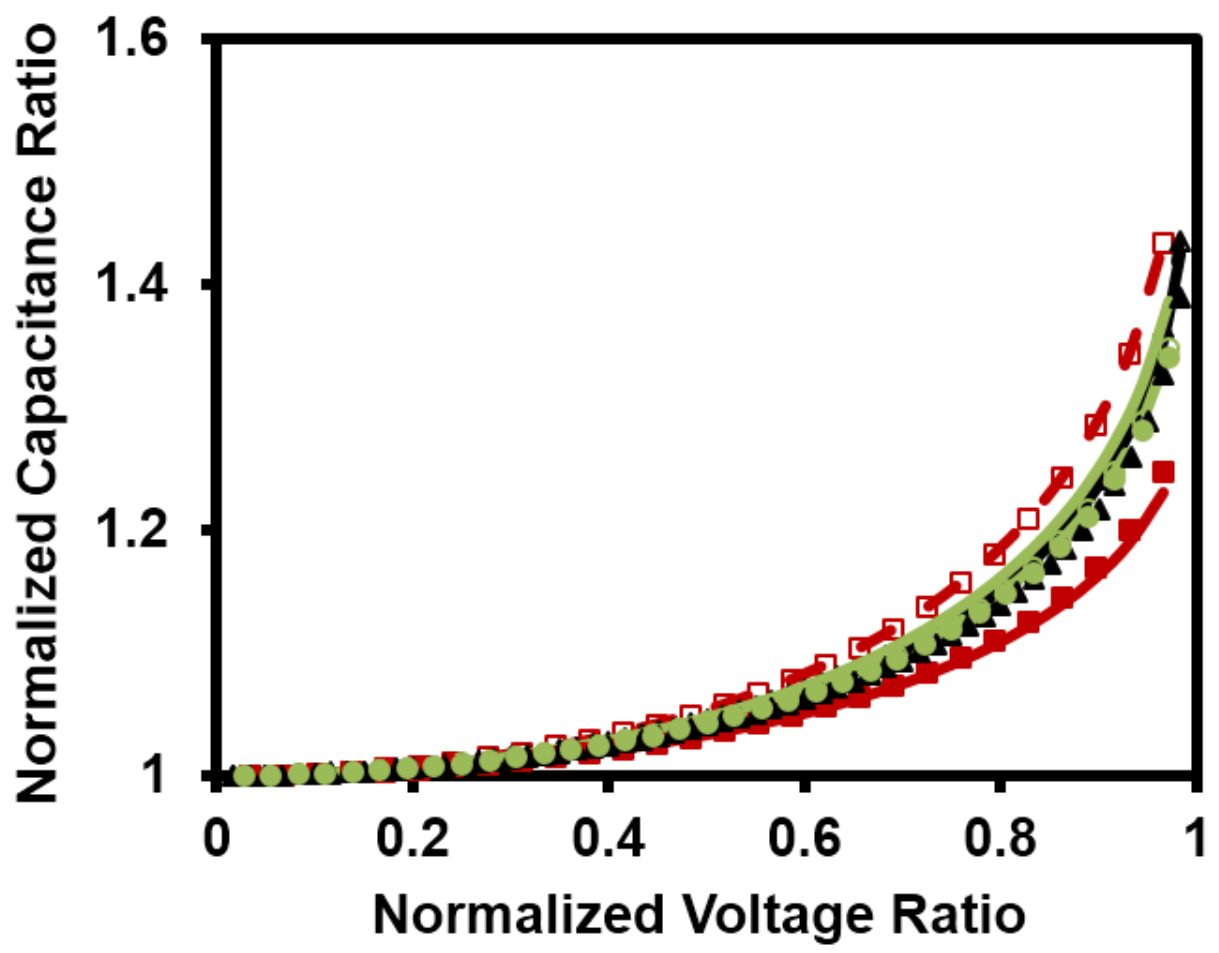

Fig. 2-7. Normalized capacitance ratio based on average deflection when stationary/movable electrode length ratio $\ell'/\ell$ = 1/10(——), 1/3(——), 1(——), maximum deflection when $\ell'/\ell$ = 1/10(-○-), 1/3(-Δ-), 1(-□-), compared with ANSYS simulation results when $\ell'/\ell$ = 1/10(●), 1/3(▲), 1 (■).

In the previous section, the effects of bending and residual stress are studied separately. In this section, the complete coupled analytical solution is derived. The differential equations considering both residual stress and bending effect are:

$$\begin{cases} EI\dfrac{d^4\Delta z(x)}{dx^4} - N\dfrac{d^2\Delta z(x)}{dx^2} = \xi, & -\dfrac{\ell'}{2} < x < \dfrac{\ell'}{2} \\ EI\dfrac{d^4\Delta z(x)}{dx^4} - N\dfrac{d^2\Delta z(x)}{dx^2} = 0, & -\dfrac{\ell}{2} < x < -\dfrac{\ell'}{2} \text{ and } \dfrac{\ell'}{2} < x < \dfrac{\ell}{2} \end{cases} \tag{2-17}$$

Apply the boundary conditions and continuity requirements for (), the solutions are as follows. For $-\dfrac{\ell'}{2} < x < \dfrac{\ell'}{2}$,

$$\Delta z\left(x,\ell'\right)=-\frac{\xi x^2}{2N}+\frac{\xi e^{-\frac{k\ell'}{2}}\cosh\left(kx\right)\left[e^{k\ell}-e^{k\ell'}+e^{\frac{1}{2}k\left(\ell+\ell'\right)}k\ell'\right]}{\left(-1+e^{k\ell}\right)k^2N}$$

$$+\frac{\xi\left[-8+8\,\text{csch}\left(\frac{k\ell}{2}\right)\sinh\left(\frac{k\ell'}{2}\right)\right]}{8k^2N} \tag{2-18}$$

$$-\frac{\xi k\ell'\left[-2k\ell+4\coth\left(\frac{k\ell}{2}\right)+k\ell'\right]}{8k^2N}.$$

For $-\frac{\ell}{2}<x<-\frac{\ell'}{2}$,

$$\Delta z(x,\ell')=\frac{\xi x\ell'}{2N}+\frac{\xi e^{-k\left(x+\frac{\ell'}{2}\right)}\left[1-e^{k\ell'}+k\ell' e^{\frac{1}{2}k\left(\ell+\ell'\right)}\right]}{2\left(-1+e^{k\ell}\right)k^2N}$$

$$+\frac{\xi e^{k\left(x-\frac{\ell'}{2}\right)}\left[e^{k\ell}-e^{k\left(\ell+\ell'\right)}+k\ell' e^{\frac{1}{2}k\left(\ell+\ell'\right)}\right]}{2\left(-1+e^{k\ell}\right)k^2N} \tag{2-19}$$

$$+\frac{4\xi\,\text{csch}\left(\frac{k\ell}{2}\right)\sinh\left(\frac{k\ell'}{2}\right)+\xi k\ell'\left[k\ell-2\coth\left(\frac{k\ell}{2}\right)\right]}{4k^2N}.$$

The solution for $\frac{\ell'}{2}<x<\frac{\ell}{2}$ case is easily obtained from (2-19) due to symmetric deflection. At $x=0,\ell'=\ell$, (2-19) provides the same solution as in [10].

Fig. 2-8(a) compares the analytical solutions for the average to maximum deflection ratio using (2-9) and (2-19) based on the same configuration. Both solutions agree with ANSYS simulation results well, indicating that (2-9) is a good approximation of (2-19) when no residual stress exists. Similarly, Fig. 2-8(b) compares the analytical solutions

using (2-15) and (2-19) based on same configuration. The discrepancy between two analytical solutions attribute to the neglect of bending effect in (2-15). Considering both residual stress and bending effect shows better agreement with ANSYS simulation results.

Finally, the bending and residual stress show similar behavior because they both have linear relationship between the mechanical restoring force and the displacement at beam center, provided the displacement is small (smaller than beam thickness).

At large displacement (greater than beam thickness), the stretching effect becomes significant. The relationship between mechanical restoring force and maximum displacement becomes nonlinear. The coupled closed-form analytical solution cannot be obtained from differential equation. The solutions are either analytical expressions based on various simplifying approximations or numerical solutions.

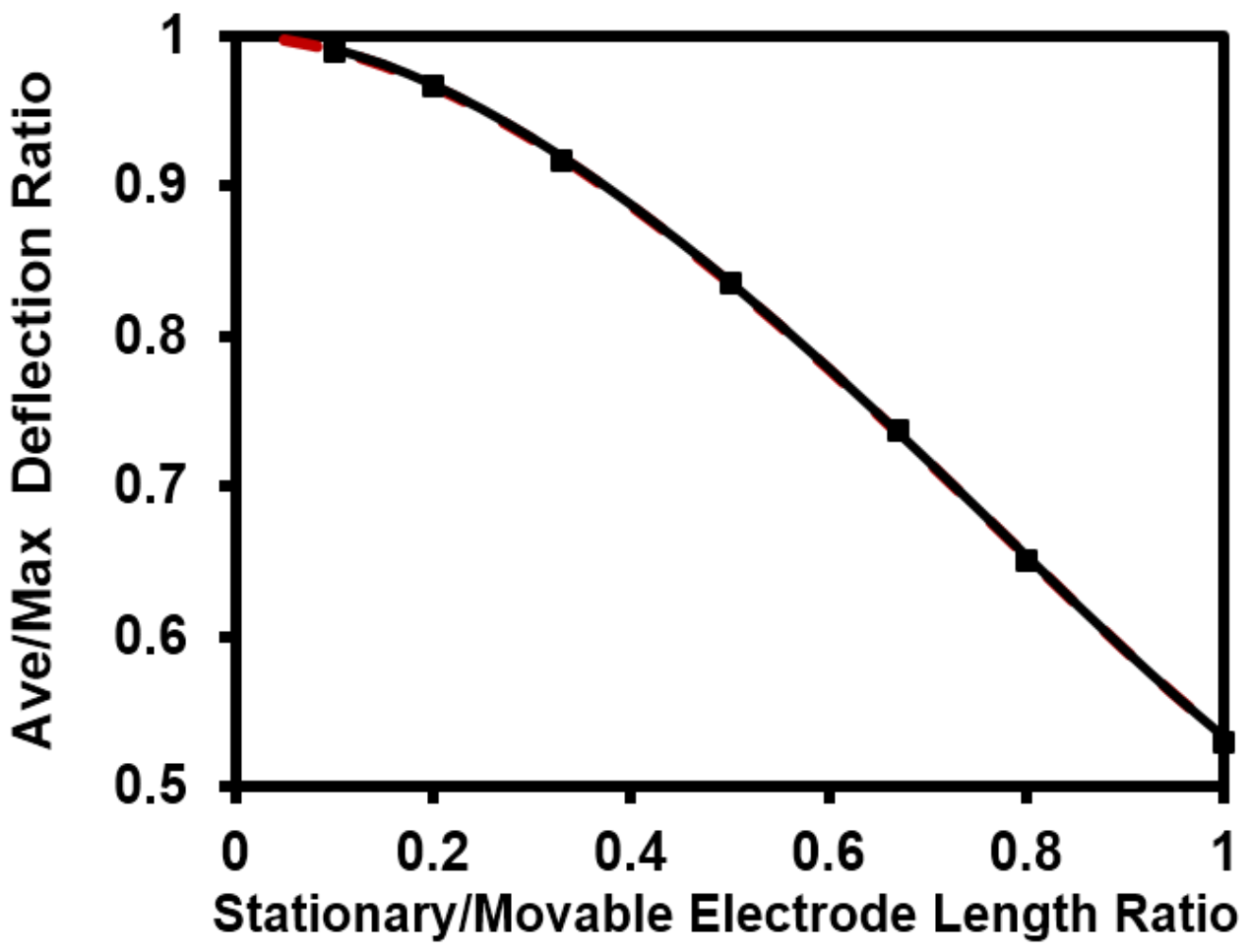

(a)

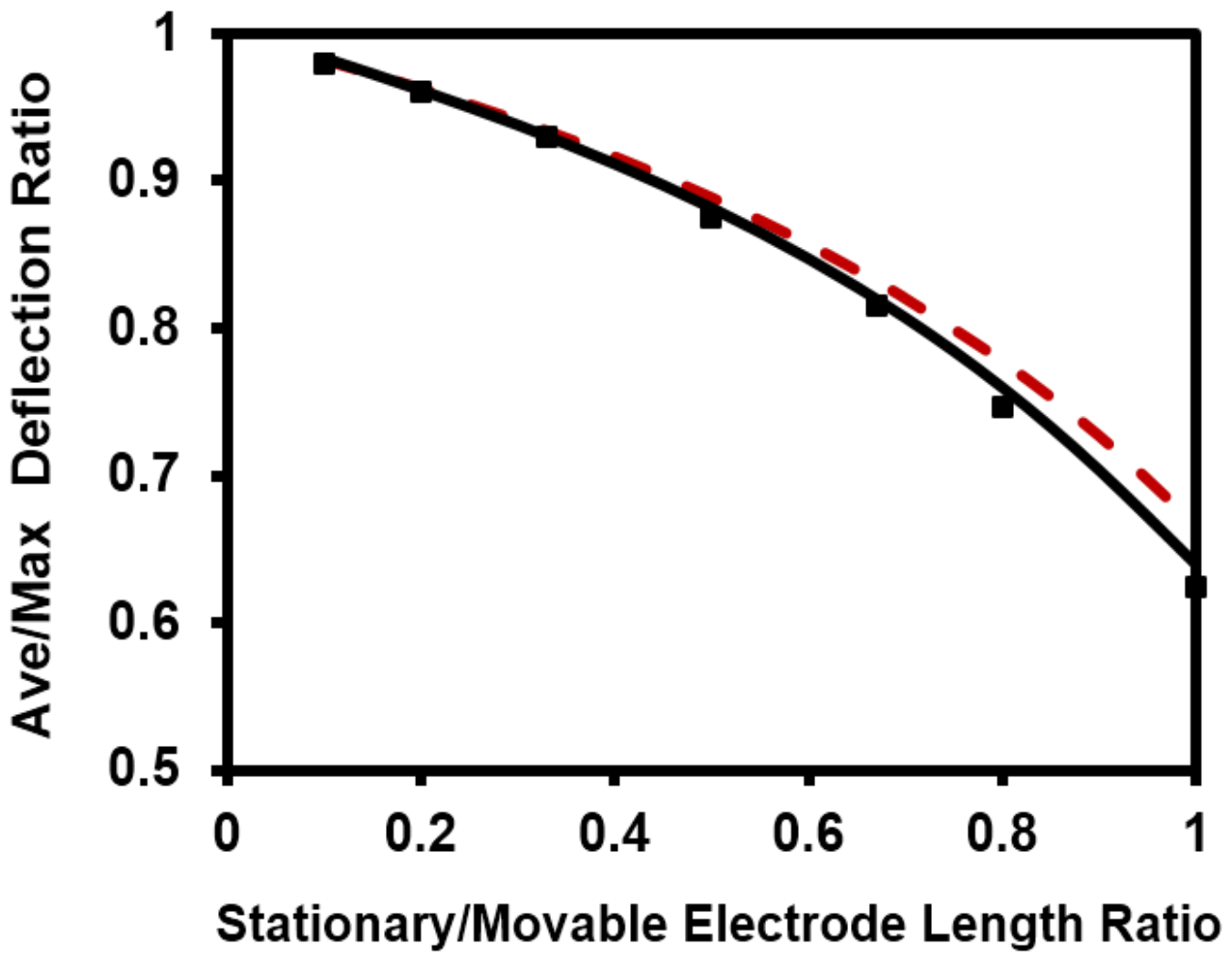

(b)

Fig. 2-8. (a) Average/maximum deflection ratio based on analytical solution considering both residual stress and bending effects (——), only bending effect (- - -), and ANSYS simulation (■). (b) Average/maximum deflection ratio based on analytical solution consider both residual stress and bending effect (——), only residual stress (- - -), and ANSYS simulation (■).

### *2.2.3 Nonlinear Elastic Restoring Force*

For a linear isotropic elastic material, the stiffness due to the bending and residual stress effects is independent of the displacement, i.e. the elastic restoring force shows linear relationship with maximum displacement. However, for a clamped-clamped structure, the arc length of the deformed structure increases if it bends. The length increase produces axial stress, which adds to the stiffness of the structure and further impacts the maximum displacement as a function of applied voltage. The nonlinear spring constant subject to stretch effects is preferred in flexural mode vibrating fixed-fixed beam RF MEMS resonators because it shifts the stiffness [16]–[17]. The applications are found in resonant strain gauges [18] and micromechanical resonator [19]. This section studies the influence of stretch effects on the maximum displacement subject to applied voltage.

Considering the stretch effects, (2-2) becomes:

$$\Delta\tilde{z} + \frac{k_S}{k_1} g_0^2 \Delta\tilde{z}^3 = \frac{1}{2}\frac{\varepsilon_0 W \ell' V^2}{k_1 g_0^3 \left(1-\Delta\tilde{z}\right)^2} \tag{2-20}$$

where $\Delta\tilde{z} = \dfrac{\Delta z}{g_0}$, $k_1$ is the effective linear spring constant caused by bending and residual stress and $k_S$ is considered a nonlinear spring constant caused by the stretch effects. Similar to (2-2), (2-20) is also based on the parallel-plate assumption except the nonlinear stretching effects are considered. From (2-8) and (2-14), applying the superposition principle, $k_1 = \frac{32EW\left(\frac{t}{\ell}\right)^3}{2-2\left(\frac{\ell'}{\ell}\right)^2+\left(\frac{\ell'}{\ell}\right)^3} + \frac{8\sigma w\left(\frac{t}{\ell}\right)}{2-\left(\frac{\ell'}{\ell}\right)}$. Assuming the beam shape is described by (2-13), the individual axial stress due to the nonlinear stretching is given by $\Delta\sigma = \frac{8E\Delta\tilde{z}^2\left(3-2\frac{\ell'}{\ell}\right)}{3\ell^2\left(2-\frac{\ell'}{\ell}\right)^2}$, which agrees very well with $\Delta\sigma = \frac{\pi^2 E\Delta\tilde{z}^2}{4\ell^2}$ when $\ell' = \ell$ in [20].

Correspondingly, $k_s = \frac{64EWt\left(3-2\frac{\ell'}{\ell}\right)}{3\ell^3\left(2-\frac{\ell'}{\ell}\right)^3}$.

It can be seen that $\frac{k_s}{k_1} g_0^2$ is the key parameter that determines the displacement due to the applied voltage. The analytical solutions for (2-20) are compared with ANSYS simulation results. The ANSYS simulation employs a fixed-fixed aluminum beam plate capacitor with $\ell$ = 300 μm, $W \to \infty$ (plane strain), $t$ = 0.6 μm, with gap height $g_0$ = 3 μm. The bottom stationary electrode has a length $\ell'$. Assume bottom electrode with zero thickness and only air between the top and bottom electrode. The residual stress $\sigma$ varies from 0 to 150 MPa, so the $\frac{k_s}{k_1} g_0^2$ ranges from 14 to 0.1.

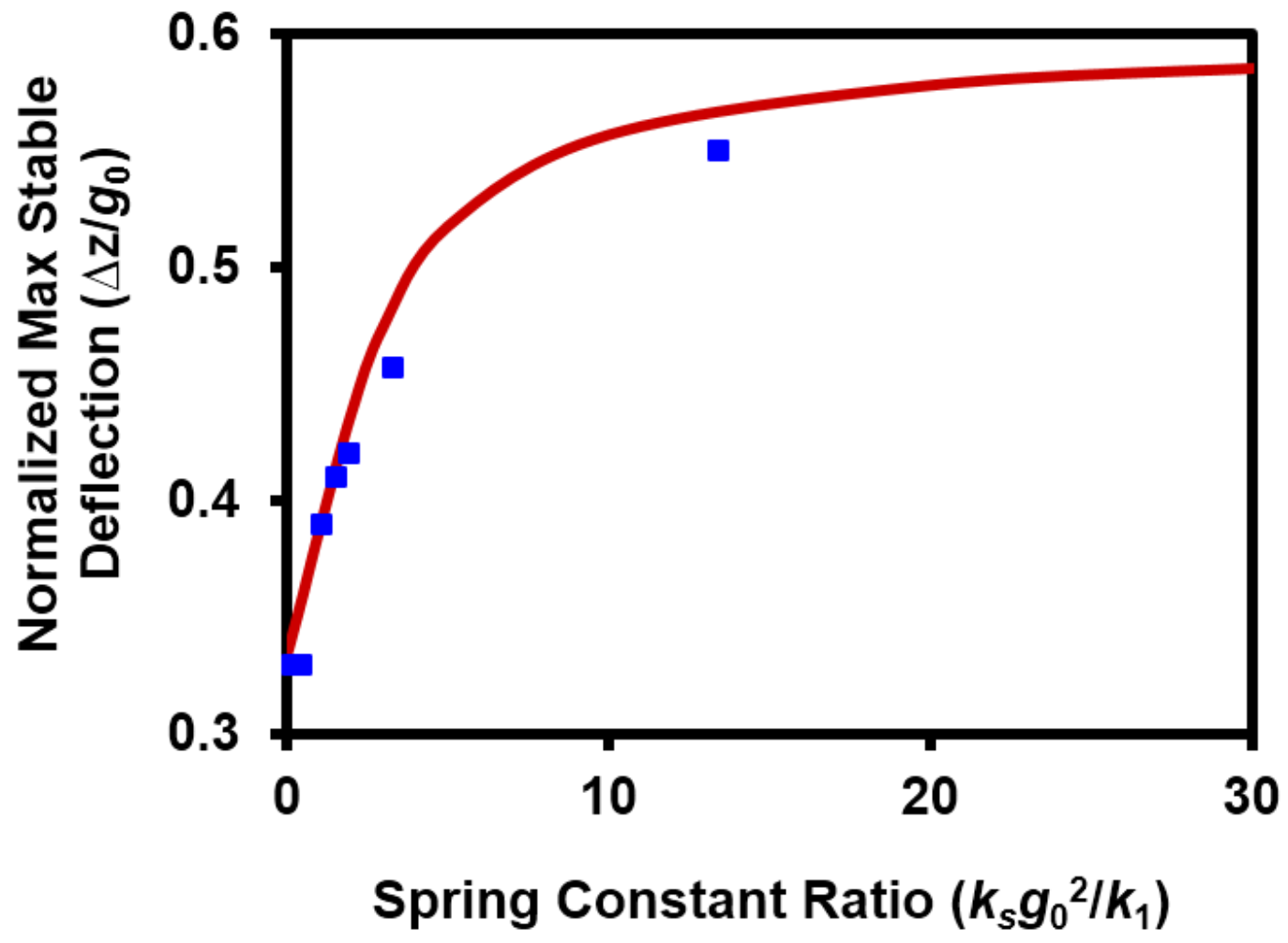

Fig. 2-9. Normalized maximum stable deflection as a function of spring constant ratio ($k_s g_0^2/k_1$). The solid curve is analytical solution and symbols are from ANSYS simulation results.

Solving (2-20) for $\frac{\Delta V}{\Delta \tilde{z}} = 0,$ the analytical solution for normalized maximum stable deflection is obtained. Fig. 2-9 shows normalized maximum stable deflection as a function of spring constant ratio ($\frac{k_s}{k_1} g_0^2$). Both analytical solution and ANSYS simulation results demonstrate a similar trend. As $\frac{k_s}{k_1} g_0^2$ approaches zero, the linear spring constant $k_1$ dominates the beam deformation behavior and the maximum deflection before pull-in is 1/3 of the gap height $g_0$ [7]. However, as $\frac{k_s}{k_1} g_0^2$ increases, the nonlinear spring constant extends the maximum stable beam travel range before pull-in occurs. For example, in the most extreme case ($k_1 \ll k_s g_0^2$), the maximum deflection can reach 3/5 of the gap height $g_0$ [21]. The extension of the maximum travel range is preferred in varactors, micro mirrors and resonators [8].

Solving (2-20) for $\Delta\tilde{z}$, the solution for displacement under applied voltage is obtained. However, the beam maximum displacement and electrostatic force relationship for fixed-fixed beam MEMS switches cannot be generally modeled accurately by solving (2-20). The main reason is the fixed-fixed beam deflects with a non-uniform curved displacement, significantly violates the parallel-plate electrostatic assumption [22],[23]. Comparing (2-20) with dimensionless ANSYS simulation results for the beam maximum displacement, as a function of applied voltage, a series of simple correction factors is obtained. The relationship between the correction factor ($\alpha$) and $\ell'/\ell$ is obtained by curve fitting.

For example an expression for $\alpha$ as a function of $\ell'/\ell$ is

$$\alpha = 0.232\tanh\left[1.524\left(\frac{\ell'}{\ell}\right)\right]+1 \tag{2-21}$$

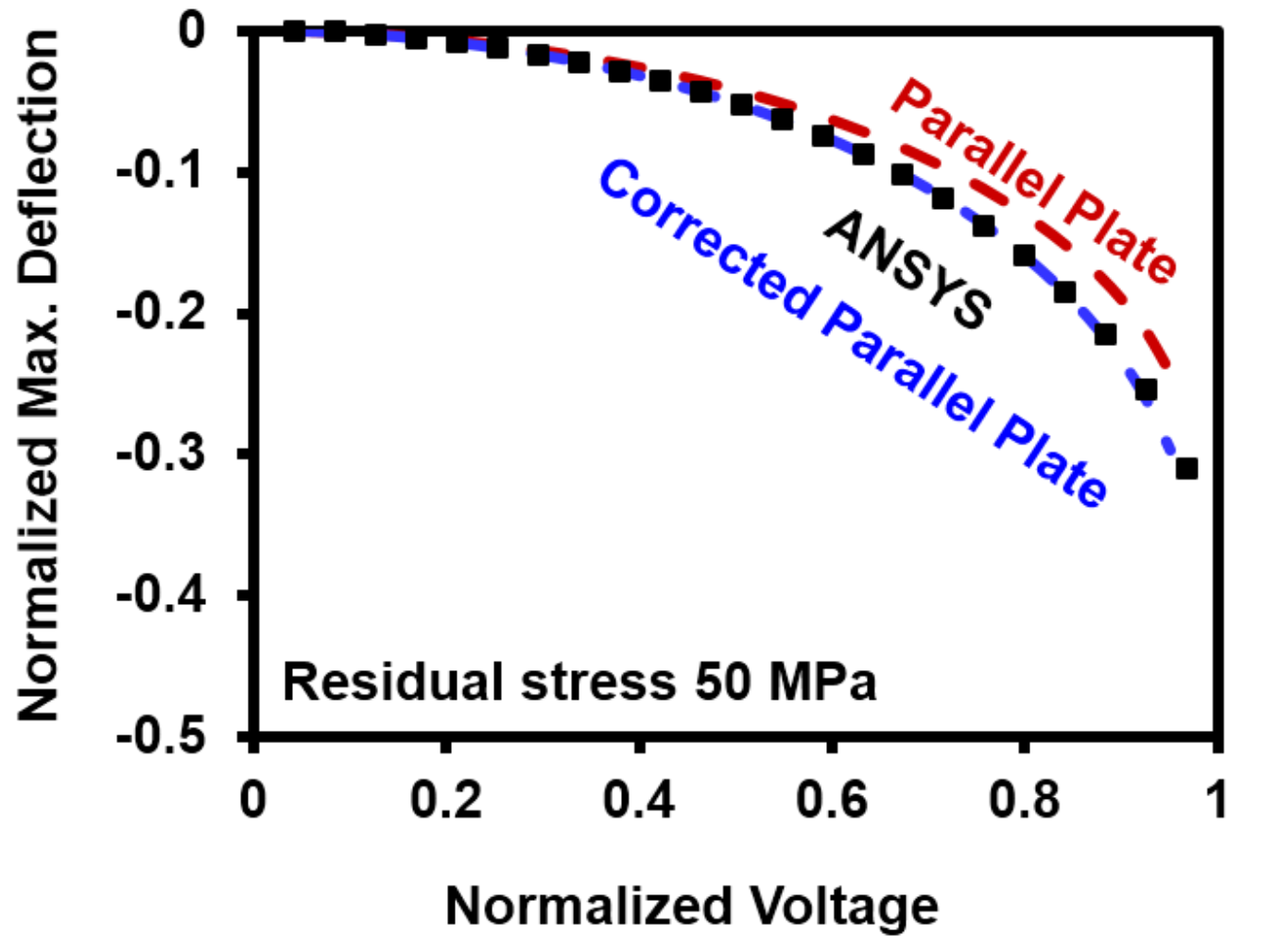

(a)

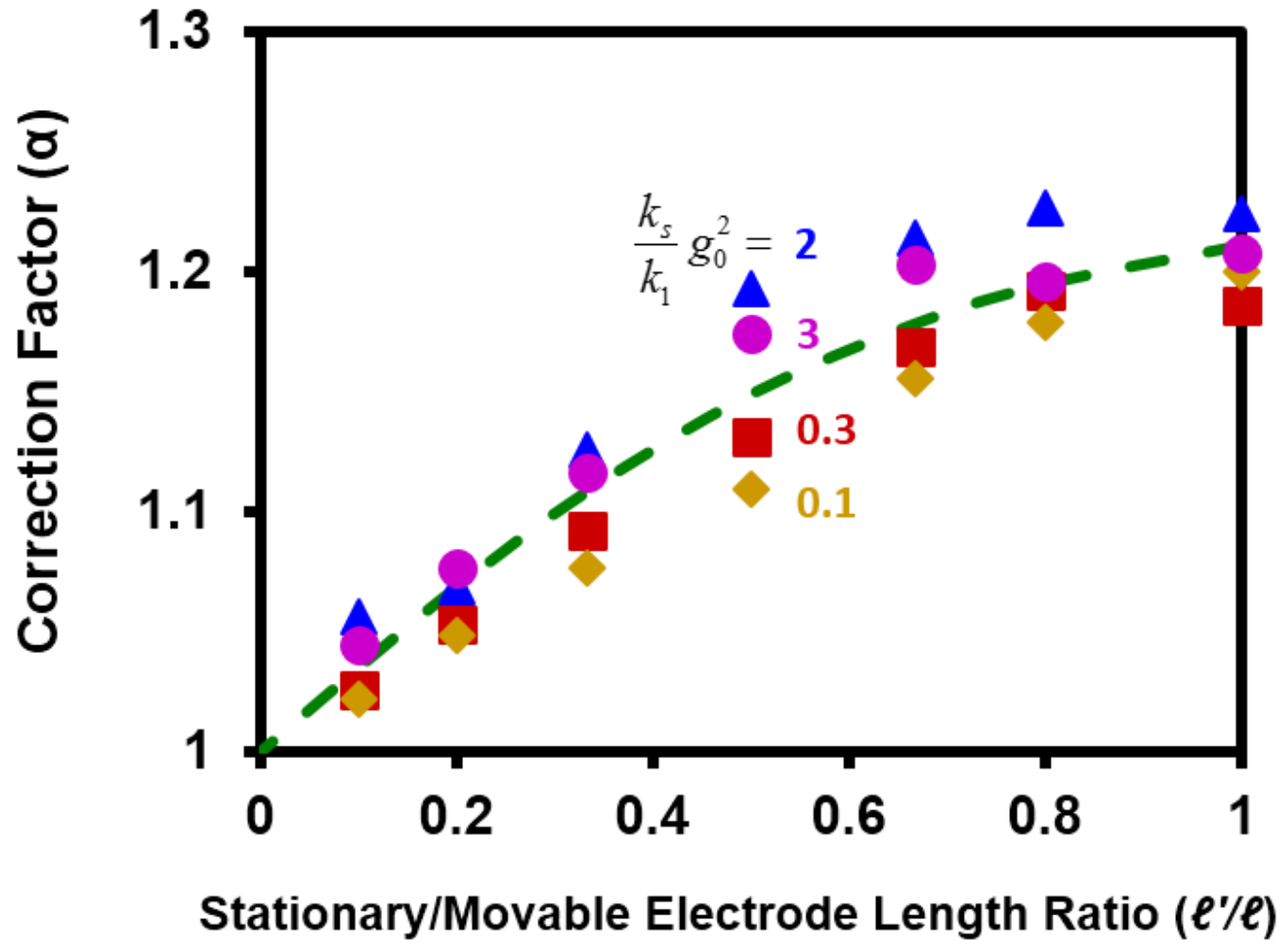

(b)

Fig. 2-10. (a) Normalized maximum deflection as a function of voltage when stationary/movable electrode length ratio $\ell'/\ell$ = 1/1. The symbols are from ANSYS simulation results and dash curve are from analytical solution for parallel plate assumption with and without correction factor ($\ell'/\ell$). (b) Correction factor ($\alpha$) as a function of stationary to movable electrode length ratio. $k_s/k_1g_0^2$ values are 2,3,0.3,0.1 and the corresponding residual stresses are 10, 5, 50, and 150 MPa. The symbols are from ANSYS simulation results and the dashed line is fitting expression.

Fig. 2-10(a) shows that the corrected analytical solution in very good agreement with the ANSYS simulation results. The correction factor indicates a linear relationship between the solutions for (2-20) and the ANSYS simulation results. Although (2-20) does not consider the non-uniform curved displacement, it still capture the relationship between the maximum displacement and the applied voltage. Together with a simple correction factor, it can predict the maximum displacement due to applied voltage accurately. It can be seen from Fig. 2-10(b) that the correction factor is close to 1 when $\ell'/\ell$ approaches zero, which indicates that the situation is close to parallel-plate assumption. However, when $\ell'/\ell$ increase to greater than 1/3, the correction factor increases to 1.1 and saturates at 1.22, reflecting the curved nature of capacitor plate. The maximum error in the fitting expression and ANSYS simulation results is about 4%.

### *2.2.4 Geometric Design for Linear Material Behavior*

When a clamped structure bends, it becomes longer thus developing axial stress [24]–[26]. Therefore, for a fixed-fixed beam MEMS switch, it is important to make sure the axial strain in the beam is within elastic region when it deforms. To satisfy this requirement, the switch geometry, material properties, and anchor boundary conditions need to be considered. This section focuses on the relationship between the axial strain and the switch geometry. To simplify the case, the residual stress is not included in the analytical analysis and ANSYS simulation. Therefore, the beam shape subject to electrostatic force is predicted by bending effects only. The total axial strain on *x* axis is given by

$$\begin{aligned}\varepsilon_x &= \varepsilon_{bending} + \varepsilon_{stretch} \\ &= -\frac{t}{2}\frac{\frac{d^2\Delta z}{dx^2}}{\left[1+\left(\frac{d\Delta z}{dx}\right)^2\right]^{\frac{3}{2}}} + \frac{2}{\ell}\left[\int_{-\frac{\ell}{2}}^{-\frac{\ell'}{2}}\frac{1}{2}\left(\frac{d\Delta z}{dx}\right)^2 dx + \int_{-\frac{\ell'}{2}}^{0}\frac{1}{2}\left(\frac{d\Delta z}{dx}\right)^2 dx\right] \\ &\approx -\frac{t}{2}\frac{d^2\Delta z}{dx^2} + \frac{2}{\ell}\left[\int_{-\frac{\ell}{2}}^{-\frac{\ell'}{2}}\frac{1}{2}\left(\frac{d\Delta z}{dx}\right)^2 dx + \int_{-\frac{\ell'}{2}}^{0}\frac{1}{2}\left(\frac{d\Delta z}{dx}\right)^2 dx\right]\end{aligned} \tag{2-22}$$

In the configurations where bending effects is the dominant factor, the deflection is described by (2-7). Substitute (2-7) into (2-22). The beam strain on x axis is

$$\varepsilon_{bending} = \begin{cases} \dfrac{8t\Delta z_{MAX}\left[3\left(1+4\frac{x}{\ell}\right)+\left(\frac{\ell'}{\ell}\right)^2\right]}{\ell^2\left[2-2\left(\frac{\ell'}{\ell}\right)^2+\left(\frac{\ell'}{\ell}\right)^3\right]}, & -\frac{\ell}{2} < x < -\frac{\ell'}{2} \\ \dfrac{8t\Delta z_{MAX}\left[-12\left(\frac{x}{\ell}\right)^2+3\frac{\ell'}{\ell}-3\left(\frac{\ell'}{\ell}\right)^2+\left(\frac{\ell'}{\ell}\right)^3\right]}{\ell^2\left[2\frac{\ell'}{\ell}-2\left(\frac{\ell'}{\ell}\right)^3+\left(\frac{\ell'}{\ell}\right)^4\right]}, & -\frac{\ell'}{2} < x < 0 \end{cases} \tag{2-23}$$

$$\varepsilon_{stretch} = \frac{16\Delta z_{MAX}{}^{2}\left[63-105\left(\frac{\ell'}{\ell}\right)^{2}+133\left(\frac{\ell'}{\ell}\right)^{4}-96\left(\frac{\ell'}{\ell}\right)^{5}+21\left(\frac{\ell'}{\ell}\right)^{6}\right]}{105\ell^{2}\left[2-2\left(\frac{\ell'}{\ell}\right)^{2}+\left(\frac{\ell'}{\ell}\right)^{3}\right]^{2}} \tag{2-24}$$

From (2-23) and (2-24), the stretch to bending strain ratio depends on max deflection to beam thickness ratio, which indicates the stretch is not an significant factor when max deflection is smaller than beam thickness. As shown in Fig. 2-11(a) and (b), the strain on *x* axis predicted by (2-22) agrees well with ANSYS simulation when the maximum deflection (0.3 and 0.2 μm in (a) and (b)) is not greater than beam thickness (0.6 μm) [16]. However, Fig. 2-12(a) and (b) illustrate that the stretching effects change *x* axis strain distribution when max deflection (1.5 μm in (a) and (b)) is greater than beam thickness (0.6 μm). The slight deviation is because (2-23) does not consider the influence stretch factor on movable electrode curve and it becomes invalid when stretch factor is significant.

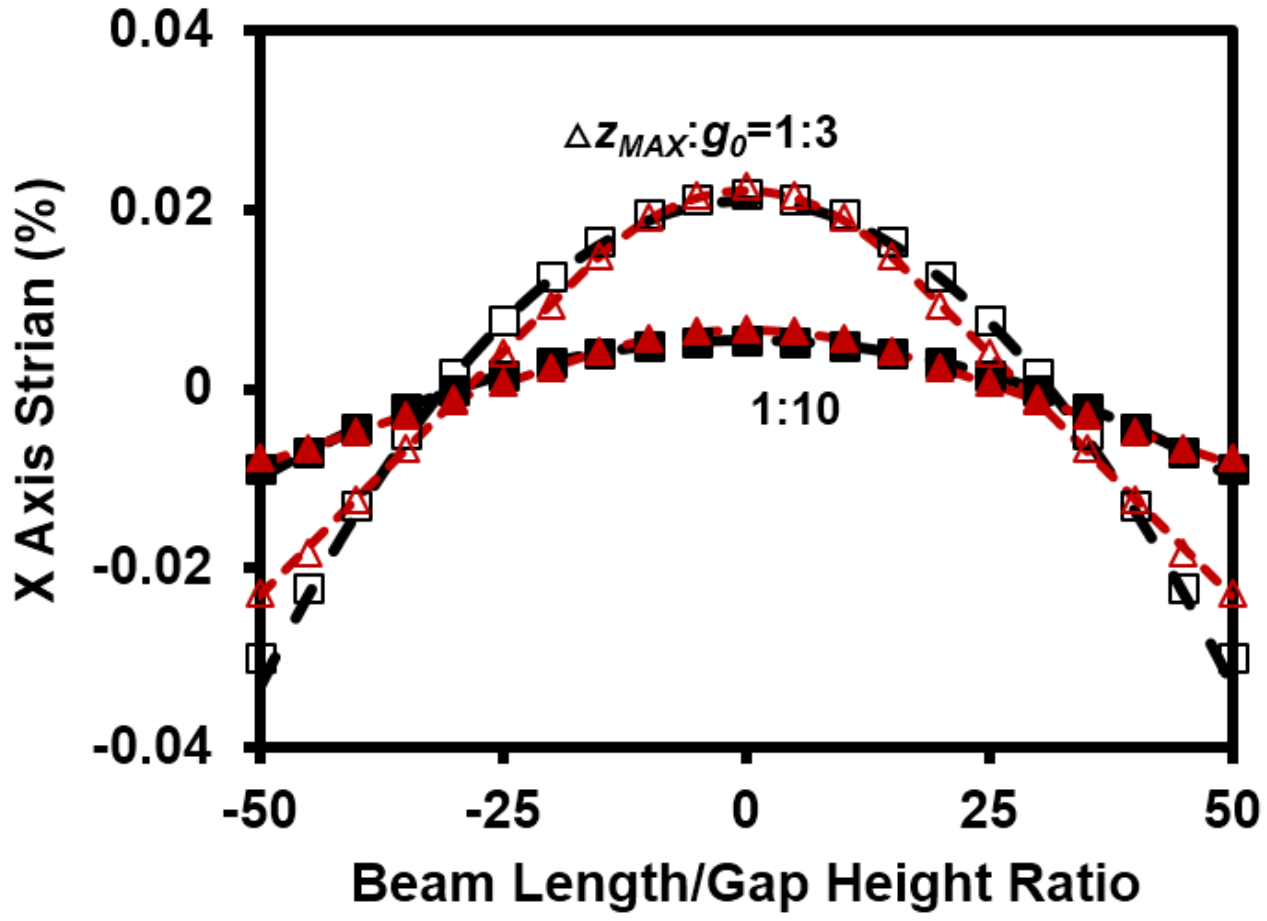

(a)

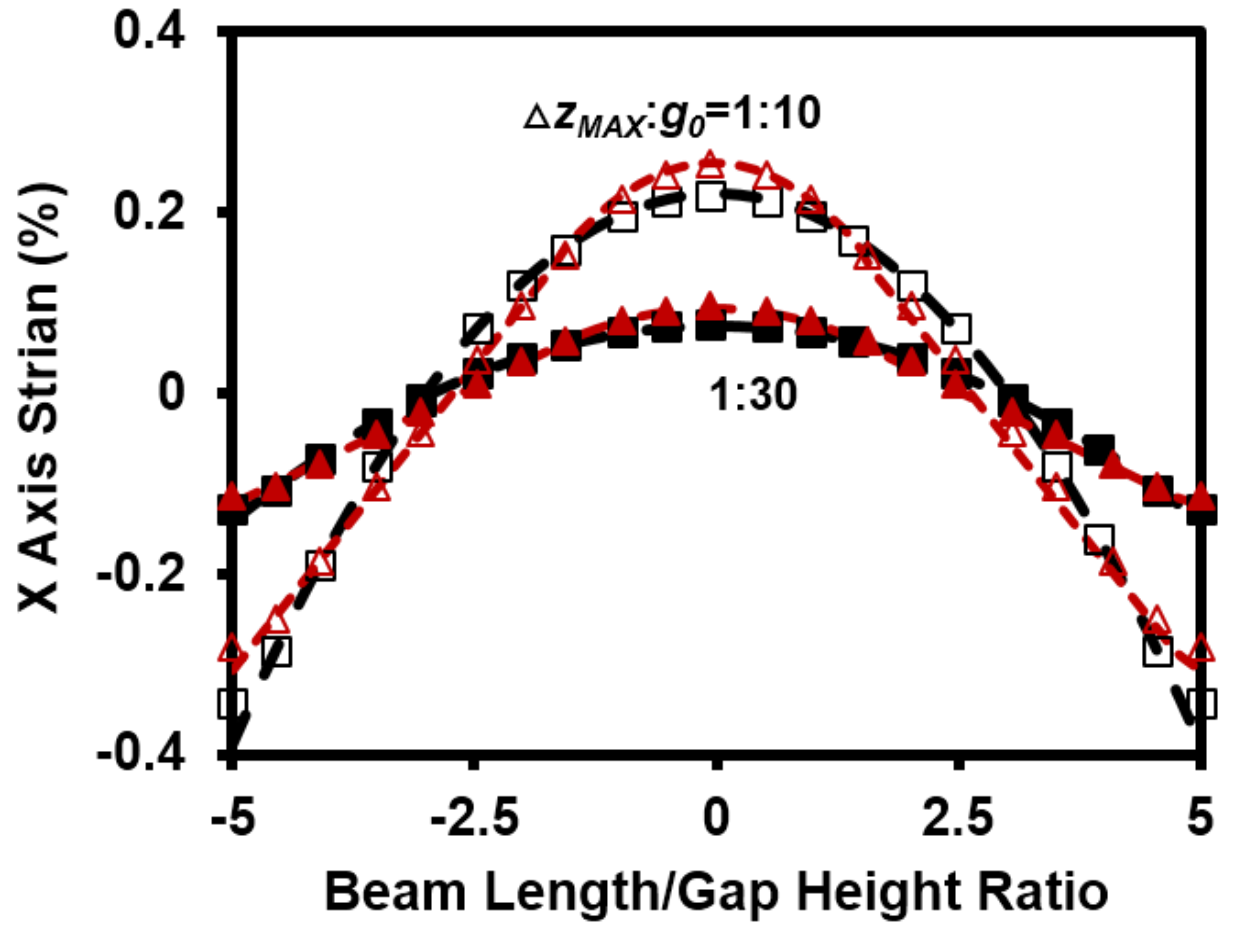

(b)

Fig. 2-11. ANSYS simulation results (symbols) of $x$ axis strain distribution when normalized beams length is (a) 100/1 and (b) 20/2. In (a), the hollow and solid symbols represent max deflection $\triangle z_{MAX}/g_0$ =1/3 and 1/10. In (b), they represent $\triangle z_{MAX}/g_0$ = 1/10 and 1/30.In both cases squares and triangles represent stationary to movable beam length ratio is 1 and 1/3, respectively. The dashed lines are analytical solutions.

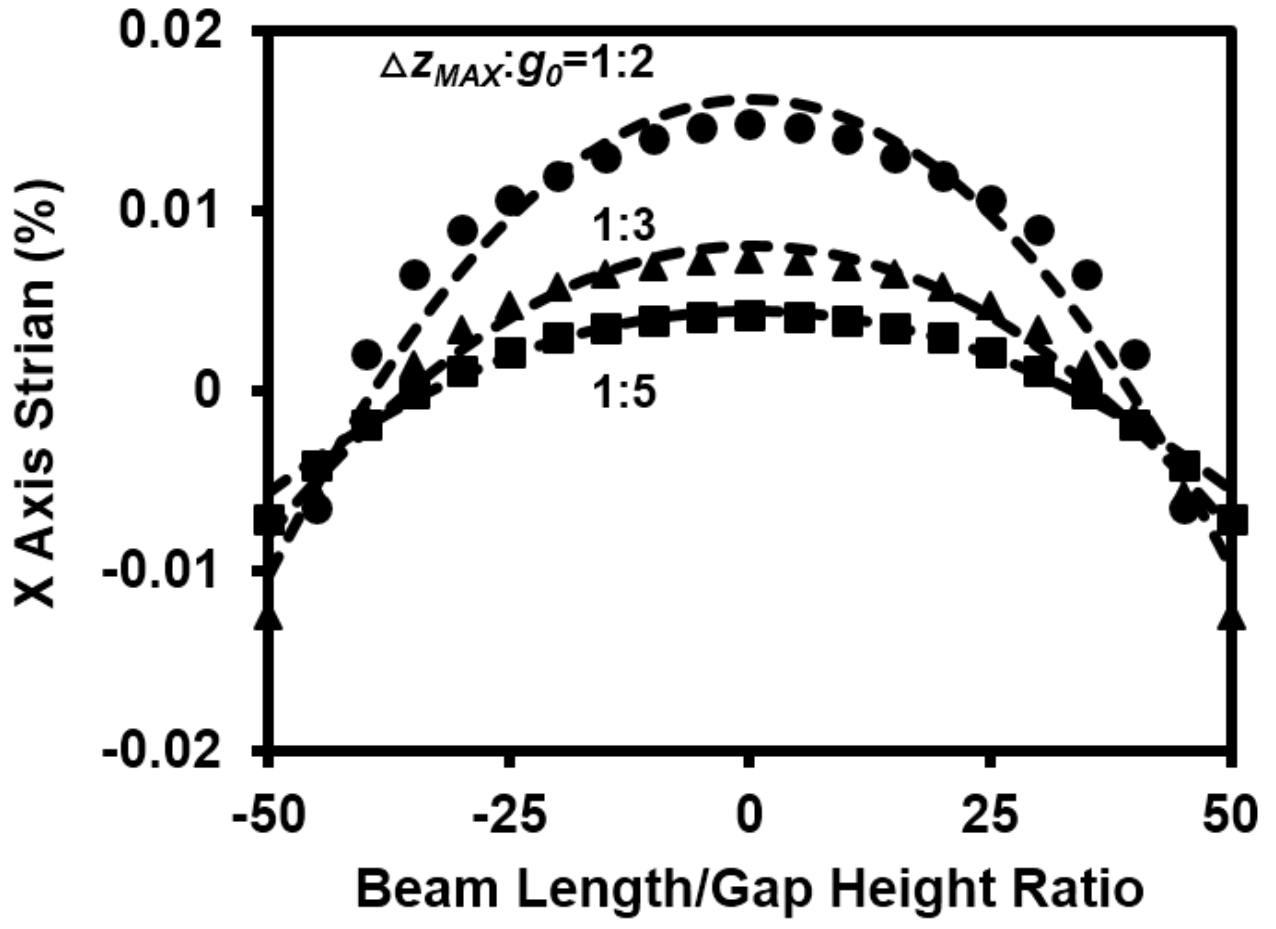

(a)

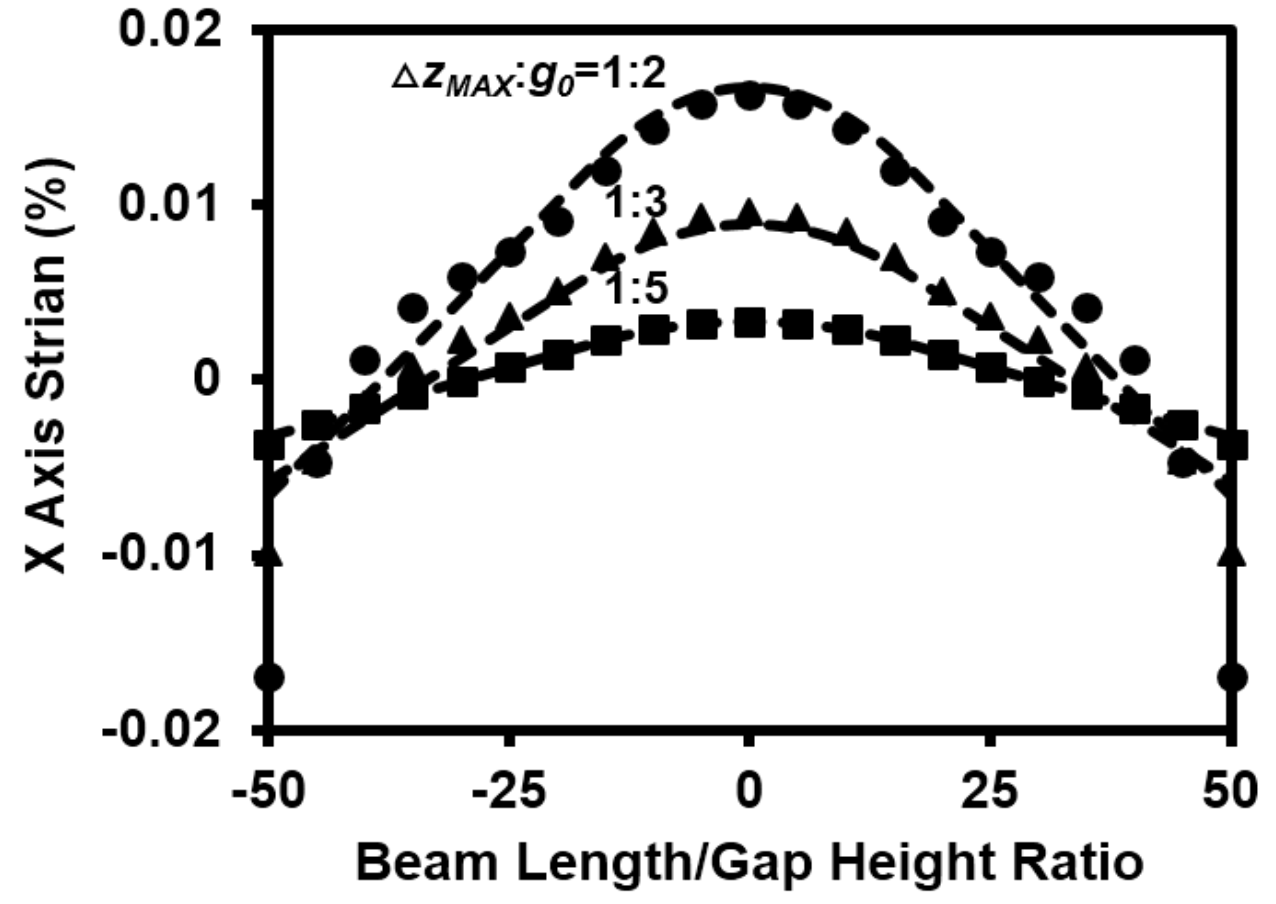

(b)

Fig. 2-12. ANSYS simulation results (symbols) of *x* axis strain distribution when stationary to movable electrode length ratio are (a) 1/1 and (b) 1/3. In both cases normalized max deflection $\triangle z_{MAX}/g_0$ = 1/2 (●), 1/3 (▲), and 1/5 (■). The dashed lines are analytical solutions.

To avoid beam material deform plastically, the *x* axis strain need to be below yield strain. From Fig. 2-11 and Fig. 2-12, the maximum strain occurs at the center and the edge of the beam. Therefore, the maximum strain due to bending and stretch are given by

$$\varepsilon_{bending,\max} = \begin{cases} \dfrac{\alpha_1 t \Delta z_{MAX}}{\ell^2}, & x = -\dfrac{\ell}{2} \\ \dfrac{\alpha_2 t \Delta z_{MAX}}{\ell^2}, & x = 0 \end{cases} \tag{2-25}$$

$$\varepsilon_{stretch} = \frac{\beta_2 \Delta z_{MAX}{}^2}{\ell^2} \tag{2-26}$$

where $\alpha_1 = \dfrac{8\left[3\left(1+4\dfrac{x}{\ell}\right)+\left(\dfrac{\ell'}{\ell}\right)^2\right]}{\left[2-2\left(\dfrac{\ell'}{\ell}\right)^2+\left(\dfrac{\ell'}{\ell}\right)^3\right]}, \alpha_2 = \dfrac{8\left[-12\left(\dfrac{x}{\ell}\right)^2+3\dfrac{\ell'}{\ell}-3\left(\dfrac{\ell'}{\ell}\right)^2+\left(\dfrac{\ell'}{\ell}\right)^3\right]}{\left[2\dfrac{\ell'}{\ell}-2\left(\dfrac{\ell'}{\ell}\right)^3+\left(\dfrac{\ell'}{\ell}\right)^4\right]}$, and

$$\beta_2 = \frac{16\left[63-105\left(\dfrac{\ell'}{\ell}\right)^2+133\left(\dfrac{\ell'}{\ell}\right)^4-96\left(\dfrac{\ell'}{\ell}\right)^5+21\left(\dfrac{\ell'}{\ell}\right)^6\right]}{105\left[2-2\left(\dfrac{\ell'}{\ell}\right)^2+\left(\dfrac{\ell'}{\ell}\right)^3\right]^2}$$

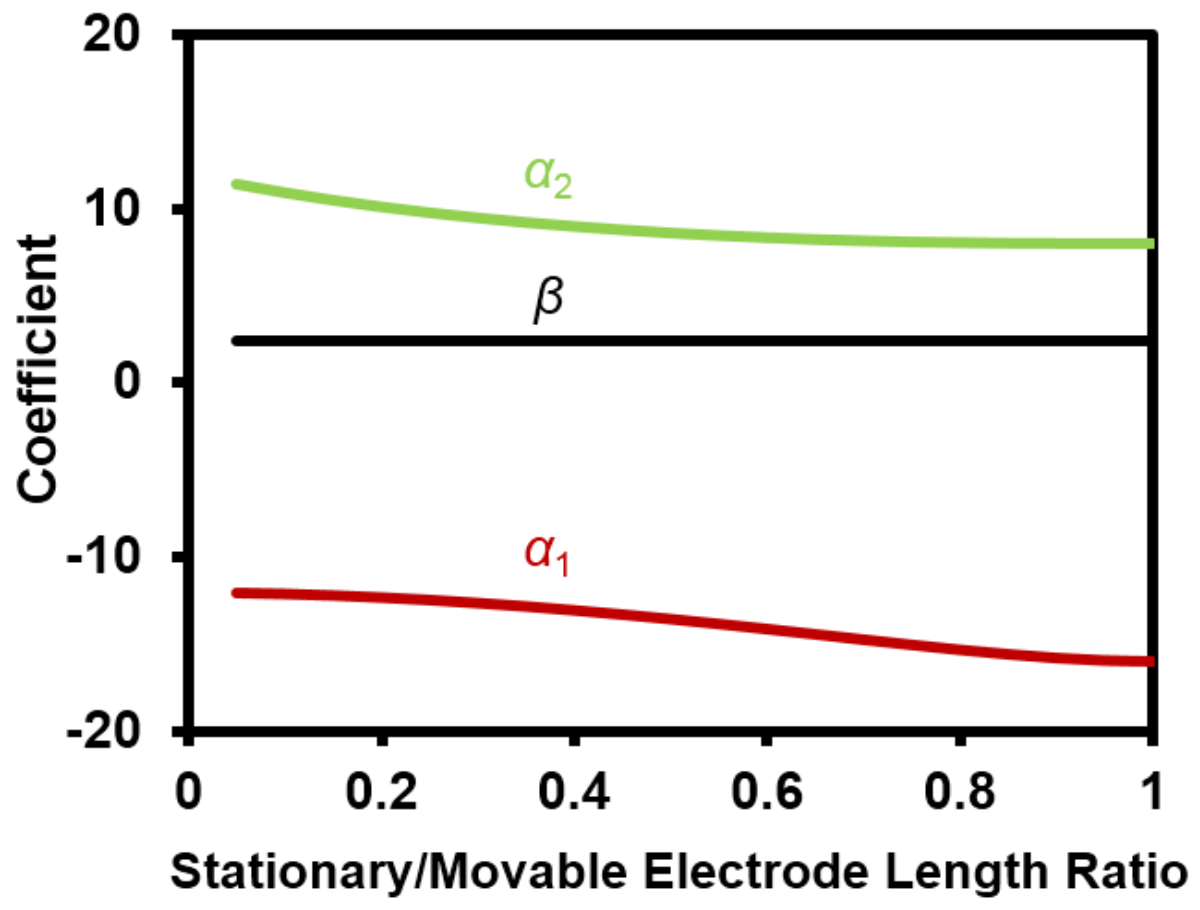

Fig. 2-13. $\alpha_1$, $\alpha_2$, and $\beta$ values as a function of stationary/movable electrode length ratio.

Fig. 2-13 shows the values of coefficient $\alpha_1$, $\alpha_2$, and $\beta$.

The overall maximum strain on *x* axis is given by

$$\varepsilon_{x,\max} = \begin{cases} \dfrac{\alpha_1 t \Delta z_{MAX}}{\ell^2} + \dfrac{\beta_2 \Delta z_{MAX}{}^2}{\ell^2} \\ \dfrac{\alpha_2 t \Delta z_{MAX}}{\ell^2} + \dfrac{\beta_2 \Delta z_{MAX}{}^2}{\ell^2} \end{cases} = \begin{cases} \dfrac{\Delta z_{MAX}{}^2}{\ell^2}\left(\alpha_1 \dfrac{t}{\Delta z_{MAX}} + \beta_2\right), & \text{edge} \\ \dfrac{\Delta z_{MAX}{}^2}{\ell^2}\left(\alpha_2 \dfrac{t}{\Delta z_{MAX}} + \beta_2\right), & \text{center} \end{cases} \tag{2-27}$$

The *x* axis maximum strain at the edge assumes idealized fixed-slope boundary condition, but the boundary condition is much more complicated in real cases and the assumption becomes invalid easily [26]. The *x* axis maximum strain at the center is

$$\frac{\beta_2 \Delta z_{MAX}{}^2}{\ell^2} + \frac{\alpha_2 t \Delta z_{MAX}}{\ell^2} < \varepsilon_{yield}, \tag{2-28}$$

$$\frac{-\alpha_2 - \sqrt{\alpha_2^2 + 4\varepsilon_{yield}\beta_2\left(\dfrac{\ell}{t}\right)^2}}{2\beta_2} < \frac{\Delta z_{MAX}}{t} < \frac{-\alpha_2 + \sqrt{\alpha_2^2 + 4\varepsilon_{yield}\beta_2\left(\dfrac{\ell}{t}\right)^2}}{2\beta_2}, \tag{2-29}$$

where $\varepsilon_{yield}$ is yield strain, $\varepsilon_{residual}$ is the strain caused by residual stress.

Since $\dfrac{\Delta z_{MAX}}{t} > 0$, (2-29) becomes

$$0 < \frac{\Delta z_{MAX}}{t} < \frac{-\alpha_2 + \sqrt{\alpha_2^2 + 4\varepsilon_{yield}\beta_2\left(\dfrac{\ell}{t}\right)^2}}{2\beta_2} \tag{2-30}$$

If the residual stress is the dominant factor in determining beam deform behaviors, (2-29) is still valid, but the strain caused by residual stress needs to be subtract from the yield strain and the coefficient needs to be modified correspondingly. In that case, (2-30) becomes $0 < \frac{\Delta z_{MAX}}{t} < \frac{-\alpha_2 + \sqrt{\alpha_2^2 + 4\left(\varepsilon_{yield} - \varepsilon_{residual}\right)\beta_2\left(\frac{\ell}{t}\right)^2}}{2\beta_2}$, and solving (2-13), (2-22),

$$\alpha_2 = \frac{4}{\left[2\frac{\ell'}{\ell} - \left(\frac{\ell'}{\ell}\right)^2\right]}, \quad \beta_2 = \frac{8\left(3 - 2\frac{\ell'}{\ell}\right)}{3\left(-2 + 2\frac{\ell'}{\ell}\right)^2}$$

The quadratic relationship between axial strain and $\frac{\Delta z_{MAX}}{\ell}$ indicates axial strain increases significantly as $\frac{\Delta z_{MAX}}{\ell}$ increase. This may become a serious problem for miniature switches, which often has large $\frac{\Delta z_{MAX}}{\ell}$ ratio compared to standard switches. Large $\frac{\Delta z_{MAX}}{\ell}$ ratio is beneficial to reduce switches temperature dependence and achieve faster switch time, but it increases the axial strain and potentially reduces the life cycle of the switch. These competing influence needs to be considered.

Actually, (2-30) is valid only before pull-in occurs, but it cannot predict the *x* axis strain when the deflection is comparable to gap height. In addition, it becomes invalidate when stretch effects outweighs bending effects at large displacement. However, (2-30) still provides the guides to design switches geometry to avoid the beam over-stretched in the suspended state.

If the microstructures are fabricated using materials that can sustain large strain (e.g. conductive polymers [27]), the yield strain ($\varepsilon_{yield}$) in (2-29) needs to be changed accordingly. For beams made by those materials, they can be elastic at larger displacement.

## *References*

[1] S. D. Senturia, A. Narayan , and J. White, "Simulating the behavior of MEMS devices: computational methods and needs." *IEEE Comput. Sci. Eng.*, vol. 4, no. 1, pp. 30–43, Jan. 1997.

[2] S. D. Senturia, "CAD challenges for microsensors, microactuators, and microsystems." *Proc. IEEE*, vol. 86, Aug. 1998, pp. 1611–1626.

[3] ANSYS Mechanical APDL Coupled-Field Analysis Guide, ANSYS, Inc., Canonsburg, PA, 2013, pp. 1–4.

[4] K. Slava and S. Shimon, "Higher order correction of electrostatic pressure and its influence on the pull-in behavior of microstructures," *J. Micromech. Microeng.*, vol. 16, no. 7, pp. 1382–1396, Jun. 2006.

[5] J. Teva, G. Abadal, Z.J. Davis, J. Verd, X. Borrisé, A. Boisen, F. Pérez-Murano, and N. Barniol. "On the electromechanical modelling of a resonating nano-cantilever-based transducer," *Ultramicroscopy*, vol. 100, no. 3, pp. 225–232, Aug. 2004.

[6] S. Chowdhury, M. Ahmadi, and W. C. Miller, "A comparison of pull-in voltage calculation methods for MEMS-based electrostatic actuator design." in *Proc. 1st Int. Conf. Sensing Tech.*, Nov. 2005, pp. 112–117.

[7] H. C. Nathanson, W. E. Newell, R. A. Wickstrom, and J. R. Davis, "The resonant gate transistor," *IEEE Trans. Electron Devices*, vol. 14,no. 3, pp. 117–133, Mar. 1967.

[8] S. Timoshenko, History of Strength of Materials. New York, NY: McGraw-Hill, 1953, pp. 30–35.

[9] P. M. Osterberg and S. D. Senturia, “M-TEST: A test chip for MEMS material property measurement using electrostatically actuated test structures,” *J. Microelectromech. Syst.*, vol. 6, pp. 107–118, Jun. 1997.

[10] S. D. Senturia, *Microsystem Design*. Boston: Kluwer academic publishers, 2001, pp. 230–231.

[11] P. Osterberg, H. Yie, X. Cai, J. White, and S. Senturia, "Self-consistent simulation and modeling of electrostatically deformed diaphragms," in *Proc. IEEE Int. Conf. Microelectromech. Syst.*, Jan. 1994, pp. 28–32.

[12] G. M. Rebeiz, RF MEMS *Theory*, *Design*, and *Technology*. Hoboken, NJ: Wiley, 2003, pp. 23–27.

[13] D. J. Ijntema and H. A. C. Tilmans, “Static and Dynamic Aspects of an Air-gap Capacitor,” *Sensors Actuators A*, vol. 35, no. 2, pp. 121–128, Dec. 1992.

[14] H. A. C. Tilmans and R. Legtenberg, “Electrostatically driven vacuum-encapsulated polysilicon resonators: Part II. Theory and Performance,” *Sensors Actuators A*, vol. 45, no.1, pp. 67–84, Oct. 1994.

[15] C. Palego, J. Deng, Z. Peng, S. Halder, J. C. M. Hwang, D.Forehand, D. Scarbrough, C. L. Goldsmith, I. Johnston, S. K.Sampath, and A. Datta, “Robustness of RF MEMS capacitive switches with molybdenum membranes,” *IEEE Trans. Microwave Theory & Tech.*, vol. 57, pp. 3262-3269, Dec. 2009.

[16] J. E. Mehner, L. D. Gabbay, and S. D. Senturia, "Computer-aided generation of nonlinear reduced-order dynamic macromodels. II. Stress-stiffened case," *J. Microelectromech. Syst.*, vol. 9, no. 2, pp. 270–278, Jun. 2000.

[17] K. Van Caekenberghe, "Modeling RF MEMS Devices," *IEEE Microwave*, vol. 13, no. 1, pp. 83–110, Feb. 2012.

[18] C. Gui, R. Legtenberg, H. A. C. Tilmans, J. H. J. Fluitman, and M. Elwenspoek, “Nonlinearity and hysteresis of resonant strain gauges,” *J. Microelectromech. Syst.*, vol. 7, no. 1, pp. 122–127, Mar.1998.

[19] T. Veijola, “Compact models for squeezed-film dampers with inertial and rarefied gas effects,” *J. Micromech. Microeng.*, vol. 14, no. 7, pp. 1109–1118, Jun. 2004.

[20] S. Pamidighantam, R. Puers, K. Baert, and H. A. C. Tilmans, "Pull-in voltage analysis of electrostatically actuated beam structures with fixed–fixed and fixed–free end conditions," *J. Micromech. Microeng*., vol. 12, no. 4, pp. 458–464, Jul. 2002.

[21] E. S. Hung and S. D. Senturia, "Extending the travel range of analog-tuned electrostatic actuators," *J. Microelectromech. Syst.*, vol. 8, no. 4, pp. 497–505, Dec. 1999.

[22] J. I. Seeger and B. E. Boser, "Charge control of parallel-plate, electrostatic actuators and the tip-in instability," *J. Microelectromech. Syst.*, vol. 12, no. 5, pp. 656–671, 2003.

[23] K. B. Lee, "Closed-form solutions of the parallel plate problem," *Sensors and actuators A*:*Physical*, vol. 133, no. 2, pp. 518–525, Jun. 2007.

[24] J. E. Mehner, L. D. Gabbay, and S. D. Senturia, “Computer-aided generation of nonlinear reduced-order dynamic macromodels II: Stress-stiffened case,” *J. Microelectromech*. *Syst*., vol. 9, pp. 270–278, Jun. 2000.

[25] C. O'Mahony, M. Hill, R. Duane, A. Mathewson, "Analysis of electromechanical boundary effects on the pull-in of micromachined fixed-fixed beams," *J. Micromech. Microeng.* vol. 13, no. 4, pp. S75–S80, Jun. 2003.

[26] Y.C. Hu, P.Z. Chang, W.C. Chuang, "An approximate analytical solution to the pull-in voltage of a micro bridge with an elastic boundary," *J. Micromech. Microeng*., vol. 17, no. 9, pp. 1870–1876, Aug. 2007.

[27] A. Huang, V. T. S. Wong and C-M. Ho, “Conductive silicone based MEMS sensor and actuator,” in *IEEE Int. Conf. on Solid-State Sensors, Actuators and Microsystems Tech. Dig.*, Jun. 2005, pp 1406–1411.

# Chapter 3 Theory and Hyperbolic Models

## 3.1 Hyperbolic Model

### 3.1.1 Limitations on Parallel-plate Approximation

In fact, as the fixed-fixed top movable electrode starts to deform under the electrostatic force, the vertical displacement is a function of horizontal location (not a constant), which invalidates the parallel-plate assumption. With the help of ANSYS finite element software, the limitations of parallel-plate assumption are explored.

In ANSYS finite element software, the Direct Coupling method is employed as the computational approach in this paper. It gives a single solution by using a coupled formulation when the coupled-field involves strongly coupled-physics [1]. The Dell T7600 workstation with 16 cores and 64 GB memory is used for ANSYS simulation. The typical computation time for one MEMS configuration is 5 minutes.

Neglecting the fringing capacitance, the capacitance of RF MEMS switches based on parallel-plate model for two perfectly flat plates before pull-in is:

$$C = \varepsilon_0 w \ell' \Big/ \left[ g \left( 1 + z' \right) \right] \tag{3-1}$$

where $C$ is the capacitance, $\varepsilon_0$ is vacuum permittivity, w is beam width, $z'$ is the normalized beam deflection.

To study the charge density distribution on the beam, the MEMS switch beam is evenly divided into 200 sections and apply parallel-plate assumption individually. The charge density calculated from this piecewise parallel-plate assumption is given by

$$Q' = CV' = \varepsilon_0 A_{unit} V' \Big/ \left\{ g \left[ 1 + z'(x) \right] \right\} \tag{3-2}$$

where Q′ is the charge density, z′ (x) is the normalized beam deflection in each section, *Aunit* is the unit area, and V′ is 1 V.

The switch structures employed are with $\alpha$ ranges from 1/100 to 1/10, $t$ = 0.6 µm, $g$ = 2 µm, $\beta$ = 1, $z'(0) \approx 1/3$. The $z'(x)$ used in (3-2) is from ANSYS simulation results.

The switches parameters are defined as follows:

$$\begin{aligned} &\alpha = g/\ell; \quad \beta = \ell'/\ell; \quad \gamma = t/\ell; \quad \delta = k_S g^2 / k_0; \\ &x' = x/\ell; \quad -0.5 \le x' \le 0.5; \\ &z' = z/g; \quad -1 \le z' \le 0 \end{aligned}$$

where $k_S$ is the nonlinear spring constant, $k_0$ is the linear spring constant.

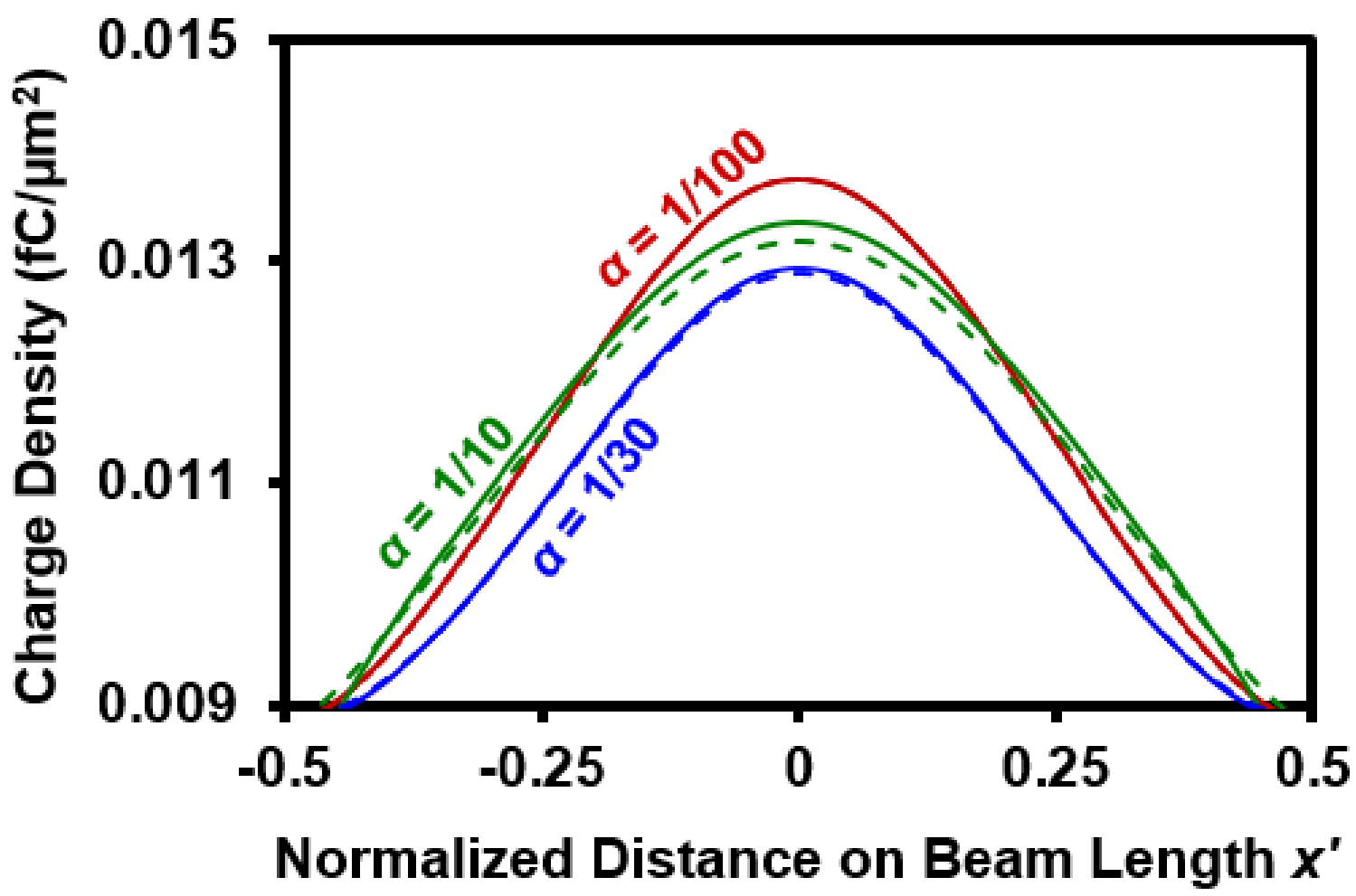

Fig. 3-1. Charge density calculated by using piecewise parallel plate model (dashed curves) and ANSYS simulation (solid curves).

Fig. 3-1 shows that the charge density at the beam center is about 44% higher than that at the edge. When α is equal to 1/100, the piecewise model predicts charge distribution accurately. But as α increase to 1/10, the difference at center point is about 2% even each section length is only 1/200 of overall length. This is mainly because the parallel-plate approximation cannot predict the slope of the movable electrode, which becomes significant at high α.

Under the parallel-plate approximation, the effective linear spring constant $k_0$ is given by [9]:

$$k_0 \approx k_0' + k_0'' = 32Ew\gamma^3 / \left[2 - 2\beta^2 + \beta^3\right] + 8\sigma(1-\nu)w\gamma / (2-\beta). \tag{3-3}$$

where $E$ is Young's modulus, $\sigma$ is the residual stress, $\nu$ is Poisson's ratio, $k_0'$ and $k_0''$ are used to designate terms attributed to bending and residual stress effects, respectively.

As the deflection increases toward its maximum value, the beam is stretched, increasing the in-plane stress beyond the stresses determined from simple bending behavior. The increase in stress is quadratic in the deflection z and adds an extra nonlinear term to the restoring force. However, for high-aspect-ratio MEMS, the stretching restoring force is negligible for small $z$ at unactuated state [2]. Without considering the nonlinear spring constant caused by the stretching effects, the predicted pull-in voltage is $\sqrt{8k_0 g^3 / (27\varepsilon_0 W \ell')}$. In the case of large deflections ($z(0)/t > 3$), it gave only one-fourth the values when compared with finite element simulation results [3]–[4]. In addition, the nonlinear stretching effects can extend the maximum travel range of the beam and make the capacitance-voltage relationship linear, which parallel-plate assumption cannot predict.

### *3.1.2 Hyperbolic Model and Electrostatic Field*

To overcome the limitations of parallel-plate assumption, the hyperbolic model considering stretching effect is proposed. The hyperbolic function satisfies the Laplace equation and its shape is similar to the MEMS beam shape under electrostatic force. This model, together with conformal mapping techniques, can calculate the electric field force, charge distribution, and corresponding capacitance analytically.

The general form of hyperbolic function is given by [5]:

$$\left(z'+1\right)^2 \big/ a'^2 - x'^2 \big/ b'^2 = 1 \tag{3-4}$$

where *a′=a/g*, *b′=b/ℓ*. *a′* and *b′* are coefficients that determine the hyperbolic function. The curve pass through point ($x'$, $z'$) = (0, $z'$(0)). Meanwhile, the beam anchored at points (−1/2, 0) and (1/2, 0). Substitute these points into (4), the relation is obtained:

$$b' = 1/2\left(a'^{-2} - 1\right)^{-1/2}. \tag{3-5}$$

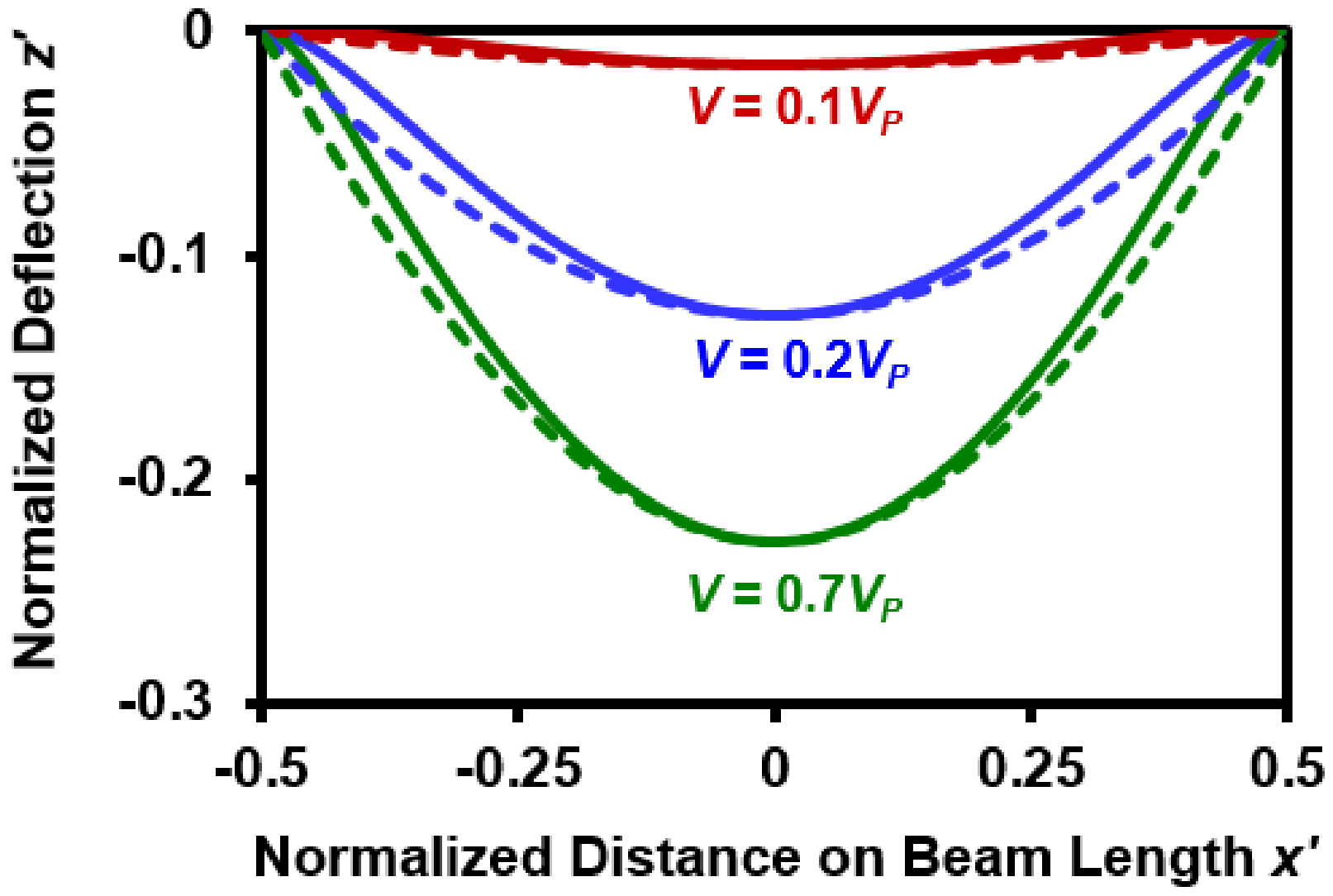

Fig. 3-2. Modeled (solid curves) versus simulated (dashed curves) beam shape for different applied voltages. ANSYS simulation configuration is aluminum beam, $\ell = \ell' = 10$ μm, $t = 0.6$ μm, $g = 1$ μm. Beam yields at $V = 0.7V_P$.

Fig. 3-2 shows the difference between modeled beam shape and ANSYS simulation results. They agree well in general and the main difference occurs at the edge of the beam, which has less influence on capacitance than the center part.

To calculate the accurate capacitance, assuming the complex function is:

$$\begin{cases} w_i = u + iv \\ z_i' = (z'+1) + ix'/\alpha \end{cases} \tag{3-6}$$

where $u$ is the potential, $v$ is the electric field, $x'$ and $z'$ are normalized beam position coordinates.

Assume transform function is:

$$w_i = \cos^{-1}\left(z_i'/r'\right) \tag{3-7}$$

From (3-7), it can be obtained

$$\begin{cases} z'+1 = r'\cos(u)\cosh(v) \\ x' = -r'\alpha\sin(u)\sinh(v) \\ r' = \sqrt{a'^2 + b'^2/\alpha^2}. \end{cases} \tag{3-8}$$

For $u$ and $v$ satisfy the Cauchy-Riemann equations [5],

$$\partial u/\partial z' = \alpha\,\partial v/\partial x', \partial v/\partial z' = -\alpha\,\partial u/\partial x' \tag{3-9}$$

$$dw_i/dz_i' = \left(\partial u/\partial z' - i\alpha\,\partial u/\partial x'\right). \tag{3-10}$$

The electric field $\xi$ is:

$$\begin{aligned} |\xi| &= U_0/g\left|\left(\partial u/\partial z'\right)^2 + \alpha^2\left(\partial u/\partial x'\right)^2\right|^{1/2} = U_0/g\left|dw_i/dz_i'\right| \\ &= U_0/g\left[r'^2\cosh^2(v) - \left(z'+1\right)^2/\cosh^2(v)\right]^{-1/2}. \end{aligned} \tag{3-11}$$

Without considering the applied voltage, for the stationary electrode at $z' = -1$, the potential is $u(z' = -1) = \pi/2$. For the point $(x', z') = (0, a'-1)$ on the movable electrode, the potential $u(0, a'-1) = \cos^{-1}(a'/r') = \tan^{-1}[b'/(\alpha a')]$. The movable electrode is equipotential, thus the movable electrode potential $u = \tan^{-1}[b'/(\alpha a')]$. The potential difference between movable and stationary electrode $\Delta V = \pi/2 - \tan^{-1}[b'/(\alpha a')] = \tan^{-1}(\alpha a'/b')$. If the voltage V is applied between two electrodes, $V = \Delta V\, U_0$, where $U_0$ is the scale factor.

### *3.1.3 Hyperbolic Model Coefficient Determination*

From (3-5), the only unknown coefficient needs to be determined is a′. It can be solved by using Rayleigh's method, which assumes the deflection shape uses only one unknown coefficient. Based on this shape, the strain energy is equated to the work done by the applied force so the unknown coefficient is solved [6].

The total work *Wq* performed on the beam by the force per unit length *q* is given by [6]:

$$\begin{cases} W_q = 1/2 \int_{-\frac{\ell}{2}}^{\frac{\ell}{2}} q \cdot z dx \\ q = 1/2\, \varepsilon_0 \xi^2 \end{cases} \tag{3-12}$$

The work $W_N$ due to the constrained beams is given by:

$$W_N = N^2 \ell / (2AE) - AE/4\ell \left[ \int_{-\frac{\ell}{2}}^{\frac{\ell}{2}} \dot{z}^2 dx \right]^2 \tag{3-13}$$

where $N = AE/(2\ell) \int_{-\frac{\ell}{2}}^{\frac{\ell}{2}} \dot{z}^2 dx$ , $A = wt$, $\dot{z} = dz/dx$..

The increase in strain energy *U*, neglecting shear effects, is given by:

$$\begin{aligned} U = N^2 \ell / (2AE) + Ewt^3/24 \int_{-\frac{\ell}{2}}^{\frac{\ell}{2}} \ddot{z}^2 dx \\ + \sigma A/2 \int_{-\frac{\ell}{2}}^{\frac{\ell}{2}} \dot{z}^2 dx \end{aligned} \tag{3-14}$$

where $\ddot{z} = d^2 z / dx^2$

The total work equals to strain energy gives:

$$W_q + W_N = U \tag{3-15}$$

The unknown coefficient $a'$ is determined by solving (3-15) and the beam shape will be determined accordingly.

After $a'$ is determined, $z'(0)$ can be obtained, so center deflection vs. voltage is obtained. Further, the capacitance vs. voltage can be derived.

$$\begin{aligned} \left|\xi\left(z'=-1\right)\right| &= U_0/g\left|\left(r'^2 + x'^2/\alpha^2\right)^{-1/2}\right| \\ &= U_0/g\left|\partial \sinh^{-1}\left[x'/\left(\alpha r'\right)\right]/\partial\left(x'/\alpha\right)\right|. \end{aligned} \tag{3-16}$$

The capacitance is calculated by using the total charge on stationary electrode ($z' = -1$) divides the potential difference between two electrodes.

$$\begin{aligned} C &= 2\int_0^{\frac{\ell'}{2}} \varepsilon_0 \xi\left(z'=-1\right)dx \bigg/ \left[U_0 \tan^{-1}\left(\alpha a'/b'\right)\right] \\ &= 2\varepsilon_0 \sinh^{-1}\left[1/2\,\beta/\left(\alpha r'\right)\right] / \tan^{-1}\left(\alpha a'/b'\right). \end{aligned} \tag{3-17}$$

### ***3.1.4 Nonlinear Spring Constant and Pull-in Voltage***

Calculating the axial strain $\varepsilon_{stretch}$ associated with stretching effects based on the hyperbolic shape gives:

$$
\begin{aligned}
\varepsilon_{stretch} &\approx 1/\ell \int_{-\frac{\ell}{2}}^{0} \left(dz/dx\right)^2 dx \\
&= 2\alpha^2\left(1-a'^2\right) - 2\alpha^2 a'\sqrt{1-a'^2}\cdot\tan^{-1}\left(\sqrt{1-a'^2}\Big/a'\right) \\
&\approx 2\alpha^2\left(1-a'^2\right) - 2\alpha^2 a'\sqrt{1-a'^2} \\
&\cdot\left[\sqrt{1-a'^2}\Big/a' - 1\Big/\left(3a'^3\right)\left(1-a'^2\right)^{3/2}\right] \\
&= 2\alpha^2\big/3\left(1/a'-a'\right)^2 \\
&\approx 2\alpha^2\big/3\left\{1\big/\left[1+z'(0)\right]-1-z'(0)\right\}^2 \\
&= 8\alpha^2 z'(0)^2\big/3.
\end{aligned} \tag{3-18}
$$

The axial stress $\sigma_{stretch}$ associated with stretching effects is given by:

$$\sigma_{stretch} = E\varepsilon_{stretch} = 8E\alpha^2 z'(0)^2\big/3 \tag{3-19}$$

which is the same as in [7]. Calculating nonlinear spring constant $k_S$ gives:

$$
\begin{aligned}
k_S &= F_{stretch}\big/z(0)^3 \\
&= 8E\varepsilon_{stretch} wtz(0)\Big/\left[\ell(2-\beta)z(0)^3\right] \\
&= 64/3\,Ewt\Big/\left[\ell^3(2-\beta)\right]
\end{aligned} \tag{3-20}
$$

where $F_{stretch}$ is the stretching restoring force.

Superposing the linear spring force and nonlinear spring forces gives:

$$z' + k_s g^2 z'^3\big/k_0 = z' + \delta z'^3 = \int_{-\frac{\ell}{2}}^{\frac{\ell}{2}} q\,dx \Big/ \left(k_0 g\right). \tag{3-21}$$

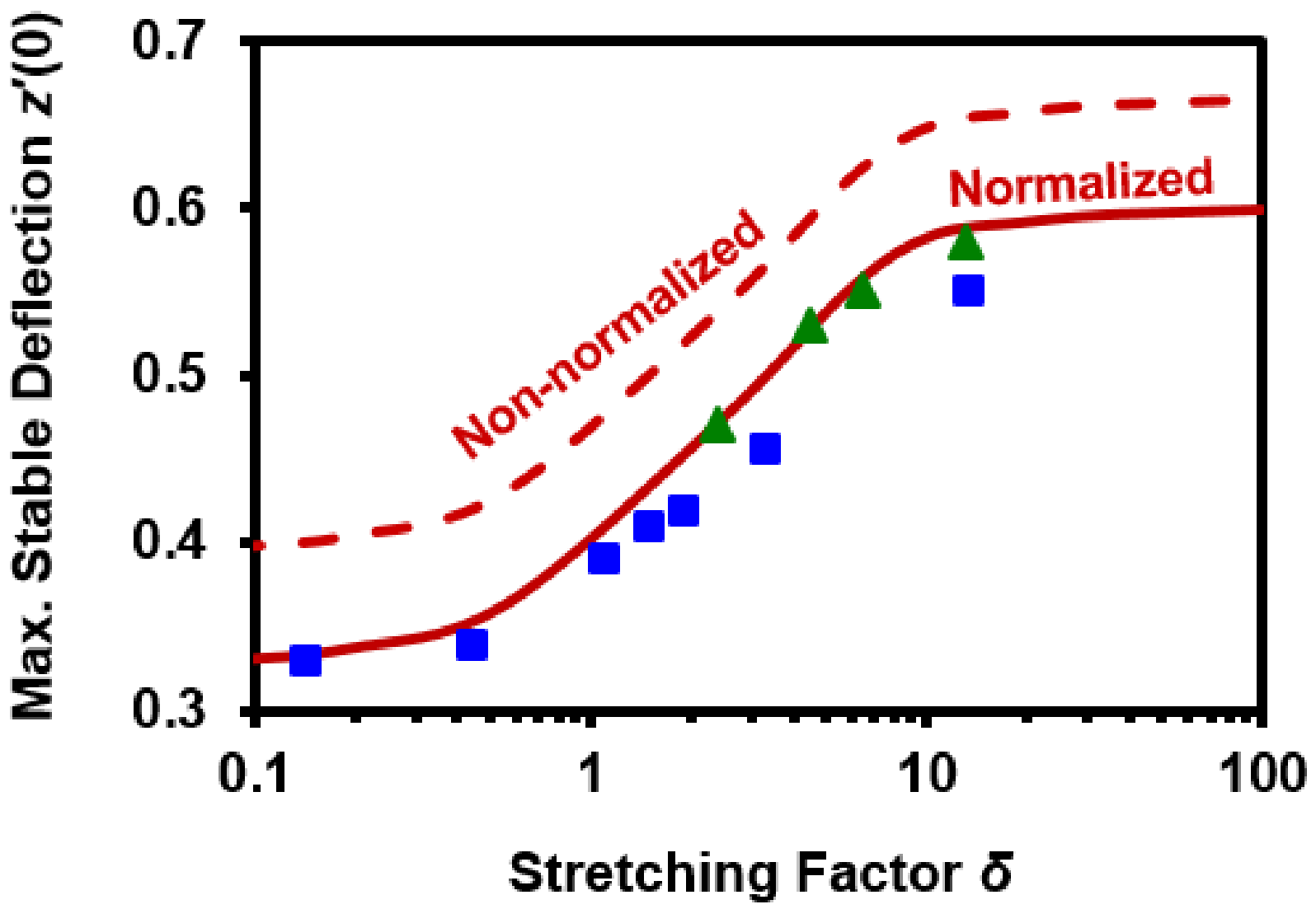

Fig. 3-3. Hyperbolic model (solid and dashed curve), ANSYS simulation results (■), and simulation results from [1] (▲) for maximum stable deflection $z'(0)$. It shows strong dependence on stretching factor $\delta$ ($k_s g^2/k_0$).

From (3-21), it can be seen that stretching factor ($\delta$) is the key parameter that determines the displacement subject to the electrostatic force. To find the relationship between maximum stable $z'(0)$ and $\delta$, the ANSYS finite element simulation employs a fixed-fixed aluminum beam plate capacitor as shown in Fig. 2-2 with $\ell = \ell' = 300$ µm, $t = 0.6$ µm, and gap height $g = 3$–$9$ µm. The residual stress σ varies from 0 to 150 MPa, so the $\delta$ ranges from 0.1 to 100. Solving (3-21) for $dV/dz' = 0$, the numerical solutions for normalized maximum stable deflection is obtained. In Fig. 3-3, all the results demonstrate a similar trend. The dashed curve is the results directly calculated from hyperbolic model.

The solid curve is the result of normalizing the dash curve to maximum stable $z'(0) = 1/3$ at $\delta = 0$. The difference between dashed and solid curves is a constant 0.06. The reason for normalizing is that as $\delta$ approaches zero, the linear spring constant $k_0$ dominates the beam deformation behavior and the maximum deflection before pull-in is $z'(0) = 1/3$. As $\delta$ increases, the nonlinear spring constant extends the maximum stable beam travel range before pull-in occurs. For example, in the extreme case ($\delta \approx 100$), the maximum deflection approaches 3/5 of the gap height $g$ [8]. The extension of the maximum travel range is preferred in varactors, micro mirrors and resonators [8]–[9].

When the stretching is important, it requires nonlinear spring forces greater than each of linear spring forces. Therefore,

$$\begin{cases} k_s z(0)^3 > k_0' z(0) \\ k_s z(0)^3 > k_0'' z(0). \end{cases} \tag{3-22}$$

Solving (3-22), it gives $z(0)/t > \sqrt{3/2}$ and $\sigma_{stretch} > \sigma(1-\nu)$.

## 3.2 *Effects of Stationary Electrode Thickness and Substrate*

In section 3.1.3, (3-17) shows that the capacitance of stationary electrode with zero thickness.

When the thickness of stationary electrode is considered, the charges distributed on the stationary electrode side and back cannot be ignored. Fig. 3-4(a) shows electric field between the stationary electrode and movable beam, which indicates the influence of charges at stationary electrode side and back. These charges influence the MEMS switch capacitance. Fig. 3-4(b) shows assumed field line in the analytical solution.

To study the impact of stationary electrode thickness on capacitance, a compact analytical solution is derived. The analytical solutions are compared with ANSYS simulation results in three sets of configurations. In all sets, the beam material is aluminum, beam thickness is 0.6 μm and no residual stress is applied. In addition, the switch $g_0$ is fixed at 1 μm. In the first set, $\ell/g_0$=100, $\ell'/\ell$=1/3, and $t'$ to $g_0$ ratio ranges from 1/10 to 1. In the second set, $\ell/g_0$=100, $t'/g_0$=0.6, and $\ell'$ to $\ell$ ratio ranges from 1/10 to 1. In the third set, $\ell/g_0$=10, $\ell'/\ell$=1, and $t'$ to $g_0$ ratio ranges from 1/10 to 1.

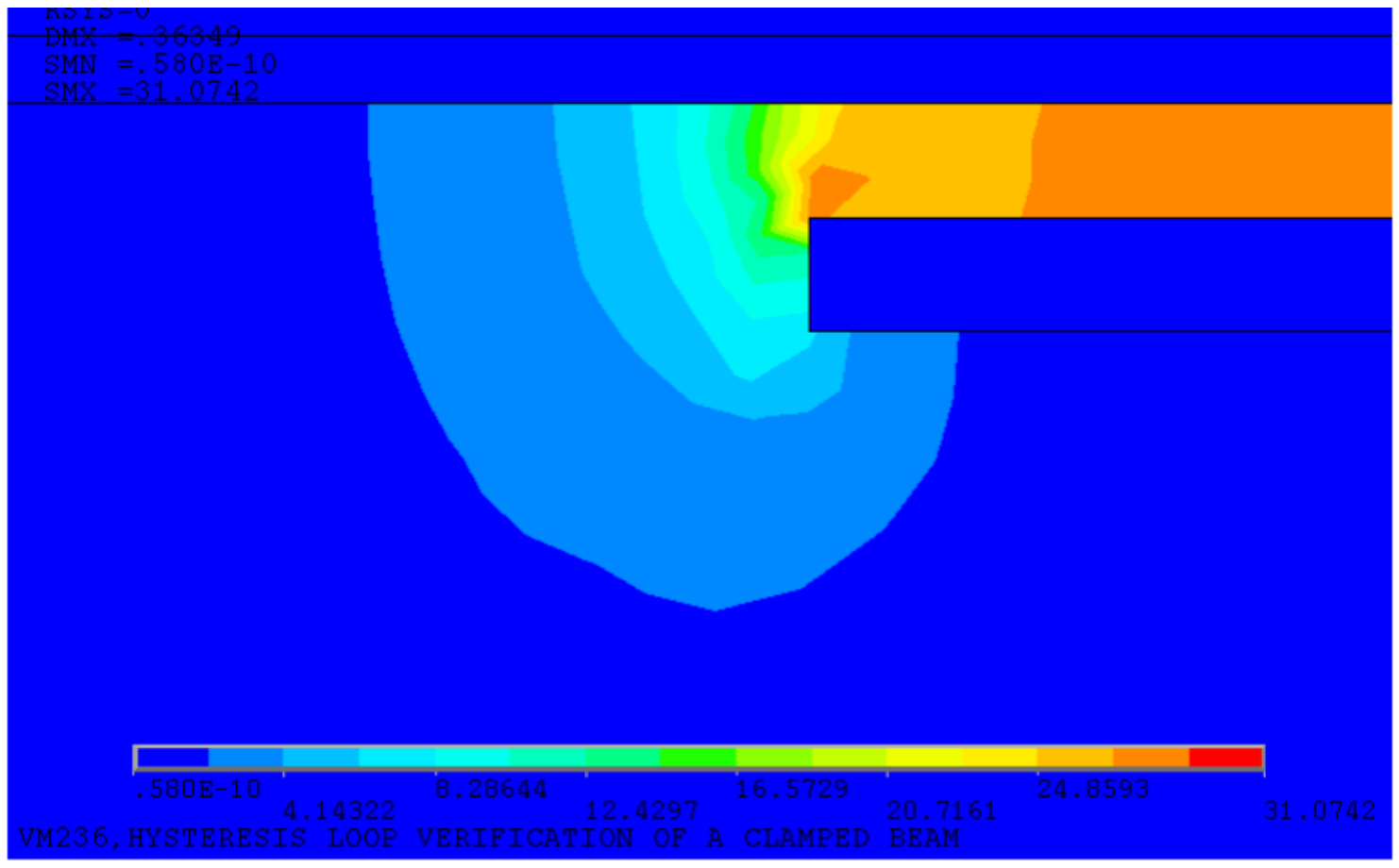

(a)

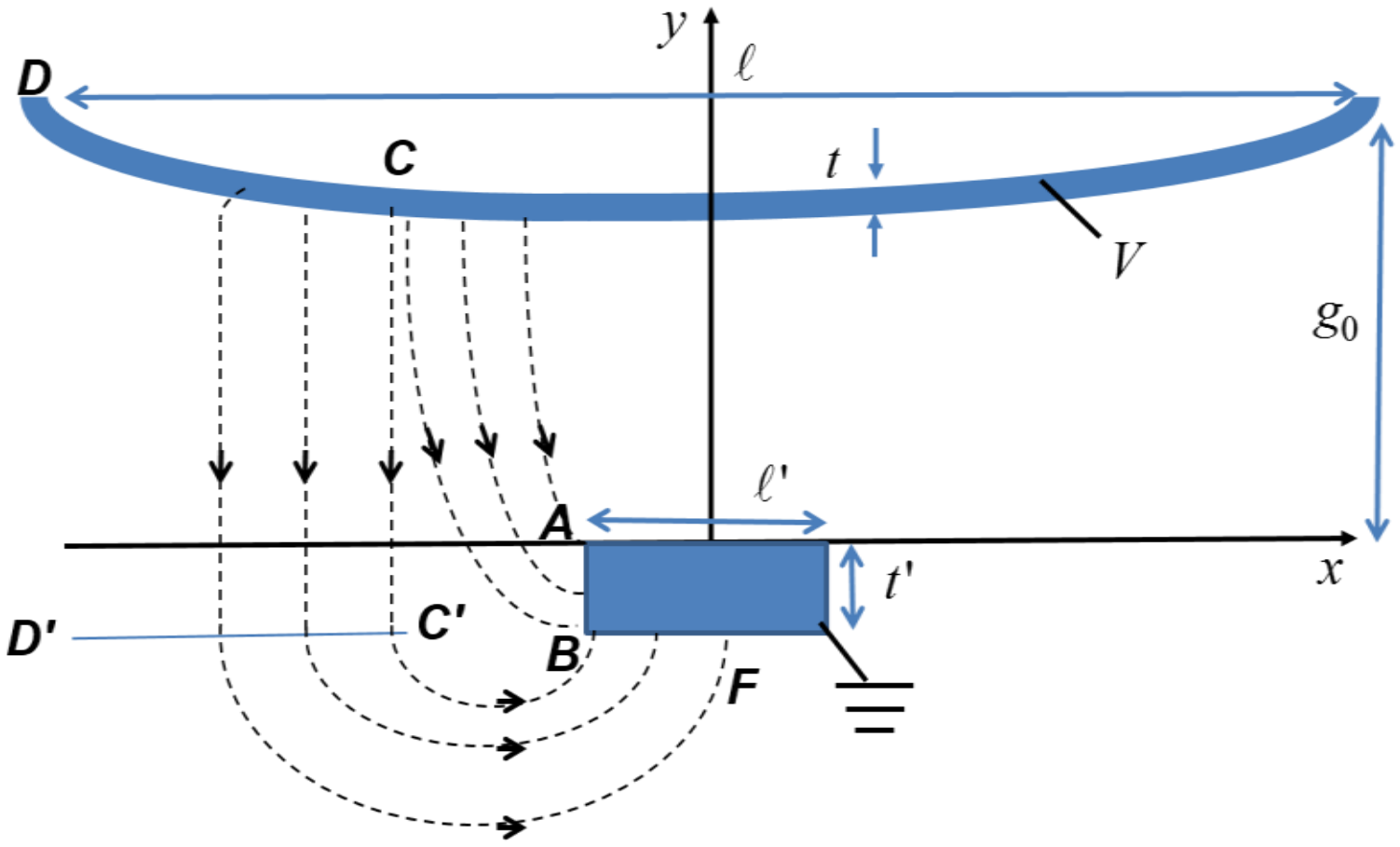

(b)

Fig. 3-4. (a) ANSYS simulation results of electric field distribution of a MEMS switch when $\ell'/\ell$=1/3, $\ell/g_0$=1/100, $t'/g_0$=1. (b) Assumed electric field line at stationary electrode side and back when $\ell'<\ell$.

To obtain the analytical solution for left side capacitance, we assume the electric filed lines between movable electrode and stationary electrode left side (AB) are represented by confocal ellipses [10]. This is shown in Fig. 3-5(a). The corresponding equipotential lines are presented by the confocal hyperbolas. Apply the transformation

$w = \arccos\left(\frac{z}{R_1}\right)$

$$\begin{cases} x = R_1 \cos(u)\cosh(v) \\ y = -R_1 \sin(u)\sinh(v) \end{cases} \tag{3-23}$$

where $R_1 = \frac{1}{2} g_0$ is the focus of the ellipse, $v$ is the electric field, and $u$ is the potential.

For the electrical field along AB:

$$E = -\frac{\partial u}{\partial x}\hat{x} - \frac{\partial u}{\partial y}\hat{y} = -\frac{\partial u}{\partial y}\hat{y} = \frac{\partial v}{\partial x}\hat{y} \tag{3-24}$$

where $\hat{x}$ and $\hat{y}$ represents the $x$ and $y$ direction. For AB, $u$ = 0, substitute (3-23) into (3-24)

$$|E| = \left| \frac{\partial \cosh^{-1}\left(\frac{x}{R_1}\right)}{\partial x} \right| \tag{3-25}$$

$$Q = \int_A^B \varepsilon_0 |E| \, dx = \varepsilon_0 \int_A^B \left| \frac{\partial \cosh^{-1}\left(\frac{x}{R_1}\right)}{\partial x} \right| dx = \varepsilon_0 \cosh^{-1}\left(\frac{B}{R_1}\right) \tag{3-26}$$

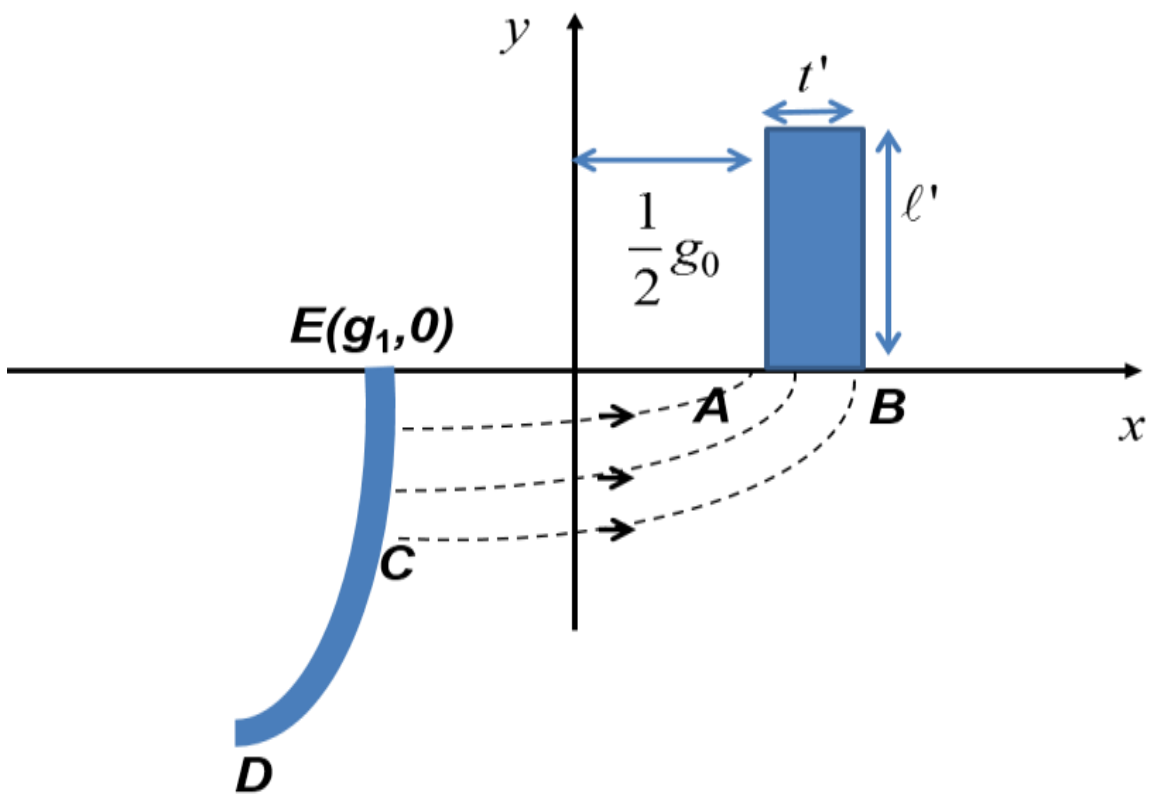

(a)

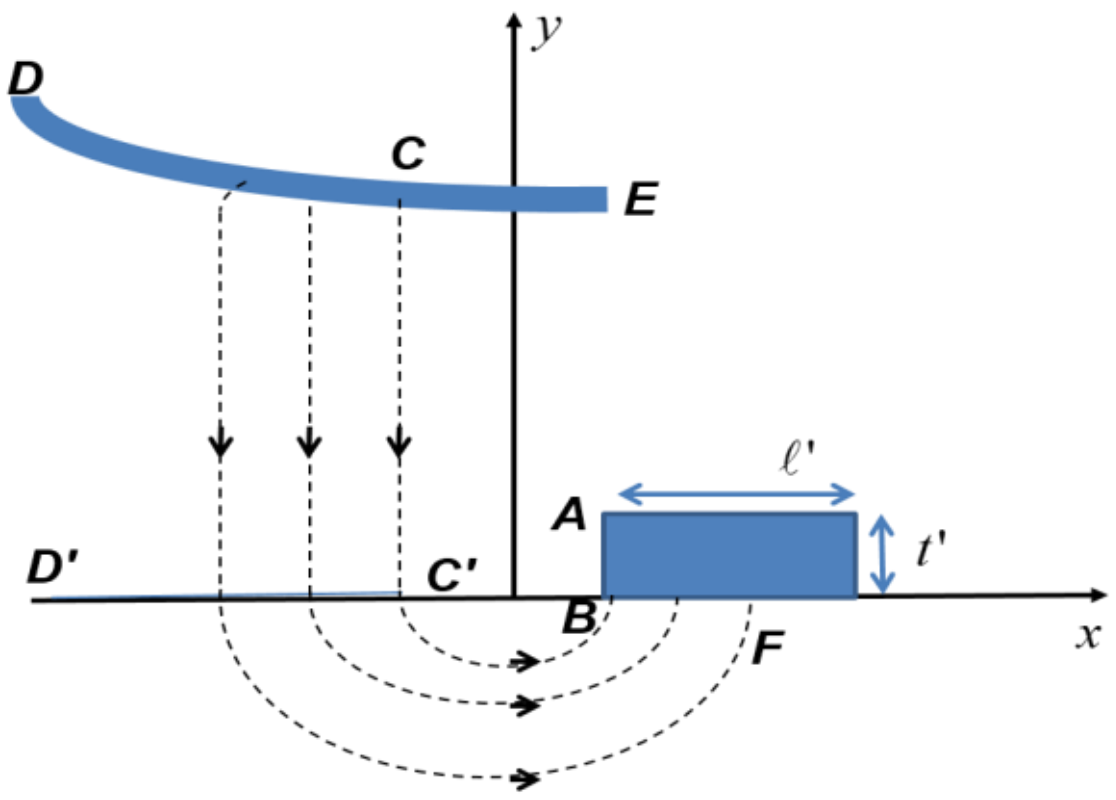

(b)

Fig. 3-5. Assume electric field line distribution of a MEMS switch between movable electrode and (a) stationary electrode side, (b) stationary electrode back when $\ell' < \ell$.

The potential difference between the movable beam and stationary electrode is needed to calculate the capacitance. At this point, there is no external voltage applied. From all points in AB locate at $y$=0 and $x$>0, the potential $u_1 = 0$. Regarding the potential on movable beam, assume one point $E(x,y)$ on beam is ($g_1$,0). From (3-23), it is obtained that $u_2 = \pi - \arccos\left(\frac{-g_1}{R_1}\right), v = 0.$ Substitute the $A\left(x = \frac{1}{2}g_0, y = 0\right)$ and

$B\left(x=\frac{1}{2}g_0+t', y=0\right)$

$$C_{side}=\frac{Q}{u_1-u_2}=\frac{\varepsilon_0}{\pi-\cos^{-1}\left(\frac{-2g_1}{g_0}\right)}\cosh^{-1}\left(1+\frac{2t'}{g_0}\right) \tag{3-27}$$

where $g_1=g_0-a\sqrt{1+\frac{\ell'^2}{4b^2}}$ .

The capacitance between the back of stationary electrode and movable electrode can be considered as capacitance *CDC′D′* in series with capacitance *BFC′D′*. This is shown in Fig. 3-5(b). For the electrical field along *BF* is described by (3-26). In this coordinate $B(x=R_2, y=0)$ and $F\left(x=R_2+\frac{1}{2}\ell', y=0\right)$

$$Q=\int_B^F \varepsilon_0 |E|\,dx=\varepsilon_0\int_B^F\left|\frac{\partial\cosh^{-1}\left(\frac{x}{R_2}\right)}{\partial x}\right|dx=\varepsilon_0\cosh^{-1}\left(\frac{F}{R_2}\right) \tag{3-28}$$

From (3-23),

$$\frac{x^2}{R_1^{\,2}\cosh^2(v)}+\frac{y^2}{R_1^{\,2}\sinh^2(v)}=1 \tag{3-29}$$

Substitute $R_1=\frac{1}{2}g_0$ and $B\left(x=\frac{1}{2}g_0+t', y=0\right)$ into (3-29), we can obtain $y=-\sqrt{g_0t'+t'^2}\ (x=0)$. Assume the position of *C* on y axis is $y=-\sqrt{g_0t'+t'^2}$ in Fig. 3-5(a), so $BC' = 2R_2 = \sqrt{g_0 t' + t'^2}$ in Fig. 3-5(b).

Since the *BF* potential is $u_1 = 0$ and *C′D′* potential is $u_2 = \pi$, the capacitance between *BF* and *C′D′* is

$$C_{BFC'D'} = \frac{Q}{u_1 - u_2} = \frac{\varepsilon_0}{\pi}\cosh^{-1}\left[1 + \frac{\min\left(BF, C'D'\right)}{R_2}\right] \tag{3-30}$$

where min(*BF*,*C′D′*) means the smaller one between *BF* and *C′D′*, $R_2 = \frac{1}{2}\sqrt{g_0 t' + t'^2}$, $BF = \frac{\ell'}{2}$, $C'D' = \frac{\ell}{2} - \frac{\ell'}{2} - R_2$.

The capacitance between *CD* and *C′D′* is taken as parallel-plate capacitance. This is not valid when *CD* and *C′D′* are not parallel and the electric field is not uniform between them. However, when $C_{BFC'D'}$ is in series with $C_{CDC'D'}$, $C_{BFC'D'}$ becomes more significant.

$$C_{CDC'D'} = \varepsilon_0\left[\frac{\min\left(BF, C'D'\right)}{g_0 + t'}\right] \tag{3-31}$$

Taking advantage of symmetry,

$$C_{back} = 2\frac{C_{CDC'D'}C_{BFC'D'}}{C_{BFC'D'} + C_{CDC'D'}} \tag{3-32}$$

Therefore, the overall capacitance is

$$\begin{aligned} C_{total} &= C_{plane} + 2C_{side} + C_{back} \\ &\approx \frac{2\varepsilon_0 \sinh^{-1}\left(\frac{\ell'}{2R}\right)}{\tan^{-1}\left(\frac{a}{b}\right)} + \frac{2\varepsilon_0 \cosh^{-1}\left(1 + \frac{2t'}{g_0}\right)}{\pi - \cos^{-1}\left(\frac{-2g_1}{g_0}\right)} \\ &\quad + \left\{\frac{2\varepsilon_0}{\pi}\cosh^{-1}\left[1 + \frac{\min\left(BF, C'D'\right)}{R_2}\right] / 2\varepsilon_0\left[\frac{\min\left(BF, C'D'\right)}{g_0 + t'}\right]\right\} \end{aligned} \tag{3-33}$$

The total capacitance is normalized by $C_0 = \dfrac{\varepsilon_0 \ell'}{g_0}$.

The (3-33) only considers the electric field in confocal ellipse shape, so correction terms should be added for the neglected electric field. The correction terms depend on $g_0/\ell'$ [11]. The empirical equation after adding correction factors becomes:

$$C_{total} \approx \frac{2\varepsilon_0 \sinh^{-1}\left(\frac{\ell'}{2R}\right)}{\tan^{-1}\left(\frac{a}{b}\right)}\left[1+\frac{g_0}{\ell'}\right] + \frac{2\varepsilon_0 \cosh^{-1}\left(1+\frac{2t'}{g_0}\right)}{\pi - \cos^{-1}\left(\frac{-2g_1}{g_0}\right)} + \left\{\frac{2\varepsilon_0}{\pi}\cosh^{-1}\left[1+\frac{\min\left(BF, C'D'\right)}{R_2}\right] / 2\varepsilon_0\left[\frac{\min\left(BF, C'D'\right)}{g_0+t'}\right]\right\} \tag{3-34}$$

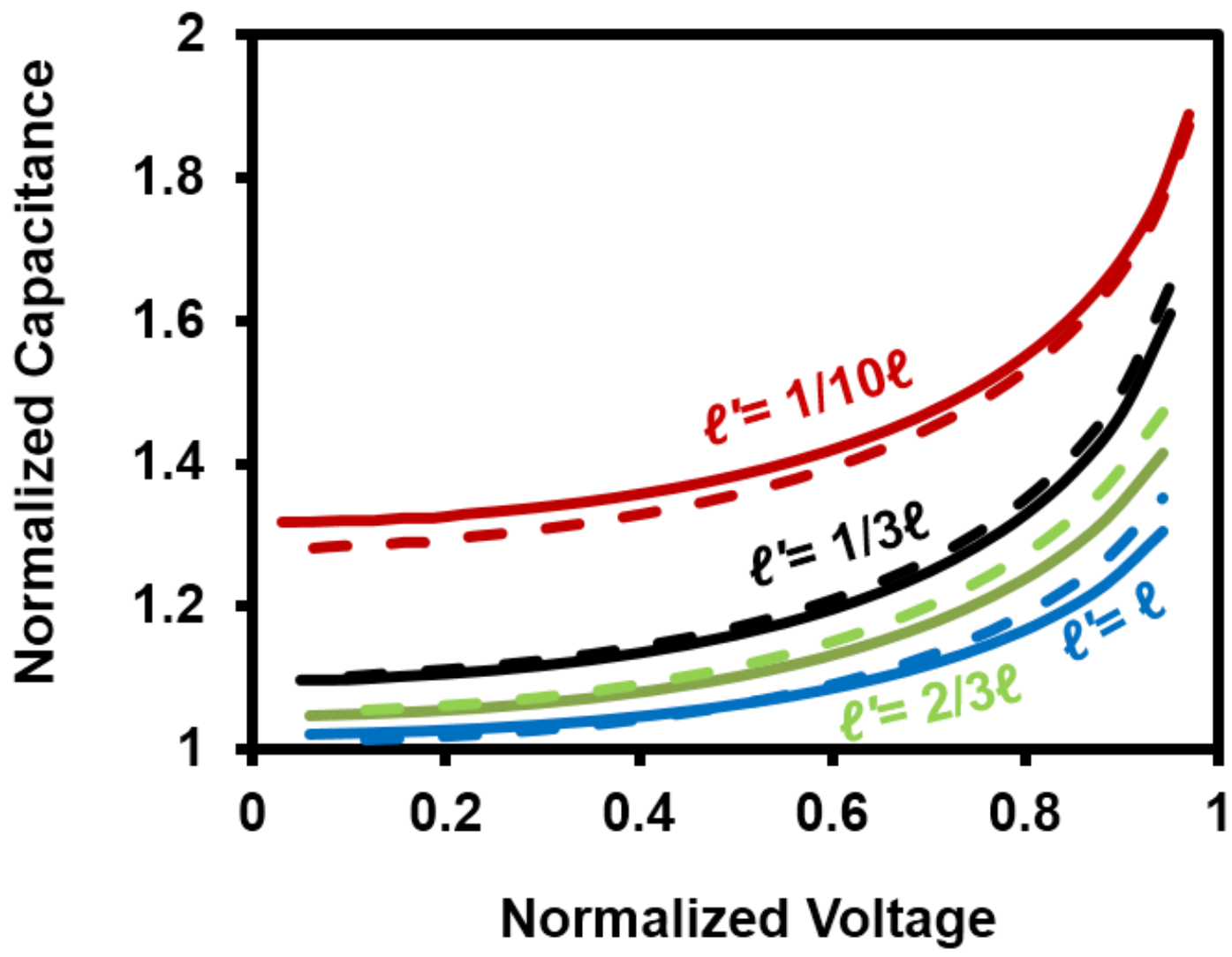

Fig. 3-6. ANSYS simulation results (solid) and analytical solutions (dash) for normalized capacitance as a function of normalized voltage at $g_0/\ell = 1/100$, $t'/g_0=0.6$. The substrate is air.

Fig. 3-6 shows that at high beam length to gap height ratio ($\ell/g_0 = 100$), fringe

capacitance influence drops to less than 10% at $\ell'/\ell = 1/3$ and it keeps dropping as $\ell'/\ell$ increases.

For the low beam length to gap height ratio ($\ell/g_0 = 10$) case, the maximum beam deflection is limited to 1/20 of gap height or the beam strain will exceed yield strain. Fig. 3-7 shows that the fringe capacitance is 20% of parallel-plate capacitance. It needs to be included in the model at low beam length to gap height ratio.

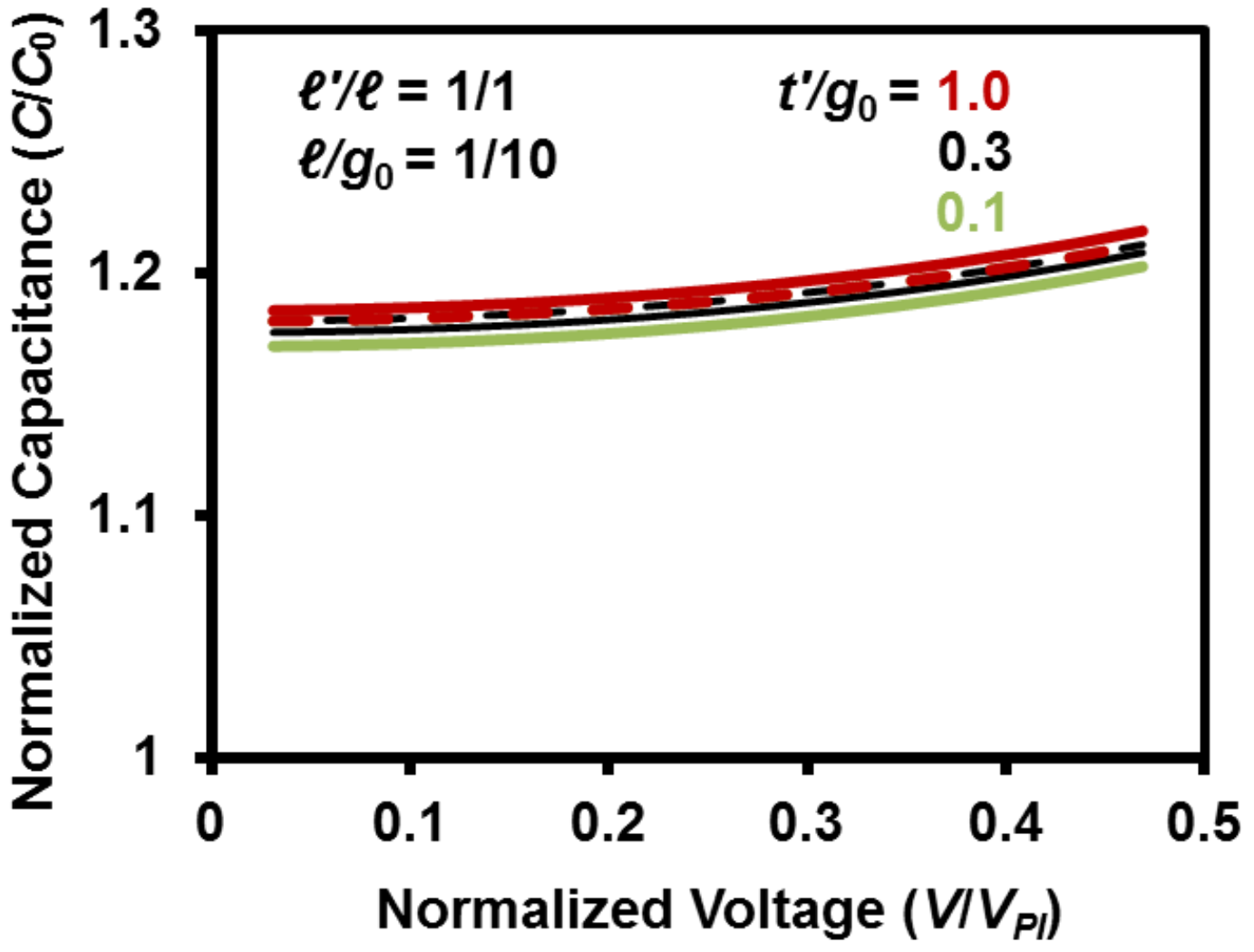

Fig. 3-7. ANSYS simulation results (solid) and analytical solutions (dash) for normalized capacitance as a function of normalized voltage when $\ell'/\ell = 1$, $g_0/\ell = 1/10$. The substrate is air.

The discussion above does consider the substrate underneath the stationary electrode. When the substrate is considered, the total capacitance becomes:

$$C_{total} \approx \frac{2\varepsilon_0 \sinh^{-1}\left(\frac{\ell'}{2R_1}\right)}{\tan^{-1}\left(\frac{a}{b}\right)}\left[1+\frac{g_0}{\ell'}+15\left(\frac{g_0}{\ell'}\right)^2\right]+\frac{2\varepsilon_0 \cosh^{-1}\left(1+\frac{2t'}{g_0}\right)}{\pi-\cos^{-1}\left(\frac{-2g_1}{g_0}\right)}$$
$$+\left\{\frac{2\varepsilon_0\varepsilon_r}{\pi}\cosh^{-1}\left[1+\frac{\min\left(BF,C'D'\right)}{R_2}\right]/2\varepsilon_0\left[\frac{\min\left(BF,C'D'\right)}{g_0+t'}\right]\right\} \tag{3-35}$$

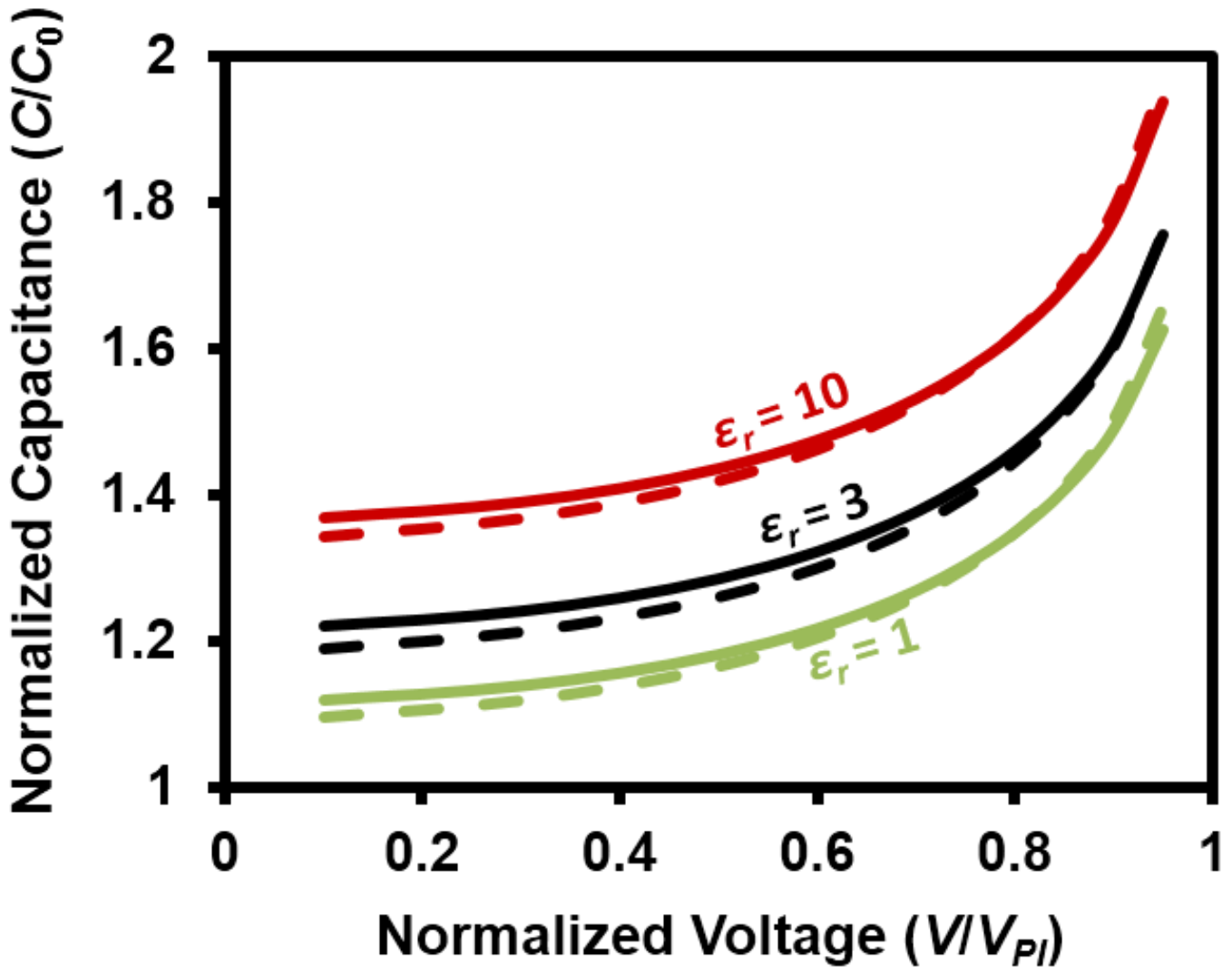

Fig. 3-8. ANSYS simulation results (solid) and analytical solutions (dash) for normalized capacitance as a function of normalized voltage. The $\ell'/\ell$ = 1/3, $g_0/\ell$ = 1/100, the colors represents different substrate dielectric constant $\varepsilon_r$.

Fig. 3-8 illustrates the normalized capacitance as a function of bias voltage after considering substrate. The higher dielectric constant introduces larger parasitic capacitance, for example, the parasitic capacitance increases from 10% of parallel-plate capacitance to about 40%. However, the parasitic capacitance increase does not improve the tuning range of unactuated state capacitance, which will limit the actuated/unactuated

capacitance ratio. The analytical solutions agree with ANSYS simulation results well.

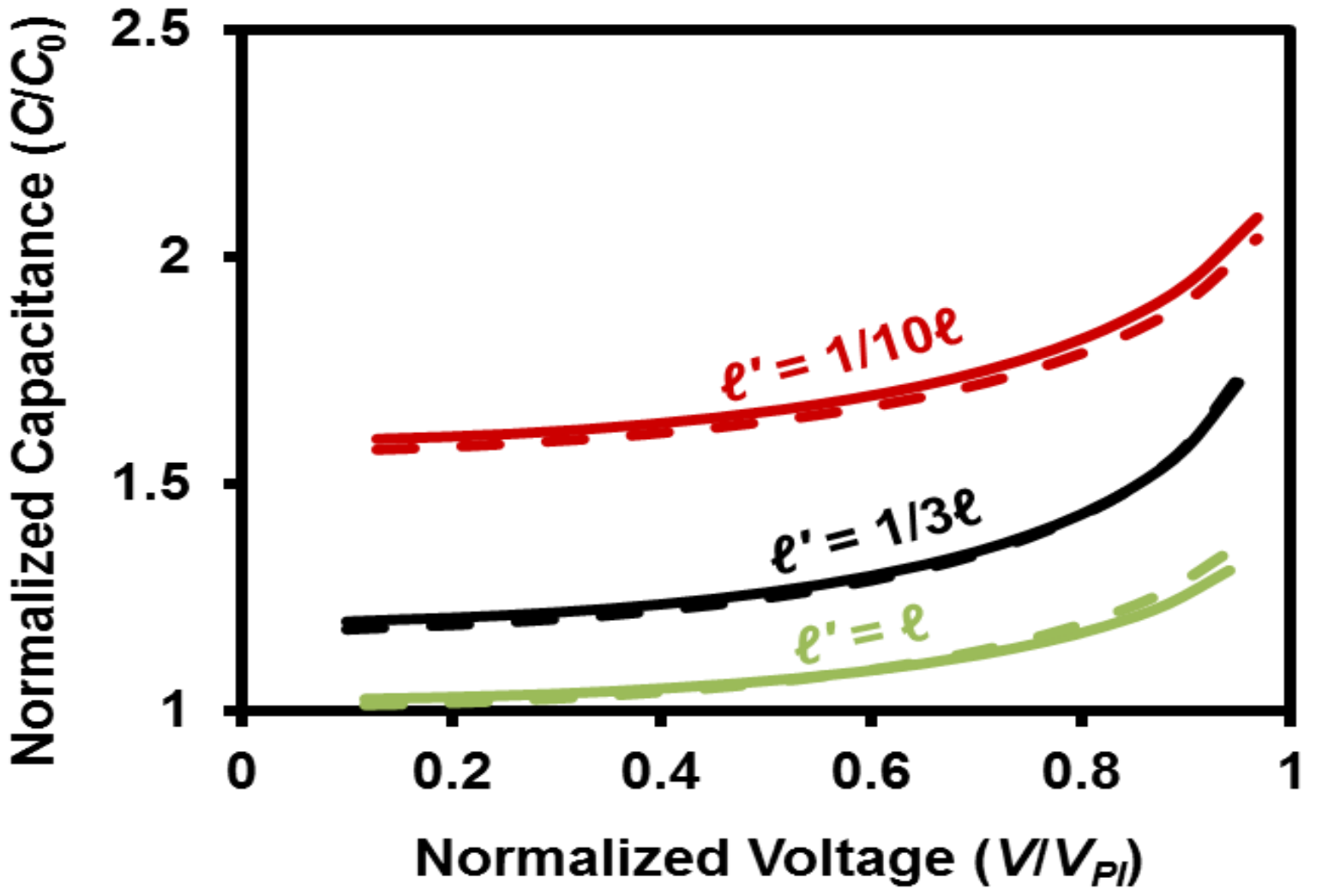

(a)

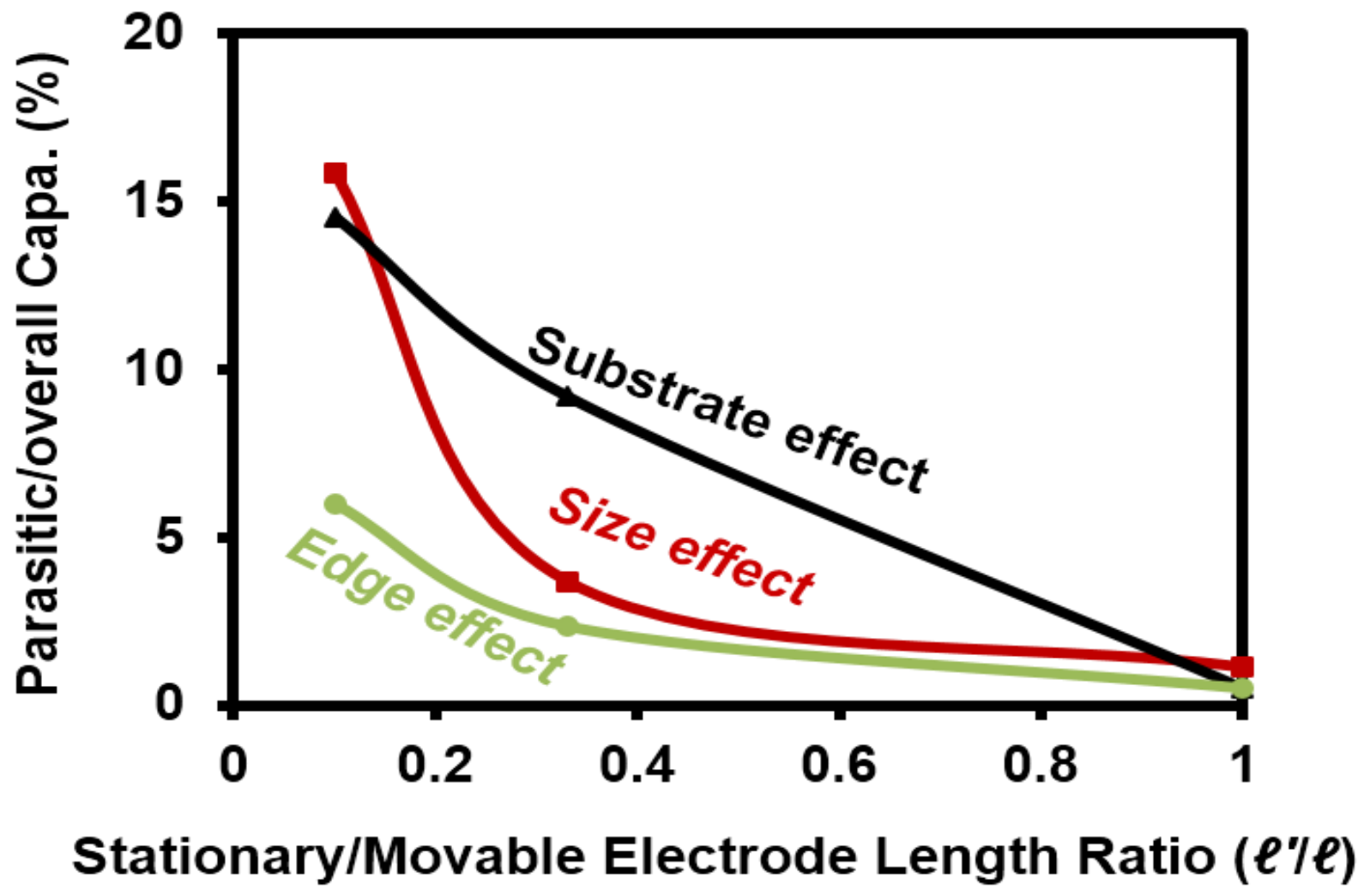

(b)

Fig. 3-9. (a) ANSYS simulation results (solid) versus analytical solutions (dash) for normalized capacitance as a function of normalized voltage. The substrate dielectric $\varepsilon_r$ =3, $g_0/\ell$ = 1/100. (b) Parasitic to overall capacitance ratio as a function of stationary to movable electrode ratio $\ell'/\ell$.

Fig. 3-9(a) shows the normalized capacitance subjects to applied voltage for various stationary/movable electrode length ratios. At $\ell'/\ell$ = 1/10, the parasitic capacitance is more than 50% of parallel-plate capacitance, and it drops as $\ell'/\ell$ increases. When $\ell'/\ell$ = 1/1, it decreases to 3%. Fig. 3-9(b) demonstrates the contribution of each parts of the parasitic capacitance subject to $\ell'/\ell$. Substrate effects and size effects contribute much more than edge effect. This is probably due to the smaller dimension of edge when compared with stationary electrode length.

Fig. 3-10 shows stationary electrode thickness influence on the normalized capacitance subject to applied voltage. It can be seen that the influence is small when t/g0 changes from 0.1 to 1, which is due to the small dimension of stationary electrode thickness.

In conclusion, a compact analytical expression for the capacitance of a fixed-fixed beam MEMS switch is derived using conformal mapping techniques. It includes fringe effects and substrate and does not rely on parallel-plate approximation. The expression can provide accurate results for simple geometry switches.

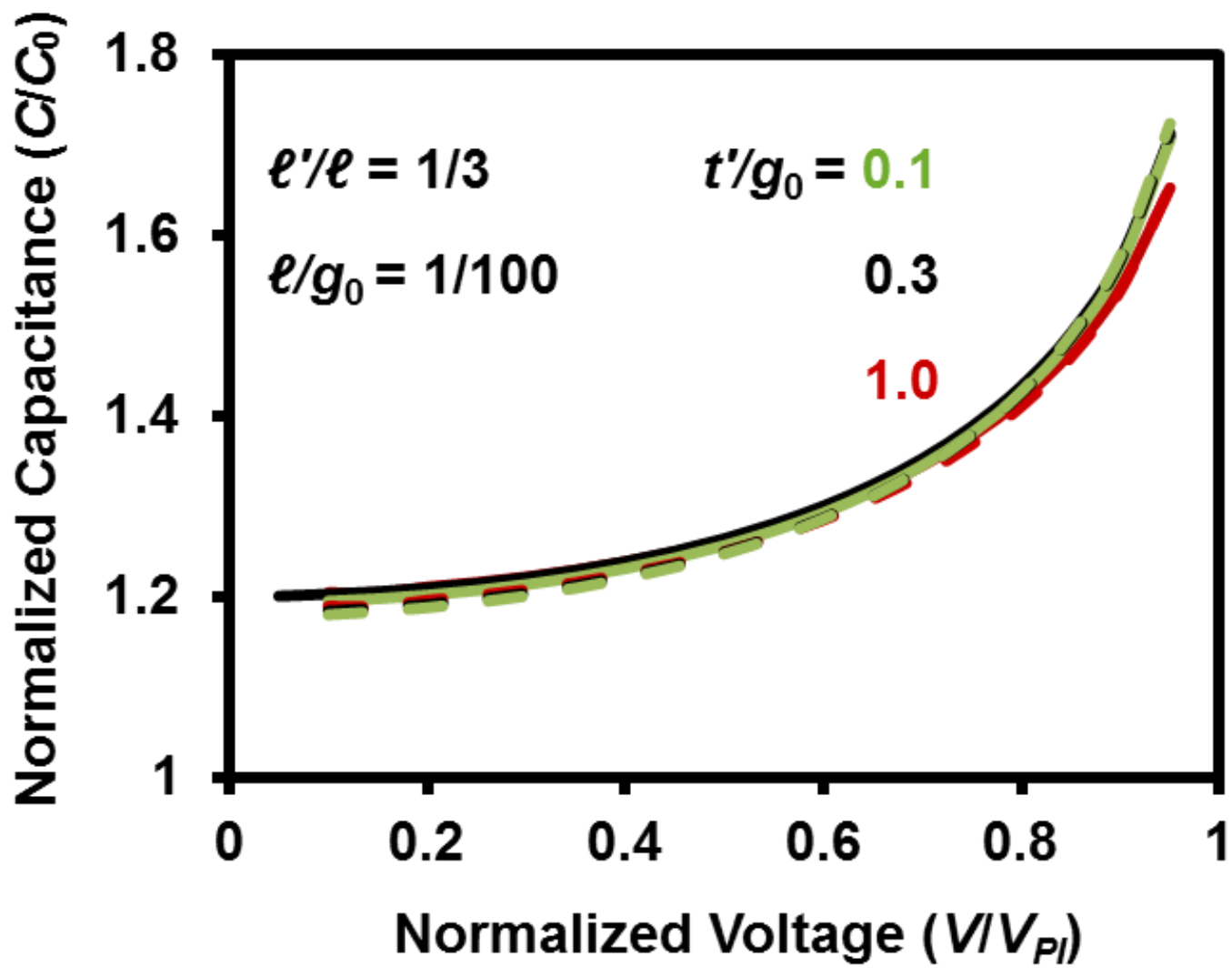

Fig. 3-10. ANSYS simulation results (solid) and analytical solutions (dash) for normalized capacitance as a function of normalized voltage. The $\ell'/\ell = 1/3$, $g_0/\ell = 1/100$.

However, this expression assumes electric field lines are confocal ellipses, but this may be invalid when switches geometry become complicated. The substrate effects and fringe capacitance add to the complexity of the switches capacitance calculation and the accurate capacitance values require finite element analysis [12]–[14].

## *References*

[1] ANSYS Mechanical APDL Coupled-Field Analysis Guide, ANSYS, Inc., Canonsburg, PA, 2013, pp. 1–4.

[2] G. M. Rebeiz, *RF MEMS Theory, Design*, and *Technology*. Hoboken, NJ: Wiley, 2003, pp. 23–27.

[3] X. Yan, W. L. Brown, Y. Li, J. Papapolymerou, C. Palego, J. C. M. Hwang, and R. P. Vinci, “Anelastic stress relaxation in gold films and its impact on restoring forces in MEMS devices,” *J. Microelectromech. Syst.*, vol. 18, no. 3, pp. 570–576, Jun. 2009.

[4] S. Chowdhury, M. Ahmadi, and W. C. Miller, "A comparison of pull-in voltage calculation methods for MEMS-based electrostatic actuator design," in *Proc. 1st Int. Conf. Sens. Technol.*, Nov. 2005, pp. 112–117.

[5] J. Lekner, “Electrostatics of hyperbolic conductors,” *Eur. J. Phys.*, vol. 25, no. 6, pp. 737–744, Nov. 2004.

[6] R. G. Budynas, Advanced Strength and Applied Stress Analysis, 2nd ed. New York: McGraw-Hill, 1999, pp. 849–850.

[7] E. S. Hung, and S. D. Senturia, "Extending the travel range of analog-tuned electrostatic actuators," *J. Microelectromech. Syst.*, vol. 8, no. 4, pp. 497–505, Dec. 1999.

[8] J. I. Seeger, and B. E. Boser, "Charge control of parallel-plate, electrostatic actuators and the tip-in instability," *J. Microelectromech. Syst.*, vol. 12, no. 5, pp. 656–671, Oct. 2003.

[9] E.R. Deutsch, J.P. Bardhan, S.D. Senturia, G.B. Hocker, D.W. Youngner, M.B. Sinclair, and M.A. Butler, "A large-travel vertical planar actuator with improved stability," in *IEEE Int. Conf. on Solid-State Sensors, Actuators and Microsystems Tech. Dig.*, Jun. 2003, pp. 352–355.

[10] A. Bansal, B. C. Paul, K. Roy, "An analytical fringe capacitance model for interconnects using conformal mapping," *IEEE Trans. Computer-Aided Design Integr. Circuits Syst.*, vol. 25, no. 12, pp. 2765–2774, Dec. 2006.

[11] R. C. Batra, M. Porfiri, and D. Spinello, "Electromechanical Model of Electrically Actuated Narrow Microbeams," *J. Microelectromech. Syst.*, vol. 15, no. 5, pp. 1175-1189, Oct. 2006.

[12] N. P. van der Meijs and J. T. Fokkema, "VLSI circuit reconstruction from mask topology," *Integr. VLSI J.*, vol. 2, no. 2, pp. 85–119, Mar. 1984.

[13] S. Chowdhury, M. Ahmadi, and W. C. Miller, "A closed-form model for the pull-in voltage of electrostatically actuated cantilever beams." *J. Micromech. Microeng.*, vol. 15, no. 4, pp. 756–763, Feb. 2005.

[14] M. Rahman and S. Chowdhury, "A Highly accurate method to calculate capacitance of MEMS sensors with circular membranes," in *IEEE Int. Conf. on Electro/Information Tech.*, Jun. 2009, pp. 178–181.

# Chapter 4 Experimental Validation and Discussion

## *4.1 Center Deflection subject to Voltage*

To validate the hyperbolic model for beam center deflection subjected to applied voltage, the experimental data obtained from [1] are used. The device is a two-layer polysilicon beam structure, whose lower beam bends when a voltage is applied to the stationary electrode. The lower beam length varies from 600 to 1200 μm to investigate displacement-voltage characteristics and the lower beam and stationary electrodes are equal length. The lower beam thickness is 1 μm, width is 20 μm, and air gap height between the lower beam and stationary electrode is 8 μm. Young's modulus $E$ = 160 GPa, the extracted residual stress $\sigma$ on the beam is 15 MPa. The interference microscopy is used to measure vertical movement of individual beam elements. If the sample is tilted, the parallel fringes are produced on the devices. The vertical displacement that corresponds to one fringe shift is half the incident wavelength [2].

Fig. 4-1 compares the hyperbolic solution to the experimental data for beam center deflection. They agree with each other well in general. As the lower beam length decreases from 1200 μm to 600 μm, the $\delta$ increases from 1.5 to 5.6. It can be seen from

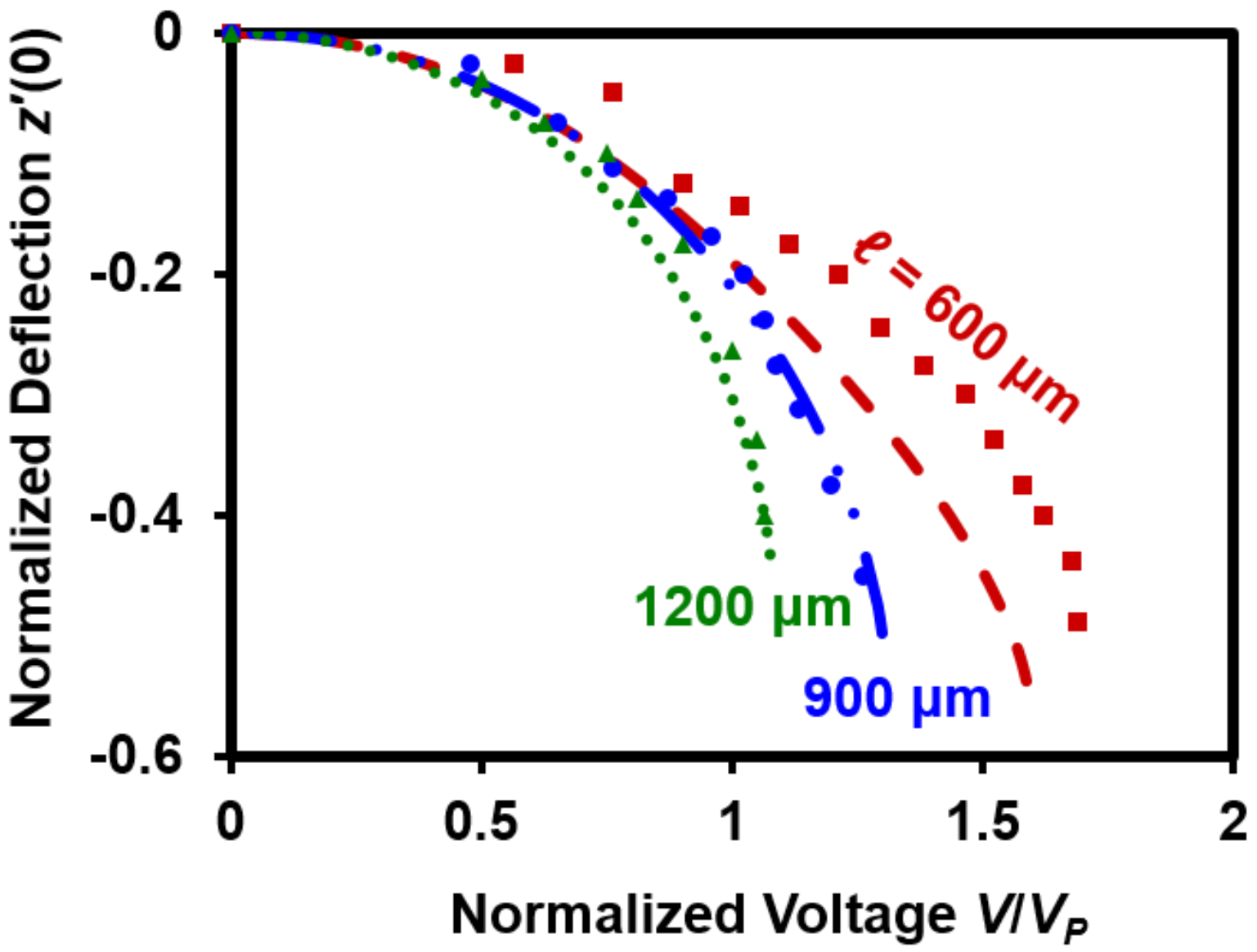

Fig. 4-1 Modeled (curves) versus measured (symbols) beam center deflection. The model predicts the deflection for beam length 600 (----), 900(—··—), and 1200 (····) µm. The measured data are from [1] and beam lengths are 600 (■), 900 (●), and 1200 (▲) µm, respectively.

Fig. 4-1 that the beam center deflection passes over the one-third of the gap without pull-in because of strong nonlinear stretching components. Their travel ranges agree with hyperbolic model prediction in Fig. 3-3. In terms of deflection-voltage characteristics, the hyperbolic model shows better agreement with experimental data when lower beam lengths are longer. The residual stress to bending spring constant ratio $k_0''/k_0'$ is given by:

$$k_0''/k_0' = \sigma(1-\nu)/(4E\gamma^2). \tag{4-1}$$

For the same material and residual stress, reducing lower beam length increases $\gamma$, thus $k_0''/k_0'$ decreases. This indicates the bending effects become more significant. The bending effects add the bending moment at the fixed-fixed anchor, which the hyperbolic function model cannot predict it well. This explains why hyperbolic model shows better prediction for longer beam length. The limitation of the hyperbolic model on bending moment is discussed in 4.3 section.

## *4.2 Capacitance-Voltage Characteristics*

The experimental capacitance-voltage characteristics of MEMtronics capacitive switches are used to compare with hyperbolic model results. These electrostatically actuated capacitive switches are based on a movable aluminum membrane electrode approximately 300μm-long, 100μm-wide, and 0.3μm-thick. The movable membrane is anchored on both ends to the ground conductors of a 50Ω coplanar transmission line [4]. The typical residual stress for these switches is around 60 MPa.

Fig. 4-2 shows a comparison between hyperbolic function and the measurement data. Excellent agreement was achieved. For the real measurement data, the fringe capacitance is difficult to predict because it varies according to the different devices geometry and material properties. However, the fringe capacitance does not change the capacitance change with respect to the voltage. The hyperbolic model shows good

prediction after compensating the fringe capacitance.

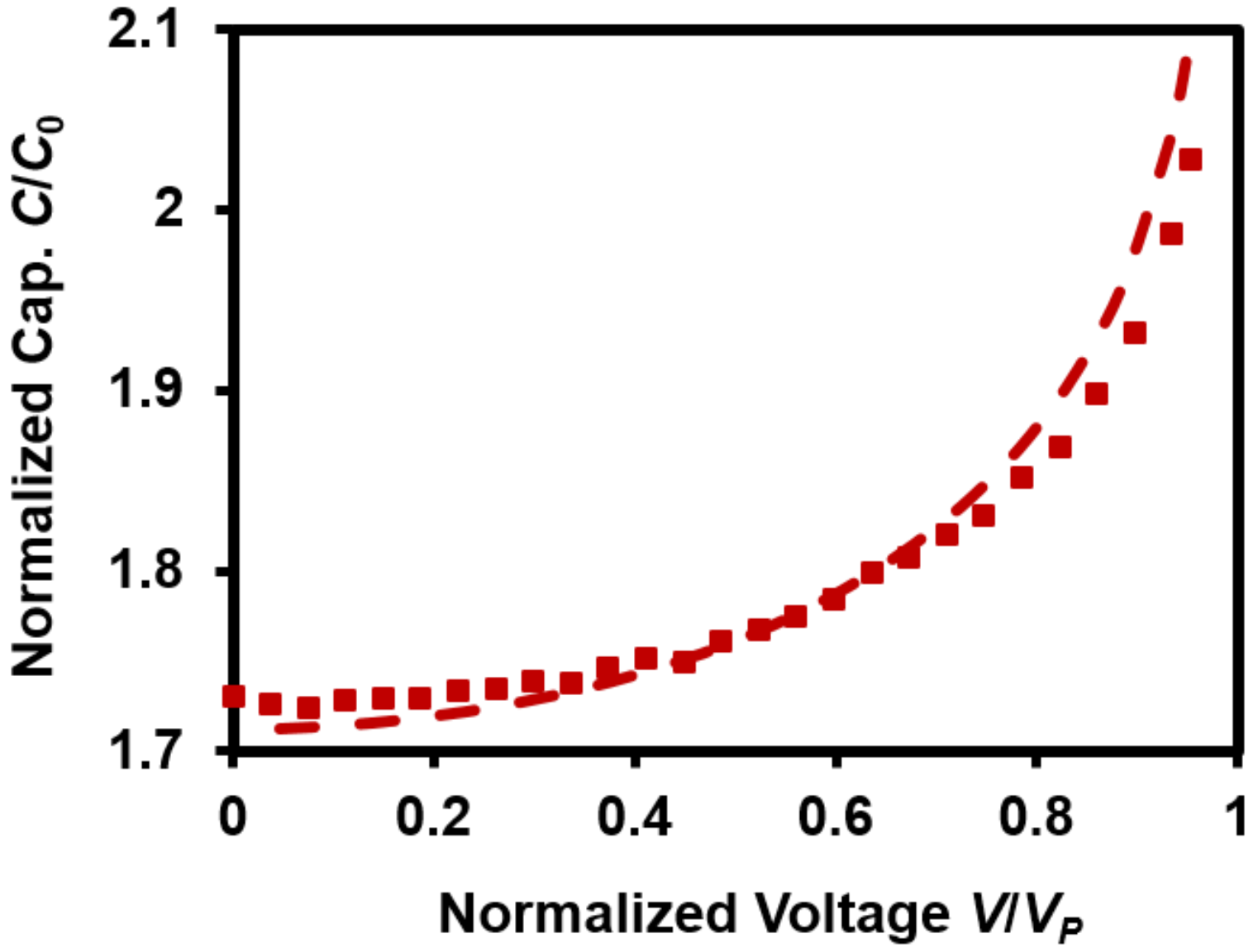

Fig. 4-2 Modeled (dashed curves) versus measured (symbols) capacitance-voltage characteristics for MEMtronics switches.

## *4.3 Anchor Condition for Hyperbolic Model*

As mentioned in Chapter 3.1.3 hyperbolic model coefficient determination, the hyperbolic model needs to calculate the work due to the constrained beams and strain energy to determine the unknown coefficient. For a fixed-fixed beam, the bending moment is generated at the anchor when it bends. The bending moment contributes to the work due to the constrained beams and strain energy. For a rigid fixed-fixed anchor condition, it requires $z(-\ell/2) = z(\ell/2) = 0$ and $dz/dx|_{z=-\ell/2} = dz/dx|_{z=\ell/2} = 0$. Nevertheless, the hyperbolic model does not satisfy the second requirement, so that it does not include the bending moment. Therefore, the hyperbolic model is no longer valid when the bending components are significant. The residual stress and stretching effects do not create bending moments at anchor so the hyperbolic model is valid in those cases. In typical MEMS capacitive switches, the residual stress is the dominant factor. In NEMS devices with a ultrathin beam, the stretching effects dominate. The hyperbolic model works well in those cases.

Fig. 4-3(a) shows the voltage difference between the ANSYS simulation results and hyperbolic model when the beam center deflection reaches one-third of the gap height. The ANSYS simulations employ a structure of aluminum beam with $\ell = \ell' = 100$–300 µm, $t = 0.6$ µm, and gap height $g = 1$–3 µm. The residual stress $\sigma$ varies from 5 to 150 MPa. The voltage differences show a strong dependence on $k_0''/ k_0'$. The increase of $k_0''/ k_0'$ indicates the residual stress component become stronger and the voltage difference becomes smaller. After $k_0''/ k_0'$ greater than 10, the error is less than 10%. The prediction agrees with measured data from [1] well.

Fig. 4-3(b) shows the $\delta$ dependence of voltage difference between the ANSYS

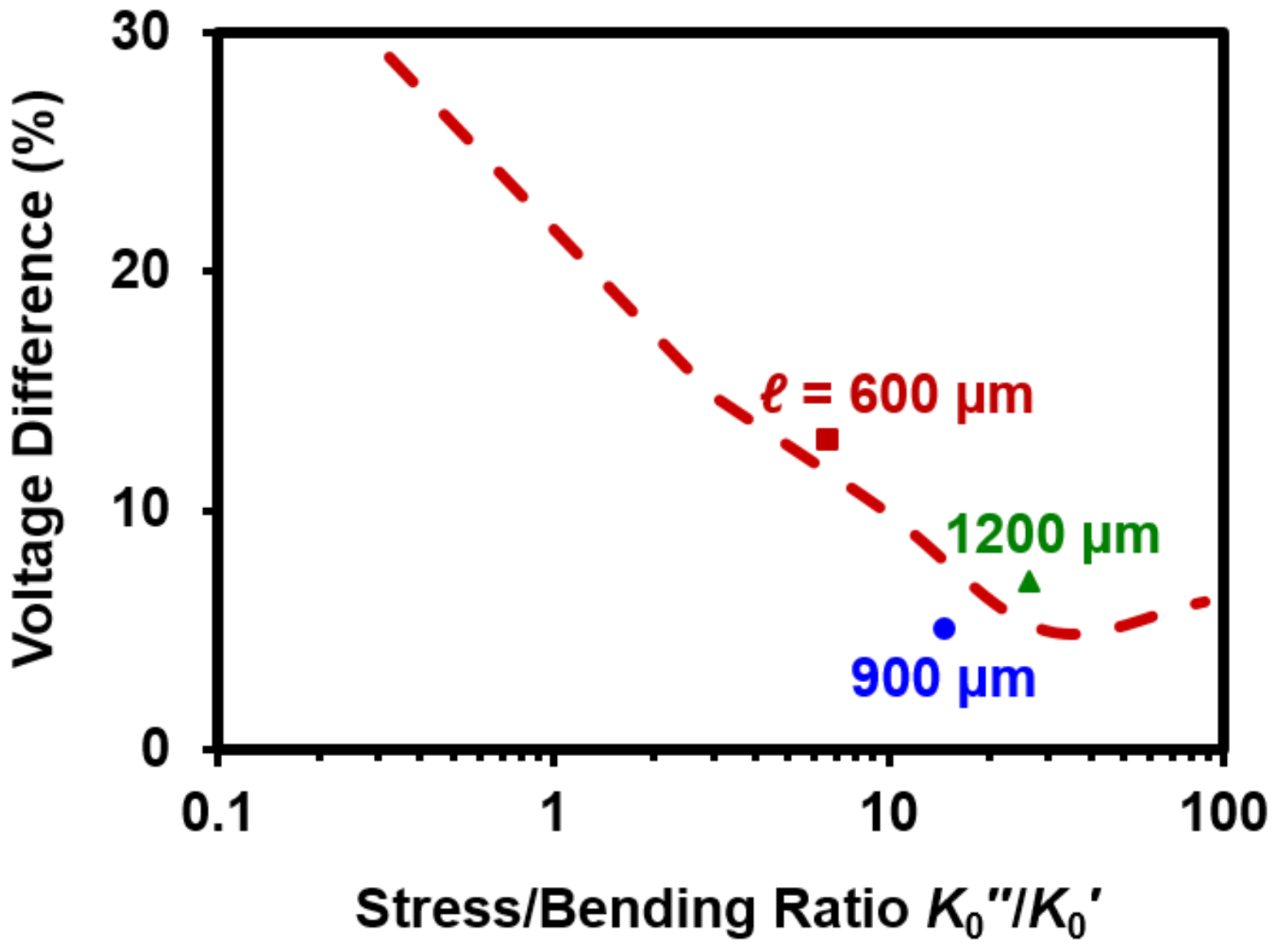

(a)

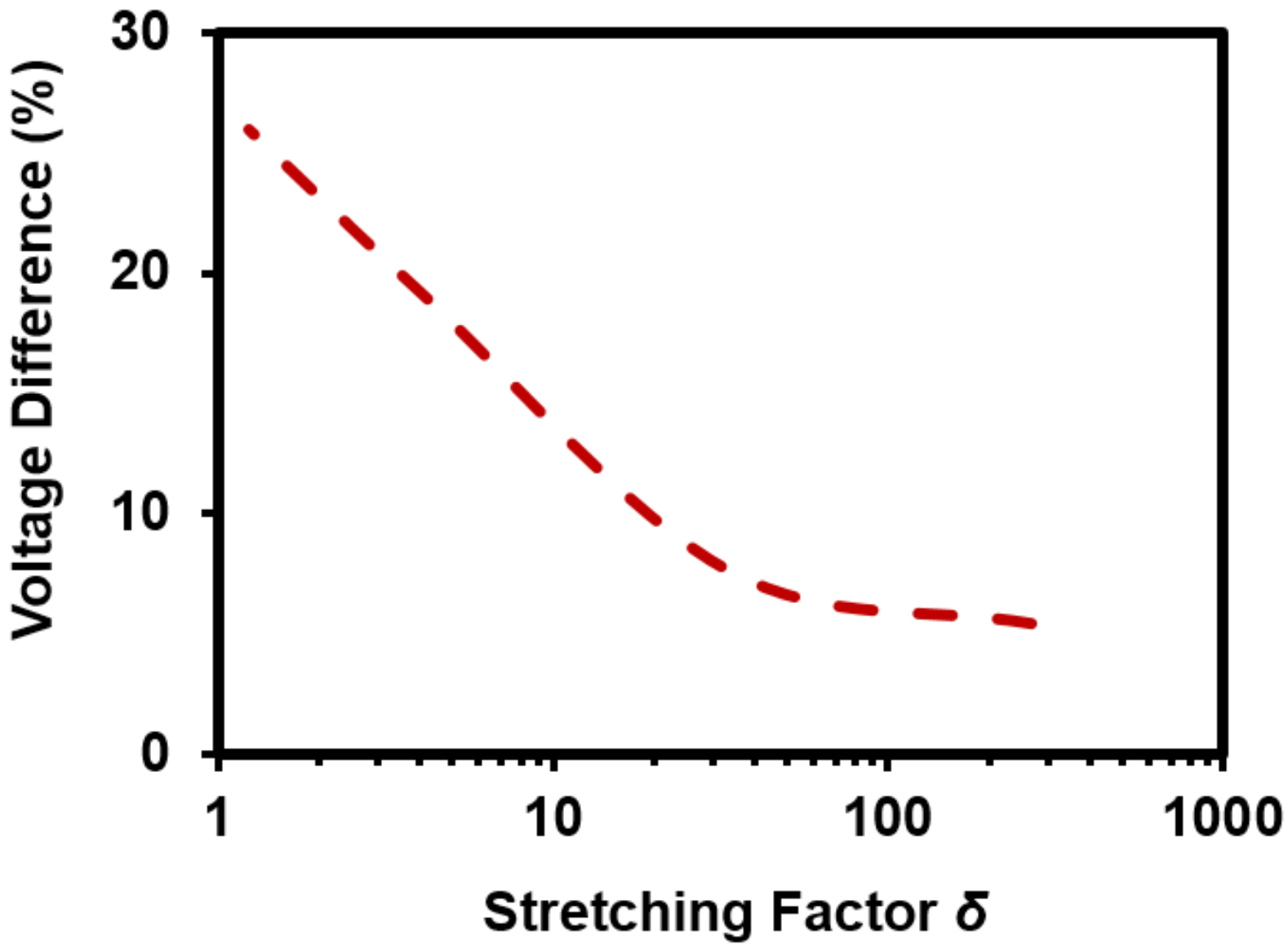

(b)

Fig. 4-3. Voltage difference between hyperbolic model results and ANSYS simulation results when the beam center deflection reaches 1/3 of gap height. The voltage difference depends on (a) stress/bending ratio ($K_0''/K_0'$) and (b) stretching factor $\delta$ ($k_s g^2/k_0$). Symbols in (a) are voltage difference between experimental data from [1] and hyperbolic model for beam lengths 600 (■), 900 (●), and 1200 (▲) μm.

simulation results and hyperbolic model. In order to study the stretching dominant case, the residual stress is not included in the simulation. ANSYS simulations employ structure of aluminum beam with $\ell = \ell' = 10$ µm, $t = 0.3$ to 0.05 µm, and gap height $g = 1$ µm, so $\delta$ ranges from 1 to 300. The voltage difference also decreases as $\delta$ increases, which indicates hyperbolic model works well when stretching effects are strong. When $\delta$ is greater than 10, the voltage difference is less than 10%.

## *4.4 Plastic Deformation Limit for Down Scaling*

For a fixed-fixed beam MEMS switch, the axial strain in the beam should be within the elastic region when it deforms. To satisfy this requirement, the switch geometry, material properties, and anchor boundary conditions need to be considered. This section focuses on the relationship between the axial strain and the switch geometry. The beam shape subject to an electrostatic force is predicted by hyperbolic functions. The total axial strain $\varepsilon_x$ on the *x*-axis is given by [3].

$$\begin{aligned}\varepsilon_x &= \varepsilon_{bending} + \varepsilon_{stretch} \\ &\approx -t\, d^2\Delta z / dx^2 + 1/\ell \int_{-\frac{\ell}{2}}^{0} \left(d\Delta z / dx\right)^2 dx\end{aligned} \tag{4-2}$$

where $\varepsilon_{bending}$ is the bending induced strain and $\varepsilon_{stretch}$ is the stretching strain.

From Fig. 4-4, the maximum beam strain on *x*-axis occurs at the edge and the center.

This is because the bending term peaks at the edge and the center while the stretching term is approximately uniform. However, the maximum beam strain on *x*-axis at the edge assumes idealized fixed-slope boundary condition, but the real boundary condition is much more complicated and the assumption becomes invalid easily [5]–[6]. This paper focuses on the strain at beam center. The hyperbolic function cannot predict the strain at the anchor well because it does not satisfy rigid fixed-fixed anchor condition. Nevertheless, it still provides accurate results of strain at the beam center.

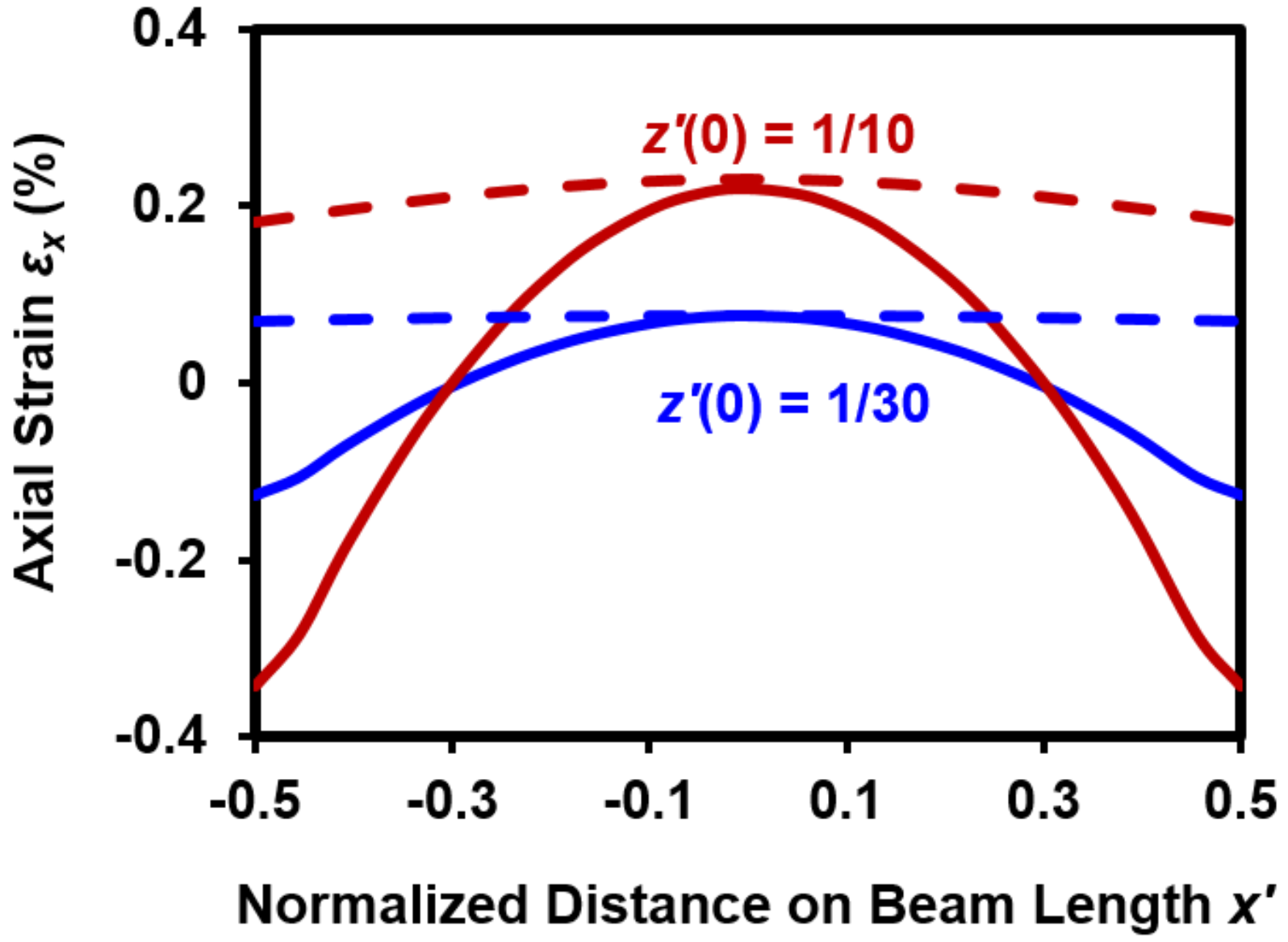

Fig. 4-4. Modeled (dashed curves) versus measured (symbols) capacitance-voltage characteristics for MEMtronics switches.

Substituting the hyperbolic function into (4-2), the strain at beam center ($x$ = 0) $\varepsilon_{x=0}$ is given by:

$$\varepsilon_{x=0} = -8\alpha\gamma z'(0) + 8\alpha^2 z'(0)^2 / 3 \\ = -8gtz'(0)/\ell^2 + 8g^2 z'(0)^2 / (3\ell^2). \tag{4-3}$$

From (4-3), $\varepsilon_{x=0}$ is proportional to $\alpha$, $\gamma$, and $z'(0)$. Therefore, low-aspect-ratio (high $\alpha$) switches with a thick beam (high $\gamma$) suffer from high strain at certain center deflection. The quadratic relationship between axial strain and $1/\ell$ indicates axial strain increases significantly as $\ell$ decrease. This may become a problem for miniature switches, whose $\ell$ is much smaller than that of standard switches. Fig. 4-4 demonstrates that for low-aspect-ratio, the material yield strain ($\varepsilon_{yield}$ = 0.2%) limits the maximum center deflection to 1/10 of gap height.

To avoid beam material deform plastically, the maximum beam strain on $x$-axis adds initial strain $\varepsilon_{ini}$ should be smaller than the yield strain $\varepsilon_{yield}$. Thus, it requires:

$$0 < |z(0)/t| < 3 - \sqrt{9 + 3(\varepsilon_{yield} - \varepsilon_{ini})/(2\gamma^2)} / 2. \tag{4-4}$$

According to (4-4), the beam deflection is limited by $\varepsilon_{yield}$, $\varepsilon_{ini}$, and $\gamma$. If the microstructures are fabricated by using materials that can sustain large strain (e.g. conductive polymers [7], graphene [8] ), the offset yield strength (0.2%) in (4-4) needs to be changed accordingly. In graphene-based NEMS, the yield strain can be 1% and the break strain is 25% [9]. For beams made by those materials, they can be used for large-displacement, low actuation voltage devices and wide range frequency tuning resonator. For example, the pull-in voltage calculated for few-layer graphene beam electromechanical switch is 1.85 V [10]. A graphene NEMS resonator can achieve electrostatic frequency tuning of up to 400% [9].

## *4.5 Graphene NEMS Resonator Design*

The deflection induced strain (stretching effects) can change the nonlinear spring constant (spring constant hardening) and in turn increases the resonant frequency. When beam deflection is greater than beam thickness and the stretching strain is greater than the initial strain, the nonlinear stretching restoring force dominates deflection-voltage behavior. Since a graphene beam is ultrathin and can withstand ultrahigh strains, the graphene NEMS devices demonstrate the potential of extremely wide range frequency tuning by electrostatic forces. For NEMS devices, the beam thickness is a monolayer atom thin (0.33 nm). Thus, the gap height to beam thickness ratio is large (on the order to $10^2$ or $10^3$). In typical MEMS switches, the gap height to beam thickness ratio is relatively low (on the order of 10) so the deflection to beam thickness ratio is limited. Therefore, the nonlinear stretching effect is dominant in NEMS devices but is often negligible for typical MEMS switches.

The NEMS device geometry and resonant frequency $f_{res}$ relationship is given by:

$$\begin{aligned} f_{res} &= 1/2\pi\sqrt{k/m} \\ &= \sqrt{0.81Et^2/\rho\ell^4 + 0.2E\left[\varepsilon_{ini} + 2.7\alpha^2 z'(0)^2\right]/\rho\ell^2} \end{aligned} \tag{4-5}$$

where $k$ is the overall spring constant, $m$ is the mass, $\rho$ is the density.

In (4-5), the first term is attributed to bending effects, the remaining terms reflect $\varepsilon_{ini}$ and deflection induced strain, respectively. Because of atomic thinness, the bending rigidity of graphene is extremely small [9]. For a given structure, the strain associated with stretching is controlled by the bias voltage, which in turn influence the resonant frequency.

To predict the resonant frequency as a function of bias voltage, the deflection-voltage characteristic need to be known. For a graphene NEMS resonator operating in the case of strong nonlinear stretching, it is predicted by using the hyperbolic model. Assume a graphene NEMS resonator has a length $\ell$ um, $t$ = 0.33 nm, and $g$ = 232 nm (see Fig. 4-5 inset). The $\varepsilon_{ini}$ is around $4\times10^{-5}$ [12]. Fig. 4-5 shows the deflection-voltage relationship at $\ell$ = 0.3, 1, 3 um. It can be seen that the devices with $\ell$ = 0.3, 1, 3 um shows a deflection of 1%, 5%, 20% of the gap height at $V$ = 9 V. This is because the spring constant decreases with increasing beam length. For $\ell$ = 0.3 um, the deflection induced strain is below the initial strain, thus the stretching is not dominant and the deflection depends on $V^2$. As the beam length increases, the deflection increases under the same bias voltage and stretching gradually dominates at $\ell$ = 1 and 3 um. In these cases, the deflection depends on $V^{0.8}$. The findings agree with the conclusion in [11].

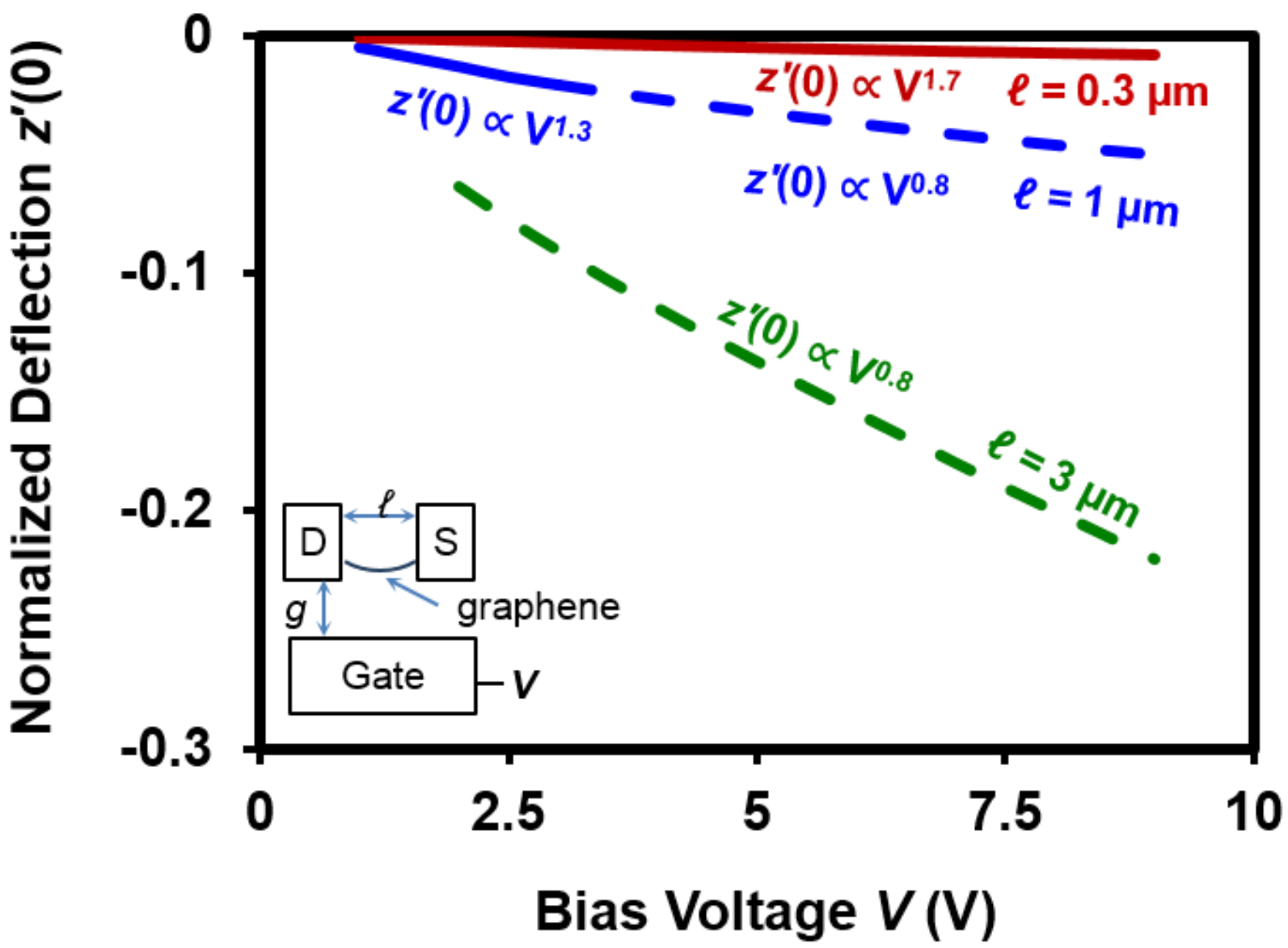

Fig. 4-5. Modeled deflection-voltage characteristics for NEMS resonators with beam length $\ell$ = 0.3, 1, 3 μm. The solid lines represents the case of stretching strain less than initial strain and the dashed lines represents stretching strain greater than initial strain. The inset is the schematic of a graphene resonator.

After the deflection-voltage characteristics are obtained, the resonant frequency is calculated by using (4-5). Fig. 4-6(a) shows voltage dependence of the resonant frequency at $\ell$ = 0.3, 1, 3 um. The device with a high spring constant at $\ell$ = 0.3 um achieves the highest resonant frequency without bias, but its deflection is limited (1%), in turn, the frequency tuning range is narrow (20% of intrinsic resonant frequency $f_0$). At $\ell$ = 1 and 3 um, the tuning range can achieve around 300% and 400% of $f_0$ respectively.

From (4-5), the deflection induced strain is proportional to $1/\ell^2$, so beam length $\ell$ strongly affects the resonant frequency range. The predicted resonant frequency in the case of $\ell$ = 1 um shows excellent agreement with measured data in [12], which demonstrates how the hyperbolic model is used to guide the design and optimization of NEMS devices.

The influence of the initial strain on resonant frequency is analyzed in Fig. 4-6(b). The NEMS device geometry is the same as above and the beam length is 1 um. It can be seen that reducing $\varepsilon_{ini}$ from $4\times10^{-5}$ to $4\times10^{-6}$ does not affects the resonant frequency at $V$ = 9 V, but increases the tuning range from 300% to 800% of $f_0$. When $\varepsilon_{ini}$ increase from $4\times10^{-5}$ to $4\times10^{-4}$, it limits the beam deflection and in turn limits the resonant frequency tune range to 16% of $f_0$. This analysis can predict the influence of $\varepsilon_{ini}$ variation on the resonant frequency and guide the NEMS devices design.

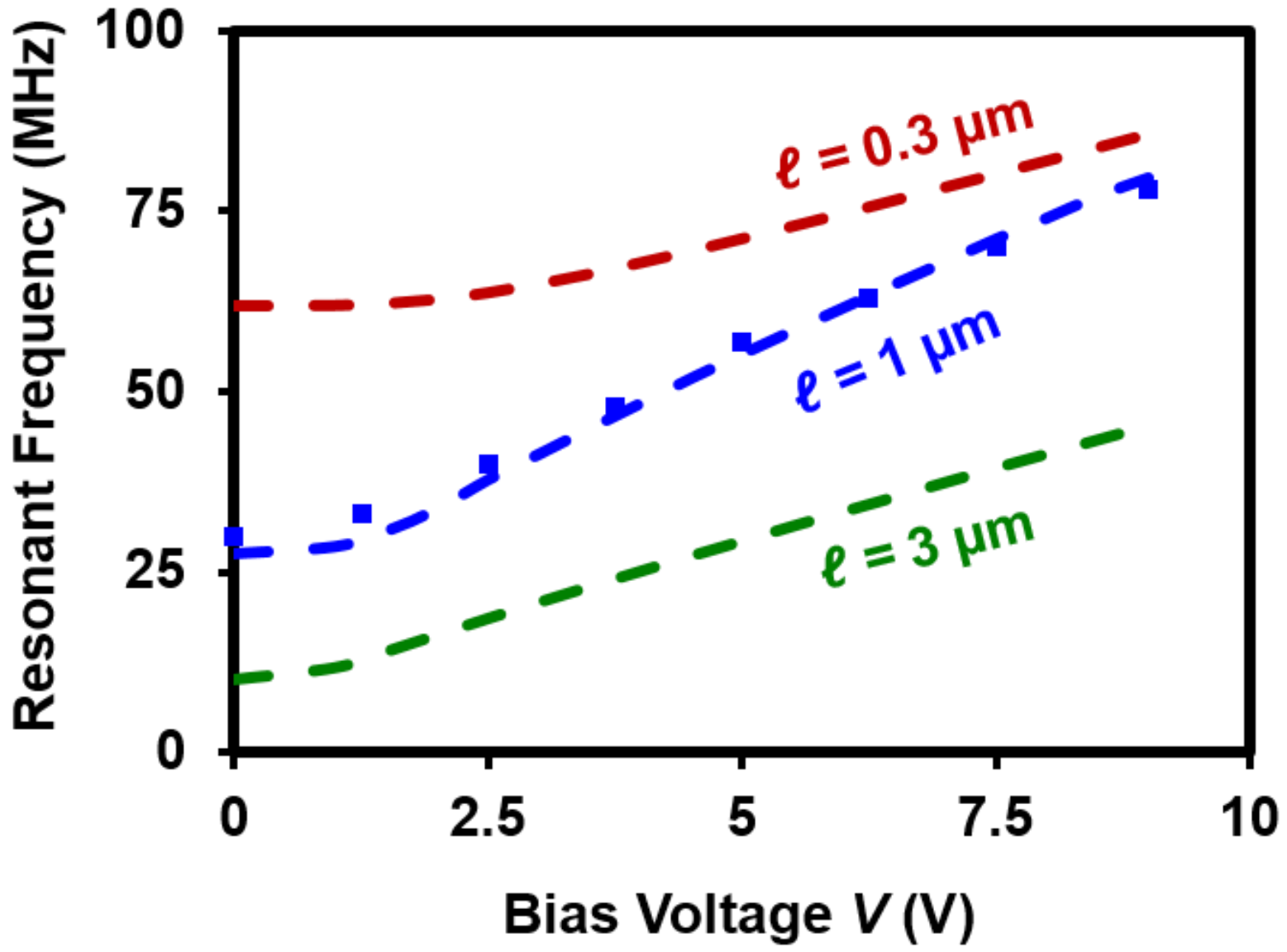

(a)

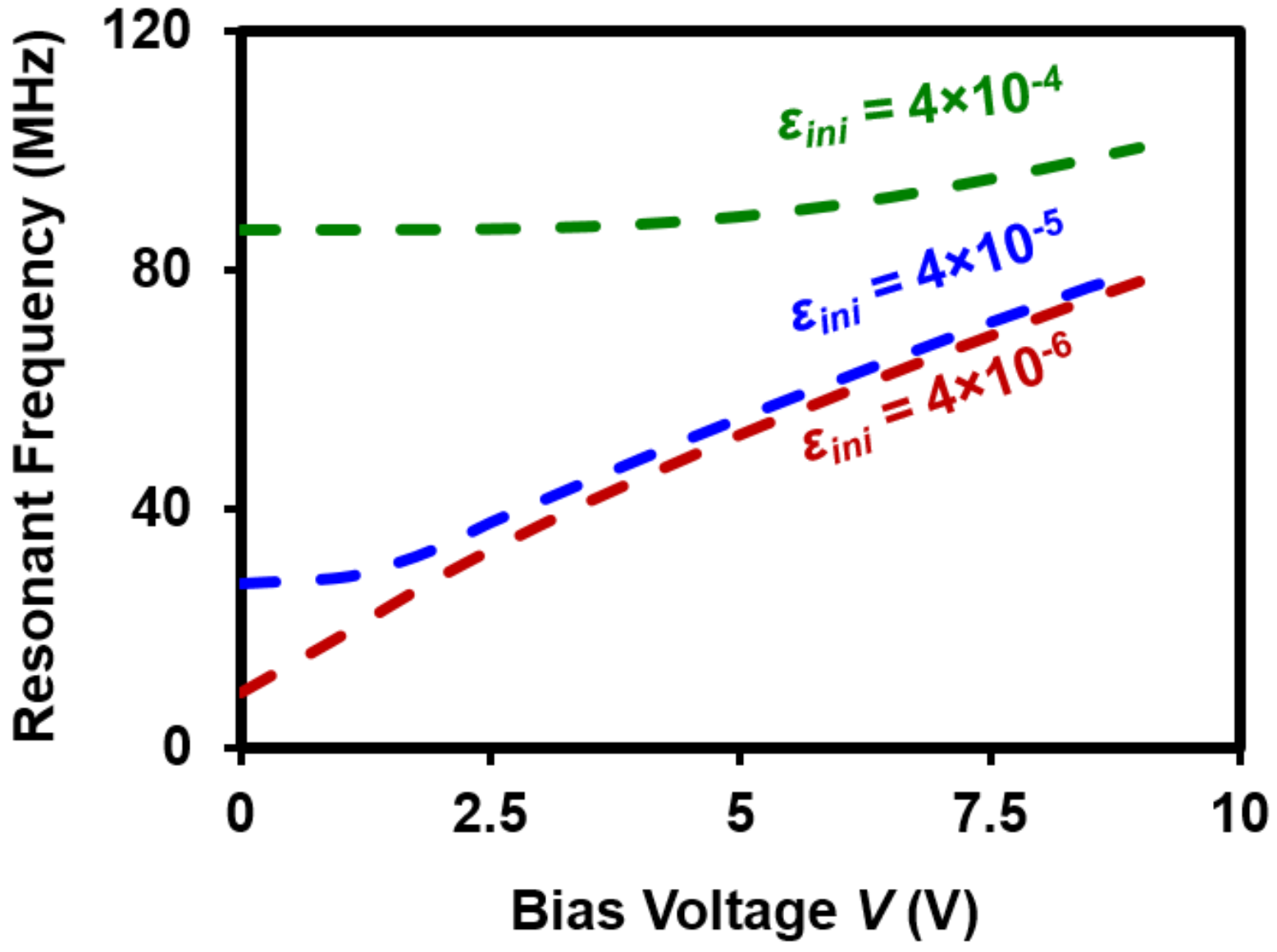

(b)

Fig. 4-6. (a) Resonant frequency for NEMS resonators with beam length $\ell$ = 0.3, 1, 3 μm. The symbols are experimental data from [12]. (b) Resonant frequency for NEMS resonators with initial strain $\varepsilon_{ini} = 4\times10^{-6}$, $4\times10^{-5}$, and $4\times10^{-4}$.

# References

[1] E.R. Deutsch, J.P. Bardhan, S.D. Senturia, G.B. Hocker, D.W. Youngner, M.B. Sinclair, and M.A. Butler, "A large-travel vertical planar actuator with improved stability," in *IEEE Int. Conf. on Solid-State Sensors, Actuators and Microsystems Tech. Dig.*, Jun. 2003, pp. 352–355.

[2] E. R. Deutsch, "Achieving Large Stable Vertical Displacement in Surface-Micromachined Microelectromechanical Systems (MEMS)," Ph.D. dissertation, Dept. Elect. Eng., MIT, Cambridge, MA, 2002.

[3] S. D. Senturia, *Microsystem Design*. Boston: Kluwer academic publishers, 2001, pp. 228–231.

[4] D. Molinero, C. Palego, X. Luo, Y. Ning, G. Ding, J. C. M Hwang, and C. L. Goldmisth, "Intermodulation distortion in MEMS capacitive switches under high RF power," in *IEEE MTT-S Int. Microw. Symp. Tech. Dig*., Jun. 2013, pp. 1–3.

[5] C. O'Mahony, M. Hill, R. Duane, and A. Mathewson, "Analysis of electromechanical boundary effects on the pull-in of micromachined fixed-fixed beams," *J. Micromech. Microeng.* vol. 13, no. 4, pp. S75–S80, Jun. 2003.

[6] Y.C. Hu, P.Z. Chang, and W.C. Chuang, "An approximate analytical solution to the pull-in voltage of a micro bridge with an elastic boundary," *J. Micromech. Microeng*., vol. 17, no. 9, pp. 1870–1876, Aug. 2007.

[7] A. Huang, V. T. S. Wong, and C-M. Ho, “Conductive silicone based MEMS sensor and actuator,” in *IEEE Int. Conf. on Solid-State Sensors, Actuators and Microsystems Tech. Dig.*, Jun. 2005, pp 1406–1411.

[8] C. Gómez-Navarro, M. Burghard, and K. Kern, “Elastic Properties of Chemically Derived Single Graphene Sheets,” *Nano Lett.*, vol. 8, no. 7, pp. 2045–2049, Jul. 2008.

[9] C. Chen and J. Hone, “Graphene nanoelectromechanical systems,” in *Proc. IEEE*, vol. 101, no. 7, pp. 1766–1779, Jul. 2013.

[10] S. M. Kim, E. B. Song, S. Lee, S. Seo, D. H. Seo, Y. Hwang, R. Candler, and K. L. Wang, “Suspended few-layer graphene beam electromechanical switch with abrupt on-off characteristics and minimal leakage current,” *Appl. Phys. Lett.*, vol. 99, no. 2, pp. 023103-1–023103-3, Jul. 2011.

[11] S. Sapmaz, Y. M. Blanter, L. Gurevich, and H. S. J. van der Zant, “Carbon nanotubes as nanoelectromechanical systems,” *Phys. Rev. B*, vol. 67, no. 23, pp. 235414-1–235414-7, Jun. 2003.

[12] C. Chen, S. Rosenblatt, K. I. Bolotin, W. Kalb, P. Kim, I. Kymissis, H. L. Stormer, T. F. Heinz, and J. Hone, “Performance of monolayer graphene nanomechanical resonators with electrical readout,” *Nat. Nanotechnol.*, vol. 4, no. 12, pp. 861–867, Dec. 2009.

# Chapter 5 Conclusions

## *5.1 Conclusions of This Dissertation*

This dissertation studies critical topics associated with RF MEMS capacitive switches, including the instability at the pull-in voltage; the switches' deformation characteristics when subjected to an electrostatic force; nonlinear stretching effect, and the capacitance calculation at small scale length scales [1]–[2]. Specifically, the accuracy of parallel-plate theory for calculating the pull-in voltage and capacitance is investigated. The study shows that applying the average displacement, rather than maximum displacement, in parallel-plate theory, results in better accuracy. This improvement increases with increasing bottom stationary electrode to moveable electrode ratio. Accuracy improves by 50% when this ratio is equal to 1. Besides the average displacement, the nonlinear stretching effect and empirical linear correction coefficients are also added to the parallel-plate model, to extend the range of the model's validity. In order to improve the life time of RF MEMS capacitive switches, a relationship between switches' geometry and membrane strain is derived. This relationship is used to avoid operating the switch beyond the elastic region.

The hyperbolic model, which can represent the deflected beam profile, is used to calculate the MEMS capacitance accurately. This is an improvement because it does not use parallel-plate assumption. By incorporating nonlinear stretching effects, this model can accurately predict the pull-in voltage and the beam's maximum stable travel range. The hyperbolic model works best for typical MEMS capacitive switches, where residual stress is dominant and NEMS devices where stretching is dominant. By comparison with the experimental data from MEMS capacitive switches and a graphene NEMS resonator, the model demonstrates that it is used to guide the design and optimization of both RF MEMS capacitive switches and NEMS devices.

## *5.2 Recommendation for Future Study*

In Chapter 3, although the 2D hyperbolic model demonstrates excellent agreement with experimental results, its two dimensional feature has many limitations when compared with that of 3D real devices. When a RF MEMS switch is considered as a 3D device, i.e. a plate, the anchor boundary condition, the residual stress distribution, are different from a 2D device and in turn the deformation shape cannot be present as a simple 2D hyperbolic function [3]. Therefore, the switch capacitance as a function of bias will change as well. Moreover, the fringe capacitance of a 3D device strongly depends on the geometry and beam deformation, so it is more challenging to predict the overall capacitance. The derivation of 2D-hyperbolic-model-estimated performance from a 3D device performance needs to be investigated by using FEM tools and experiment so that the validity range of the 2D hyperbolic model could be defined clearly.

As discussed in Chapter 3, a 2D hyperbolic function cannot represent beam deflection shape accurately if bending components are significant. However, similar to Fourier series, which is composed of infinite series of trigonometric functions, a function composed of series of hyperbolic functions may represent beam shape more accurately. This approach adds complexity to the 2D hyperbolic model, but it also extends the validity range of the model.

Besides understanding the capacitance as a function of bias for a unactuated switch, their relationship of an actuated switch is also an interesting topic. The nonlinear relationship between capacitance and applied voltage causes intermodulation distortion when RF signal passes through actuated switches. The RF MEMS switch is proven to be a highly linear device in the unactuated state, but the nonlinearity in the actuated state degrades the overall performance. For devices like RF MEMS phase shifters, which are built by RF MEMS switches, are susceptible to intermodulation distortion in both the unactuated and actuated state. When the switch is actuated, the entire beam is considered to be composed of numerous mini-beams, separated by asperities on the dielectric surface. Whether the hyperbolic model is valid in this situation, detailed study and control experiments are needed.

# *References*

[1] Wan-Chun Chuang, Hsin-Li Lee, Pei-Zen Chang and Yuh-Chung Hu, "Review on the modeling of electrostatic MEMS," *Sensors*, vol. 10, no. 6, pp. 6149–6171, Jun. 2010.

[2] R. C. Batra, M. Porfiri, and D. Spinello, "Review of modeling electrostatically actuated microelectromechanical systems," *Smart Mater*. *Struct*., vol. 16, no. 6, pp. 23–31, Oct. 2007.

[3] J. Lekner, “Electrostatics of hyperbolic conductors,” *Eur. J. Phys*., vol. 25, no. 6, pp. 737–744, Nov. 2004.

# Publications

[1] X. Luo, K. Xiong, J. C. M. Hwang, Y. Du, and P. D. Ye, “Continuous-wave and Transient Characteristics of Phosphorene Microwave Transistors,” in *IEEE MTT-S Int. Microwave Symp. Dig.*, May 2016, *Accepted for publication*.

[2] Y. Ning, X. Ma, C. R. Multari, X. Luo, V. Gholizadeh, C. Palego, X. Cheng, and J. C. M. Hwang, “Improved broadband electrical detection of individual biological cells,” in *IEEE MTT-S Int. Microw. Symp. Dig.*, Phoenix, Arizona, May 2015, pp. 1–3.

[3] X. Ma, X. Du, C. R. Multari, Y. Ning, C. Palego, X. Luo, V. Gholizadeh, X. Cheng, J. C. M. Hwang. "Broadband single-cell detection with a coplanar series gap." in *Microwave Measurement Conference Dig.*, 2015.

[4] K. Xiong, X. Luo, and J. C. M. Hwang, “Phosphorene FETs: Promising transistors based on a few layers of phosphorus atoms,” in *IEEE MTT-S IMWS-AMP Dig.*, Suzhou, China, Jul., 2015, pp. 1–3.

[5] V. Gholizadeh Y. Ning, X. Luo, C. Palego, J. C. M. Hwang, C. L. Goldsmith, "Improved compact, wideband, low-dispersion, metamaterial-based MEMS phase shifters." in *IEEE Wireless Symposium (IWS)*, 2015.

[6] Y. Ning, C. Multari, X. Luo, C. Palego, X. Cheng, J. C. M. Hwang, A. Denzi, C. Merla, F. Apollonio, M. Liberti. "Broadband electrical detection of individual biological cells." *IEEE Transactions on Microwave Theory and Techniques*, pp. 1905–1911, 2014.

[7] X. Luo, Y. Rahbarihagh, J. C. M. Hwang, H. Liu, Y. Du, P. D. Ye, "Temporal and thermal stability of Al 2 O 3-passivated phosphorene MOSFETs." *IEEE Electron Device Letters*, pp. 1314–1316, 2014.

[8] C. Palego, Y. Ning, V. Gholizadeh, X. Luo, J. C. M. Hwang, C. L. Goldsmith, "Compact, wideband, low-dispersion, metamaterial-based MEMS phase shifters" in *IEEE MTT-S Int. Microwave Symp. Dig.*, 2014.

[9] X. Luo, Y. Ning, D. Molinero, C. Palego, J. C. M. Hwang, C. L. Goldsmith, "Intermodulation distortion of actuated MEMS capacitive switches." in *Microwave Measurement Conference Dig.*, 2013.

[10] C. Palego, C. Merla, Y. Ning, C. R. Multari, X. Cheng, D. G. Molinero, G. Ding, X. Luo, J. C. M. Hwang, "Broadband microchamber for electrical detection of live and dead biological cells." in *IEEE MTT-S Int. Microwave Symp. Dig.*, 2013.

[11] Y. Ning, C. Multari, X. Luo, C. Palego, D. Molinero, X. Cheng, J. C. M. Hwang, C. Merla, "Coplanar stripline microchamber for electrical detection of live and dead biological cells." in *IEEE Microwave Conference (EuMC) Dig.*, 2013.

[12] Y. Ning, C. R. Multari, X. Luo, C. Merla, C. Palego, X. Cheng, J. C. M. Hwang, "Fast, compact and label-free electrical detection of live and dead single cells." in *Microwave Workshop Series on RF and Wireless Technologies for Biomedical and Healthcare Applications (IMWS-BIO)*, 2013.

[13] D. Molinero, X. Luo, C. Shen, C. Palego, J. C. M. Hwang, C. L. Goldsmith, "Long-term RF burn-in effects on dielectric charging of MEMS capacitive switches." *IEEE Transactions on Device and Materials Reliability*, pp. 310–315, 2013 .

[14] D. Molinero, C. Palego, X. Luo, Y. Ning, G. Ding, J. C. M. Hwang, C. L. Goldmisth, "Intermodulation distortion in MEMS capacitive switches under high RF power." in *IEEE MTT-S Int. Microwave Symp. Dig.*, 2013.

[15] D. Molinero, C. Palego, X. Luo, J. C. M. Hwang, C. L. Goldsmith, "RF burn-in of dielectric-charging characteristics of micro-electromechanical capacitive switches." in *IEEE MTT-S Int. Microwave Symp. Dig.*, 2012.

[16] C. Palego, D. Molinero, Y. Ning, X. Luo, J. C. M. Hwang, C. L. Goldsmith, "Pull-in and release transients of MEMS capacitive switches under high RF power."in *IEEE. Microwave Integrated Circuits Conference (EuMIC) Dig.*, 2012.

[17] X. Luo, S. Halder, J. C. M. Hwang, "Rugged HBT Class-C power amplifiers with base-emitter clamping." in *IEEE MTT-S Int. Microwave Symp. Dig.*, 2011.

[18] X. Luo, S. Halder, W. R. Curtice, J. C. M. Hwang, K. D. Chabak, D. E. Walker, A. M. Dabiran, "Scaling and high-frequency performance of AlN/GaN HEMTs." in *IEEE MTT-S Radio-Frequency Integration Technology (RFIT) Dig.*, 2011.

[19] D. Molinero, C. Palego, S. Halder, X. Luo, A. Hallden-Abberton, J. C. M. Hwang, C. L. Goldsmith, "Acceleration of dielectric charging/discharging by RF power in microelectromechanical capacitive switches." in *IEEE MTT-S Int. Microwave Symp. Dig.*, 2011.

# Vita

Xi Luo was born on December 10, 1983 in Yingcheng, Hubei, China. He received the B.S. degree in electronic science and technology from the Huazhong University of Science and Technology, Wuhan, China, in 2006, the M.S. degree in microelectronics and solid-state electronics from the Hebei Semiconductor Research Institute, Shijazhuang, China, in 2009, and is currently working toward the Ph.D. degree in electrical engineering at Lehigh University, Bethlehem, PA, USA.

In summer 2011, he was an Intern with RF Micro Devices Inc., where he was involved with HBT linearity characterization and compact modeling. His research interest includes HEMTs, HBTs, microelectromechanicalsystems (MEMS), and other microwave devices and circuits.